\documentclass[twocolumn,traditabstract]{aa} % for the abstract without structuration 
\usepackage{graphicx}
\usepackage{psfrag}
\usepackage{amsmath}
\usepackage{amssymb}
\usepackage{color,xcolor}
\usepackage{hyperref}
\usepackage{caption,natbib}
\hypersetup{
    colorlinks=true,
    citecolor=blue,
    filecolor=black,
    linkcolor=blue,
    urlcolor=magenta,
    linktocpage=true,
    breaklinks = true
}

\usepackage{enumitem}
\usepackage{subfigure,float}
\usepackage{placeins}
\usepackage{multirow}
\usepackage{makecell}
\usepackage[normalem]{ulem}

\usepackage{array,booktabs}
\usepackage{siunitx}
\DeclareUnicodeCharacter{2212}{\textendash}

\newcommand{\beq}{\begin{equation}}
\newcommand{\eeq}{\end{equation}}

\newcommand{\ber}{\begin{eqnarray}}
\newcommand{\eer}{\end{eqnarray}}

\newcommand{\ba}{\begin{align}}
\newcommand{\ea}{\end{align}}

\def \lcdm {$\Lambda$CDM~}

\def \dt{\Delta t}

\def \P {\mathcal{P}}

\def \lstm {{\tt LSTM}}
\def \convlstm {{\tt ConvLSTM}}
\def \convlstmd {{\tt ConvLSTM2D}}

\def \Ne {N_\text{obs,m-b}}
\def \thE {\theta_\text{E}}

\begin{document}

   \title{HOLISMOKES XVII: Detecting strongly lensed SNe Ia from time series of multi-band LSST-like imaging data}

\author{Satadru Bag\inst{1,2}\thanks{satadru.bag@tum.de}, Raoul Ca{\~n}ameras\inst{3,2,1}, Sherry H.~Suyu\inst{1,2}, Stefan Schuldt\inst{4,5}, Stefan Taubenberger \inst{1,2}, Irham Taufik Andika\inst{1,2}, Alejandra Melo\inst{1,2}
          }

   \institute{Technical University of Munich, TUM School of Natural Sciences, Physics Department,  James-Franck-Stra{\ss}e 1, 85748 Garching, Germany
         \and 
           Max-Planck-Institut f{\"u}r Astrophysik, Karl-Schwarzschild Stra{\ss}e 1, 85748 Garching, Germany
           \and
          Aix Marseille Univ, CNRS, CNES, LAM, Marseille, France
           \and
          Dipartimento di Fisica, Universit\`a  degli Studi di Milano, via Celoria 16, I-20133 Milano, Italy
          \and
          INAF - IASF Milano, via A. Corti 12, I-20133 Milano, Italy
}

%   \date{Received --; accepted --}

% \abstract{}{}{}{}{} 
% 5 {} token are mandatory

  \abstract
{
Strong gravitationally lensed supernovae (LSNe), though rare, are exceptionally valuable probes for cosmology and astrophysics. Upcoming time-domain surveys like the Vera Rubin Observatory's Legacy Survey of Space and Time (LSST) offer a major opportunity to discover them in large numbers. Early identification is crucial for timely follow-up observations. We develop a deep learning pipeline to detect LSNe using multi-band, multi-epoch image cutouts. Our model is based on a 2D convolutional long short-term memory (\texttt{ConvLSTM2D}) architecture, designed to capture both spatial and temporal correlations in time-series imaging data. Predictions are made after each observation in the time series, with accuracy expected to improve progressively as additional data are processed. We train the model on realistic simulations derived from Hyper Suprime-Cam (HSC) data, which closely matches LSST in depth and filter characteristics. In this work, we focus exclusively on Type Ia supernovae (SNe Ia). LSNe Ia are injected onto HSC luminous red galaxies (LRGs) at various phases of evolution to create positive examples of LSNe Ia time series. Negative examples include variable sources observed in HSC Transient Survey (including unclassified transients), and simulated unlensed SNe Ia in LRG and spiral galaxies. Our multi-band model shows rapid classification performance improvements during the initial few observations and quickly reaches high detection efficiency: at a fixed false-positive rate (FPR) of $0.01\%$, the true-positive rate (TPR) reaches $\gtrsim 60\%$ by the 7th observation and exceeds $\gtrsim 70\%$ by the 9th observation. If we relax the FPR to $0.1\%$, the TPR reaches close to $60\%$ as early as the 4th observation. Although single-band analysis performs reasonably well in isolation, the multi-band model significantly outperforms it, particularly in the early stages, by building a richer memory and leveraging color information. Among the negative examples, SNe in LRGs remain the primary source of FPR, as they can resemble their lensed counterparts under certain conditions. Additionally, the model detects quads more effectively than doubles, and performs better on systems with larger image separations. Although we train and test the model on HSC-like data, our approach is applicable to any cadenced imaging survey -- particularly LSST, where the higher expected cadence (5–10 times that of HSC) should further boost performance.

}
   {}
   {}
   {}
   {}


   \titlerunning{Detecting lensed SNe Ia in time series of multi-band LSST-like imaging data}
   \authorrunning{Bag et al}

   \maketitle

\section{Introduction}

Strong gravitational lensing occurs when the gravitational potential of a massive galaxy or galaxy cluster bends light from background sources and produces multiple magnified images including elongation of extended sources like another galaxy (see, e.g., \cite{1996astro.ph..6001N} for a pedagogical introduction). Acting as a cosmic telescope, this effect enables detailed studies of the initial mass function (IMF), galaxy evolution, and the nature of dark matter \citep{Mao:1997ek, Metcalf:2001ap, Dalal2002, Pooley_2009, Oguri2014, Jim_nez_Vicente_2019,shajib2024,vegetti2023}. Among its most significant applications is time-delay cosmography (TDC) \citep{Refsdal1964_1, Refsdal1964_2, Saha, Oguri_2007, 2016A&ARv..24...11T, Bonvin2017, Wong:2019kwg, 2020A&A...643A.165B, Birrer:2020jyr, Treu:2022aqp, Suyu_2024, TDCOSMO2025}, which uses time delays between multiple images of variable sources to estimate cosmological distance ratios. These measurements directly constrain the Hubble constant ($H_0$), independent of local distance ladder methods -- thus addressing the on-going $H_0$ tension \citep{Planck:2018vyg, Riess2022,valentino2021,verde2024}. They also provide crucial insights into dark energy and spatial curvature when there are multiple strongly lensed sources at different redshifts \citep[e.g.,][]{Grillo2024} and when combined with other cosmological probes \citep[e.g.][]{Linder2011, Suyu2012, Jee2016}. Time-delay measurements require time-variable sources, such as lensed quasars (QSOs) and supernovae (SNe), though other transients like kilonovae (KNe), tidal disruption events (TDEs), gamma-ray bursts (GRBs), and fast radio bursts (FRBs) are increasingly gaining attention. For further applications of strong lensing, see the review by~\citet{Treu2010}.

Lensed supernovae (LSNe) and QSOs each offer distinct advantages for TDC. LSNe are extremely rare, with only a handful known to date \citep{Kelly:2014mwa,Goobar:2016uuf,2021NatAs.tmp..164R,chen2022,Goobar2023,Frye2023,sn_encore}, but they are expected to become pivotal in the coming decade \citep{Suyu_2024}. In contrast, lensed QSOs have been the mainstay of TDC due to their abundance.  However, despite extensive monitoring by COSMOGRAIL and TDCOSMO \citep{cosmograil1, Millon2020}, only a handful have enabled precise $H_0$ measurements. The H0LiCOW team measured $H_0$ with $2.4\%$ uncertainty using six lensed QSOs \citep{Wong:2019kwg} based on two families of mass models (a power-law model and a baryon-plus-dark-matter model), where individual lenses constrained $H_0$ at the $5$–$10\%$ level. This underscores the need for significantly larger samples, especially since the uncertainty increases when standard assumptions about the lens mass density profile are relaxed \citep{2020A&A...643A.165B}.  The recent TDCOSMO results from a sample of 8 lensed quasars with relaxed assumptions on mass density profile yield $\sim5$\% uncertainty on  $H_0$ \citep{TDCOSMO2025}. To achieve percent-level precision on $H_0$ through LSNe, independent of lens model assumptions, it is essential to discover a few tens to a hundred new LSNe, thereby building a gold sample of lens systems for cosmography \citep{Suyu20, Nikki_2023}.

LSNe also offer a powerful way for probing the nature of their progenitors. For instance, thermonuclear supernovae (Type Ia, SNe Ia hereafter) are believed to be originated from the disruption of a carbon–oxygen white dwarf in a binary system, but the precise nature of their progenitor systems remains a topic of active debate \citep[e.g.,][]{maoz14}. A major limitation to probe the nature of the exploding star and its potential companion is the lack of data sufficiently close to explosion and at sufficiently blue wavelengths. LSNe provide a unique opportunity to overcome this challenge, offering critical early-time observational access. Once the first image of a LSN is detected, mass modeling can be used to predict the time delays and the appearance of future images, enabling coordinated follow-up efforts to capture the SN evolution within the first hours or days.

The Legacy Survey of Space and Time (LSST) of the Vera C. Rubin Observatory \citep{lsst1,ivezic19}, along with other wide-field  surveys like the Euclid space mission \citep{Euclid2025} and the Roman Space Telescope \citep{roman2015}, is expected to significantly expand the sample of lensed transients. LSST alone is projected to detect hundreds of LSNe \citep{Ana_2023,Nikki_2023,Bag:2024kbk} and thousands of lensed QSOs \citep{om10} over its 10-year survey. %Additionally, ongoing surveys like the Euclid mission and the Zwicky Transient Facility (ZTF) are anticipated to uncover a few new systems. 
The scientific exploitation of these future promising samples of LSNe nonetheless requires a rapid identification, ideally before peak, to enable their mass modeling and follow-up on coordinated facilities. As an example, reaching the main goals of our Highly Optimized Lensing Investigations of Supernovae, Microlensing Objects, and Kinematics of Ellipticals and Spirals \citep[HOLISMOKES,][]{Suyu20} program requires, for instance, not only spectroscopic typing and high-cadence imaging for time-delay and $H_0$ measurements, but also rest-frame UV spectroscopy in the early phases for progenitor studies. The sheer volume of alerts from wide-field surveys (e.g. $\mathcal{O}(10^7)$ per night from LSST) makes automated LSNe identification essential, with {\em machine learning} (ML) as the most practical solution.

LSNe can be detected through multiple ways. The most straightforward method is based on image multiplicity, where one directly resolves multiple images of a LSN \citep{om10}. Another approach, particularly effective for SNe Ia due to their well-understood luminosity after standardization, relies on photometry. In this case, magnification-based selection is used to identify candidates whose apparent brightness exceeds what is expected at their redshifts, even when the angular resolution is insufficient to deblend multiple images \citep[e.g.,][]{goobar17,goldstein19,sagues24}. Additionally, color–magnitude diagrams offer another selection tool for SNe Ia, as LSNe Ia tend to appear redder due to their typically higher redshifts compared to their unlensed counterparts \citep{Quimby2014}. Recently, efforts have also been made to identify unresolved LSNe using the shape of their blended light curves \citep{Bag:2020pbg,misha2021,misha2022,Bag:2024kbk}.  

Another fundamentally different strategy is catalog cross-matching. In this approach, one assembles a large samples of static strong lenses and monitors them for transients appearing at the positions of multiple lensed images of background galaxies \citep[e.g.,][]{shu18,craig24}. In previous studies, we leveraged the capabilities of supervised CNNs in selecting static galaxy-scale strong lenses from deep, wide-field surveys \citep[e.g.,][]{petrillo17,jacobs19,metcalf19}, and developed automated methods to systematically identify these systems over large footprints, with minimal false-positive rates (FPR) \citep[e.g.,][]{canameras20,schuldt25,canameras24}. Applying such algorithms to LSNe detection requires the SN host galaxy to be detectable and sufficiently deblended from the foreground lens galaxy.

 Here, we aim to extend the discovery space of LSNe to include systems without detectable hosts, expected to constitute a significant fraction (e.g., around half of LSNe Ia observed by LSST \citep{Ana_2024}), and with smaller image separations, while pushing the detection as early as possible. Time-domain surveys like LSST (and Roman in the future) will produce multi-band 2D image sequences of transients, capturing both spatial features (e.g., multiple images of lensed transients and arcs from their lensed hosts) and temporal variations (following specific light curves).
 In this work, we develop a deep learning-based pipeline to detect LSNe directly from the alert stream of the upcoming time-domain surveys.  We specifically employ Convolutional Long Short-Term Memory (\texttt{ConvLSTM}) architecture, which capture spatial and temporal correlations simultaneously across sequences of imaging data. Unlike previous approaches that convolve reference images and apply {\tt LSTM}s to the corresponding light curves \citep{Morgan2022, Morgan2023}, we employ \texttt{ConvLSTM} directly to the full image time series, capturing spatio-temporal signatures such as centroid shifts and extended residuals to distinguish LSNe from other transients.

\citet{Ramanah2022} previously explored \texttt{ConvLSTM}-based architectures on fully simulated, single-band datasets, demonstrating that using the time series of images instead of a single-epoch image significantly improves the classification performance.
Our work introduces two key improvements. First, we extend the framework to multi-band imaging, leveraging richer color and temporal information for improved early detection. Second, instead of relying solely on synthetic data, we train and validate the model on realistic simulations based on actual Hyper Suprime-Cam (HSC) observations \citep{aihara18}, which closely match LSST in depth and filter characteristics. In particular, we simulate both lensed and unlensed/normal SNe, inject them into HSC images of galaxies (LRGs and spirals), and construct the resulting time series. We further incorporate real transients from the HSC Transient Survey \citep{yasuda19}, drawing on a catalog of over $100$k variable sources identified in the same footprint by \citet{chao21} using the difference imaging pipeline of \citet{chao20}. 
This strategy trains the model to identify LSNe early using only the first few observations, even amid diverse transients and realistic image artifacts. In this initial work, we focus on detecting LSNe Ia without host light, providing a proof-of-concept architecture. While our model is trained and evaluated on HSC-like data in this study, the methodology is general and transferable to any time-domain imaging survey. 
Its design is especially compatible with LSST, whose significantly higher expected cadence (5 to 10 times that of HSC) is expected to further enhance the performance. An immediate objective is to enable real-time identification from the LSST alert stream.

The paper is organized as follows. In the next section, we describe how the datasets are constructed and how our simulations of different components used for training are anchored in real HSC observations. Section \ref{sec:model} details the model architecture and data processing. Key results are presented in Section \ref{sec:results}. Section \ref{sec:conclusions} summarizes the main conclusions of this work, and discusses the assumptions made, current limitations, and outlines future directions, particularly with an eye toward application to LSST. 

Throughout this study, we use cutout sizes of $59\times59$ pixels corresponding to $\sim$ $10\arcsec\,\times\,10\arcsec$ physical size. We adopt the flat concordant $\Lambda$CDM cosmology with $\Omega_M=0.308$ and $\Omega_{\Lambda}=1−\Omega_{\rm M}$ \citep{planck16} and with $H_0=72$ \,km\,s$^{-1}$\,Mpc$^{-1}$ \citep{Bonvin2017}.

\section{Data Acquisition and Simulation Methodology}
\label{sec:data_sim}

\subsection{The HSC Transient Survey}
\label{sec:HSC_transients}
\begin{figure}
    \centering
    \includegraphics[width=\columnwidth]{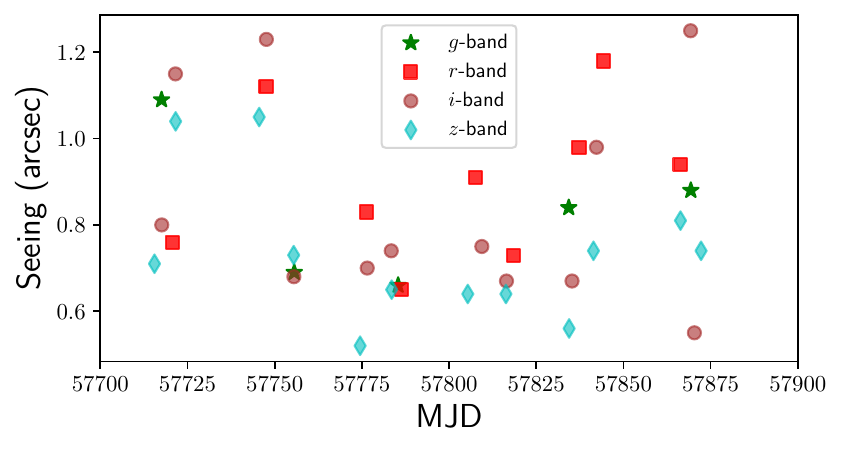}
    \caption{Observation epochs of the HSC Transient Survey in the COSMOS field across different filters during MJD 57710–57875 (Nov. 2016–Apr. 2017), used to extract image time series of variable sources (HSC variables) with highest available cadence. The number of observations in the $griz$ bands are 5, 9, 12, and 12, respectively. The observation details are taken from Table 1 of \citet{yasuda19}.}
    \label{fig:cosmos_epochs}
\end{figure}

The HSC Transient Survey \citep{yasuda19} is a cadenced survey conducted with the 8-m Subaru telescope, on the two ultra-deep fields of the HSC Subaru Strategic Program \citep[HSC-SSP,][]{aihara18}, namely COSMOS \citep{scoville07} and SXDS \citep[Subaru/XMM-Newton Deep Survey,][]{furusawa08}. In this study, we focus on the single pointing of 1.77 deg$^2$ over the COSMOS field, corresponding to tract 9813, which was observed between Nov. 2016 and Apr. 2017.

With exposure times of few 10$^3$\,sec per epoch, the HSC Transient Survey reaches median depth per epoch of 26.4, 26.3, 26.0, and 25.6 mag in $g$-, $r$-, $i$-, and $z$-bands, respectively \citep{yasuda19}. To artifically degrade the depth and get closer to LSST single-epoch image depth, we produced an alternative version of the data set based on a subset of exposures available per epoch. Images reduced and calibrated by the HSC pipeline \citep{bosch18}, divided into 9\,$\times$\,9 patches, and with a pixel size of 0.168\arcsec, were imported using the Data Archive System search form\footnote{ https://hsc-release.mtk.nao.ac.jp/das\_search/pdr3/index.html}.

We used these ``warped'' images, which are saved on a common sky grid prior to stacking, to produce shallower images per epoch. For simplicity, we restricted our data set to patches covered systematically at all epochs from the HSC Transient Survey, thereby excluding a few patches around the tract borders due to dithering. We also focused on exposures taken at MJD between 57710 and 57875 with homogeneous sets of filters. Three individual exposures of 300\,sec were stacked per epoch, in $griz$ bands. Pixels having flag values $>$1000 in any of the three exposures were masked out during stacking to avoid introducing artefacts near bright stars. The resulting frames were stitched together with SWarp \citep{bertin02} to produce the final single-epoch images over the complete footprint.

We obtained a total of $5$, $9$, $12$, and $12$ epochs of observation in $g$, $r$, $i$, and $z$ bands, respectively, over the six-month duration of the survey as shown in Figure~\ref{fig:cosmos_epochs}. As a result of using only three of the available exposures, larger regions between HSC CCDs are masked out compared to the images released in \citet{yasuda19}. Effective exposure times per pixel are variable due to the masking process.

\subsection{Positive examples: simulation of mock LSNe Ia time series}

\begin{figure}
    \centering
    \includegraphics[width=\columnwidth]{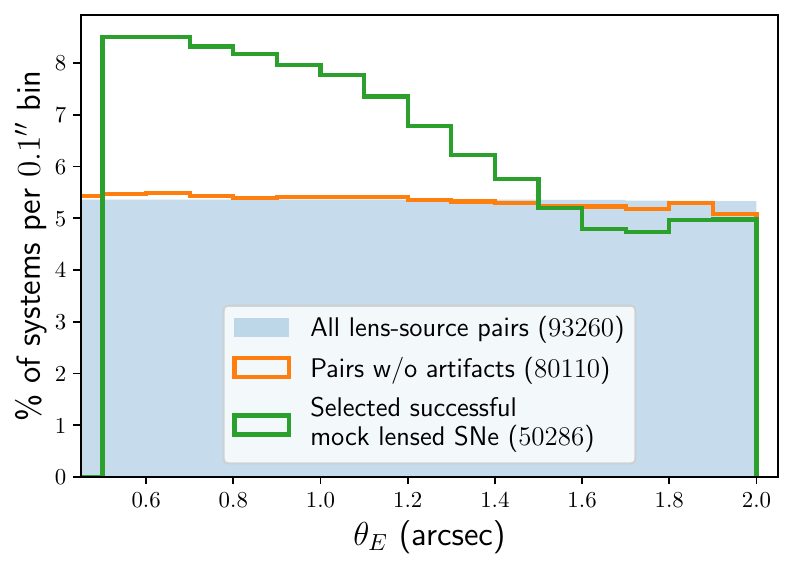}
    \caption{Einstein radius ($\thE$) distributions. The blue histogram shows the $\sim$93,000 lens–source pairs with a flat $\thE$ distribution between $0.1\arcsec$ and $2.0\arcsec$. After removing artifacts, $\sim$80,000 clean systems (orange) remain, still following a flat $\thE$ distribution. The green histogram shows the final $\sim$50,000 mock LSNe Ia passing our brightness selection and restricted to $\thE > 0.5\arcsec$, which delimits the region of interest in this study. Of these, $\sim$48,000 are used in the training, validation, and test sets with a balanced number of doubles and quads.}
    \label{fig:re_dist_CA}
\end{figure}

\begin{table*}[h]
    \centering
    \begin{tabular}{|c||c|c|c|}
        \hline
        Objects & from & \# available & \makecell{\# used in } \\ 
        \hline
        LRGs & \makecell{ wide layer\\ (co-added)} & \makecell{$\sim80,000$ (after augmentation) \\ ($griz$)} & \makecell{\textbf{LSNe Ia:} \\$\sim48,000$\\ \textbf{negatives:} \\ {SN in LRGs:} $\sim28,000$} \\ 
        \hline
        Blanks &\makecell{ HSC Transient Survey \\ in COSMOS field \\ (single epoch)} & \makecell{$g:~\sim1200$ (5 epochs) \\ $r:~\sim1200$ (9 epochs)\\ $i:~\sim1400$ (12 epochs)\\ $z:~\sim 1500$ (12 epochs) \\ \scriptsize see Figure \ref{fig:cosmos_epochs} for\\ \scriptsize the epoch details} & \makecell{all in \\ \textbf{LSNe Ia} \\ \& \\ \textbf{negatives}} \\ 
        \hline
        \makecell{HSC variables \\from \\ \cite{chao21}} & \makecell{ HSC Transient Survey \\ in COSMOS field \\ (single epoch)} & \makecell{ $\sim72,000$ (after augmentation) \\ with `good' light curves \\ ($griz$)} & \makecell{ \textbf{negatives:} \\ $\sim 28,000$} \\ 
        \hline
        Spirals & \makecell{ wide layer\\ (co-added)} & \makecell{$\sim40,000$ (unique) \\ ($griz$)} & \makecell{ \textbf{negatives:} \\ SN in spirals: $\sim 16,000$}  \\ 
        \hline
        
    \end{tabular}
    \caption{Ingredients from HSC observations.}
    \label{tab:sample_table}
\end{table*}

\begin{figure*}
    \centering
    \includegraphics[width=\textwidth]{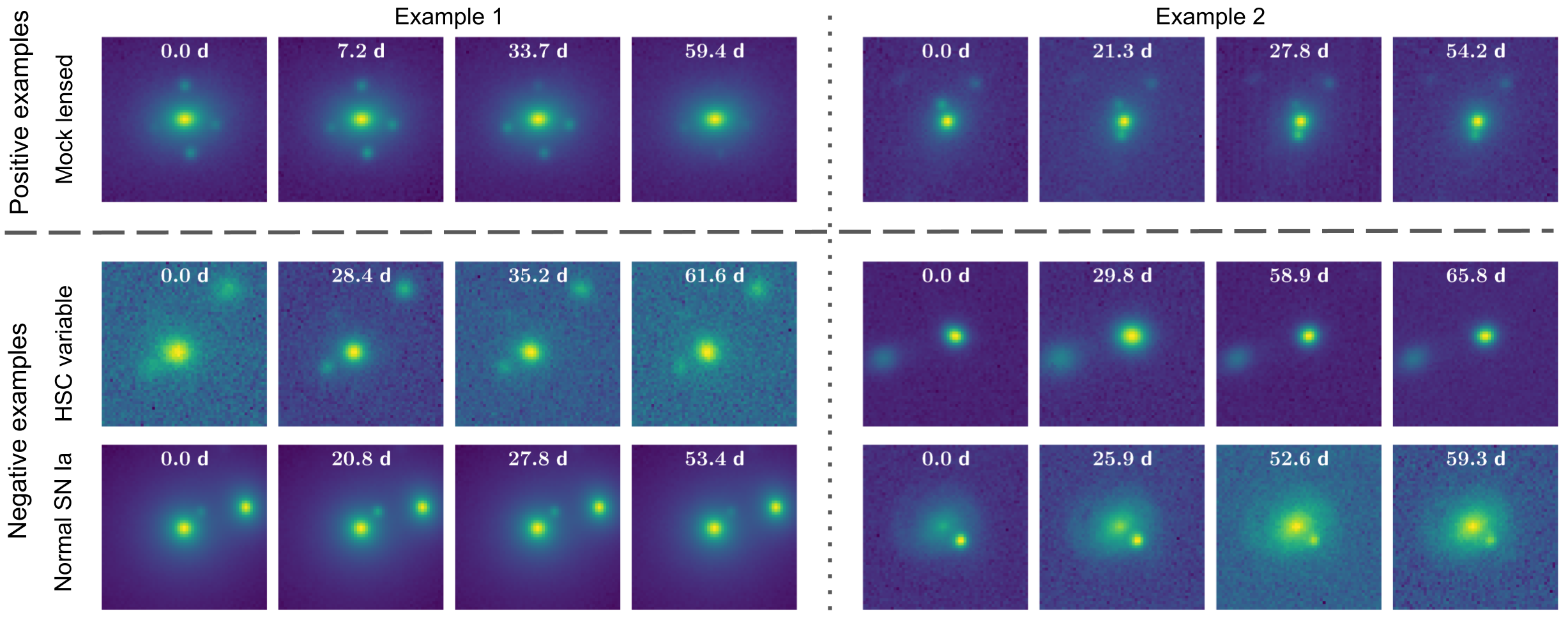}
    \caption{Example $i$-band time series of different components used for training: mock LSNe Ia (top row, forming the positive class), and variable sources and normal SNe Ia (middle and bottom rows, forming the negative class). Each row shows two representative samples. The top row includes a quadruply LSN Ia (left) and a doubly lensed one (right). The middle row shows two unrelated HSC variables, where the central objects exhibit variability over time. The bottom row shows an SN Ia in an LRG (left) and in a spiral galaxy (right). Note that, only the time series of the HSC variables in the middle row include PSF variation over time. Timestamps in each frame indicate days since the first detection, which is always set to zero. The classification task is binary: distinguishing LSNe Ia from all other types of transients and bogus detections. For illustration, we show only single-band ($i$-band) time series here. The multi-band time series corresponding to the quadruply LSN Ia shown in the top-left panel is presented in Figure~\ref{fig:multi-band_ts}.}
    \label{fig:images}
\end{figure*}

To simulate galaxy-scale LSNe, we inject point-like SN images into real HSC images of LRGs and place them in blank regions of the HSC Transient Survey. We use LRGs as foreground deflectors, given their high lensing cross sections \footnote{ We have not included spiral lenses, another class of galaxy deflector, for several reasons. They are not straightforward to model, for example via velocity dispersion. Compared to LRGs, they are typically less massive, thereby producing smaller image separations and magnifications, and are intrinsically rare. For these reasons, including them in the training set may not be worthwhile and could even increase false positives. Even without explicit training on spiral lenses, the model is expected to identify some SNe lensed by spirals using other indicators, such as lensing-like configurations, color, and temporal evolution. \label{fn:two}}. In this work, we focus on Type Ia SNe, as their intrinsic luminosities and light curves are well understood and straightforward to model. Extending the pipeline to other SN types and transients is conceptually straightforward and will be pursued in later phases by incorporating the appropriate light curve models during injection.

We adopt the redshift distributions of lenses and sources based on the Oguri \& Marshall (2010, hereafter OM10) simulation framework \citep{om10, Oguri2018}. First, we ensure that the redshift distribution of our selected LRG sample, used as lenses, broadly aligns with the lens redshift distribution for LSNe Ia in OM10. Instead of following the realistic Einstein radius distribution as in OM10, we adopt a flat distribution between $0.1\arcsec$ and $2.0\arcsec$ to ensure a more balanced training set. Based on these choices, we construct a lens–source matched catalog by assigning a source redshift to each lens, which determines at what redshift the SN Ia will later be injected. In Figure \ref{fig:re_dist_CA}, we show the Einstein radius ($\thE$) distribution of the lens–source catalog before and after removing entries with artifacts in the LRG cutouts, represented by the blue and orange histograms, respectively. Both distributions follow the intended flat trend across the $\thE$ range. In this initial study, we focus on Einstein radii of $0.5\arcsec \leq \thE \leq 2.0\arcsec$, thereby excluding low-separation systems below the LSST’s resolution limit. The range shown in the figure reflects this selection. Note that restricting to $\theta_E \geq 0.5\arcsec$ reduces the detectable LSNe Ia rates in LSST only modestly, by a factor of $\sim 2–3$ depending on the detection criteria \citep{Wojtak2019,Nikki_2023,Bag:2024kbk}.

The next step involves determining the optimal lensing configuration (image positions $\boldsymbol{\theta}$, time delays $\dt$, and magnifications $\mu$) for each lens-source pair. We then inject the SN Ia images at different phases of the explosion, placing the SN at the assigned source redshift and applying the lensing configuration accordingly to generate a time series capturing its evolution. The top panels of Figure \ref{fig:images} show time series of two examples: a quadruply lensed system on the left and a doubly lensed system on the right. More examples, systematically arranged by different $\thE$, are shown in Figure~\ref{fig:mock_lensed_examples} in Appendix~\ref{app:mock_sim}.

A detailed description of the simulation process is provided in Appendix \ref{app:mock_sim}. A few key points:
\begin{itemize} 

\item In rare cases, we fail to find a viable lensing configuration within a reasonable number of trials that produces sufficiently bright images for detection; such lens-source pairs are discarded. 

\item Here, we also restrict ourselves to mock LSNe Ia with sufficient image brightness for detection, exceeding $5\sigma_{\rm B}$, where $\sigma_{\rm B}$ is the standard deviation of the background noise. This selection criterion,  combined with the previously mentioned cut on angular separation ($0.5\arcsec \leq \thE \leq 2.0\arcsec$), yields a total of approximately 50,000 successful mock LSNe Ia, whose $\thE$ distribution is shown by the green histogram in Figure \ref{fig:re_dist_CA}. Since this sample contains a slight mismatch in the number of doubles and quads, we include a subset of $\sim$48,000 systems with equal representation in our final datasets, as shown in Table~\ref{tab:sample_table}.
Note that systems with higher $\thE$ tend to correspond to sources at higher redshifts, which are more likely to fail the brightness threshold and hence are slightly underrepresented in the selected sample. This explains the decline in the higher $\thE$ bins. However, this bias is not necessarily unfavorable. As we will discuss in the results section, systems with smaller $\thE$ are intrinsically harder to detect, so their slight over-representation may help the model learn to better identify these challenging cases.

\item We do not include the SN host galaxy at this stage, regardless of whether it would be lensed or detectable. This reflects a significant fraction of cases (e.g., roughly half of LSST cases) where the host galaxy of an LSNe Ia is either not strongly lensed or too faint to be observed in the surveys. In future work, we will systematically evaluate the impact of detectable lensed hosts by examining different SN-to-host center offset distributions and detectability criteria.
\end{itemize}

\subsection{Adding blanks to the cutouts at different epochs}
\label{sec:adding_blanks}
For the synthetic time series (i.e., for mock LSNe and normal unlensed SNe exploding in galaxies, explained later in this section), the images vary over time due to the injected LSN images or the SN itself. The outer pixels remain unchanged, leading to identical noise across epochs. To address this, we add white noise (based on the sensitivity of the telescope) at each epoch. For this, we also collect single-epoch cutouts (referred to as ``blanks" in Table \ref{tab:sample_table}) from the empty regions in the HSC COSMOS field at different epochs, separately for the four bands of interest. These blanks are added to all epochs of the time series for all components, including the fully observed time series of HSC variables, to ensure consistent noise characteristics across the entire dataset.

In reality, there would be Poisson noise associated with the measurements.
In our case, the added white noise at each time step is of the order of Poisson noise near the bright central part and dominates at the outskirts. Therefore, for simplicity, we do not add Poisson noise to the coadded layer, and leave this for future work.

Since the standard deviation of the sum of $N$ normal distributions is $\sigma_{\rm sum} = \sqrt{\Sigma_{i}^{N} \sigma_i^2}$, by adding the blanks at different epochs, we reduce the SNR by a factor of $\sqrt{2}$. This reduction is unavoidable if we want to ensure that the noise in the synthetic time series (of lensed and normal SNe) varies across different epochs without separating the signal from the noise at the pixel level. 
We stress on the fact that adding blanks is important because, without this step, the cutouts would have identical noise pattern in the outskirts at different time steps, which is unphysical and could cause the model to learn from this repeated noise rather than solely from the underlying signal.

We consider different types of objects as the negative examples. These components are described in the next two sections. 

\subsection{Negative examples: variable sources observed in the HSC Transient Survey}

\citet{chao21} compiled a catalog of 101,353 variable sources based on multi-band photometry from the HSC Transient Survey to run their lensed QSO search algorithm. These sources, referred to as `HSC variables' hereafter, are unclassified and may include a diverse mix of real transients (e.g., variable stars, SNe, GRBs, KNe, TDEs, and rare lensed events) as well as bogus or non-astrophysical detections. We aim to put these HSC variables into our negative samples. However, as we need multiple epochs from HSC Transient Survey, not all of them could be used. We collected all the variables that are present in our HSC warps.  Figure \ref{fig:cosmos_epochs} shows the HSC Transient Survey observation dates in different bands. Details of the extraction process are provided in Section \ref{sec:HSC_transients}. The key points are summarized below:
\begin{itemize}
    \item We concentrate on HSC variables from the 2017 window, where the cadence is most favorable, though still not ideal. This yields a maximum of $5$, $9$, $12$, and $12$ usable epochs in the $g$, $r$, $i$, and $z$ bands, respectively. However, not all HSC variables have good-quality cutouts in every epoch. We retain only those with sufficiently dense sampling, i.e., ``good cadence", as defined in Appendix \ref{app:cadence_matching}.

    \item After applying these selection criteria, we obtain approximately $9,000$ HSC variables with good multi-band cadence.  To further increase the dataset size, we apply data augmentation by performing four $90^\circ$ rotations and two mirror flips for each rotated instance, resulting in an $8\times$ augmentation per original sample. This gives us a final dataset of roughly $72,000$ time series for the HSC variables, sufficient for our training, validation, and test sets.
\end{itemize}
As noted in Section~\ref{sec:adding_blanks}, we also add blanks (primarily white noise) to all epochs of the HSC variable time series to maintain consistency across all components. The $i$-band time series of two unclassified HSC variables are shown in the middle row of Figure~\ref{fig:images} as examples, where the centrally located objects exhibit variability over time. 

\subsection{Negative examples: simulated normal SNe Ia exploding in LRGs and Spirals}
Since the HSC variables discussed above are unlabelled, i.e. they are not further classified into specific subcategories, they could potentially include lensed and unlensed (normal) SNe, among other types of transients. However, one of the most significant sources of confusion in distinguishing LSNe Ia arises from normal SNe. To improve the model's training and enhance its ability to differentiate these cases, we increase the number of normal SNe in the negative samples by explicitly incorporating systems where an SN explodes in LRGs and spiral galaxies.

\subsubsection{SNe Ia in LRGs}
We have approximately 28,000 LRGs that are not used in creating the mock LSNe. We use these as hosts for unlensed SN Ia. To place the supernovae on the galaxy cutouts, we require: (i) the galaxy redshifts, which are also used as the supernova redshifts, (ii) point spread functions (PSFs), (iii) the distance from the galactic center, and (iv) the corresponding light curve. Since these LRGs are already cross-matched with the SDSS catalog, we have their spectroscopic redshifts. We also have their HSC PSFs, which will be used to paint the point-like supernovae onto the galaxy cutouts. For simplicity, we use the Hsiao template \citep{Hsiao2007} to vary the supernova brightness over time for a given redshift and filter. Regarding the distance from the galactic center, we distribute the supernovae uniformly within two times the Kron radii of the LRGs. We adopt a uniform distribution in SN-host center offset to ensure equal representation across the training set, analogous to the uniform sampling in Einstein radius ($\thE$) used for the mock LSNe samples, while spanning the range of realistic SN locations within host galaxies.\footnote{We sample SN-host center offsets uniformly to enable the model to recognize SNe across all offsets without introducing bias. While astrophysically motivated priors based on stellar or light density profiles exist \citep{Lokken_2023}, their exact dependence on galaxy type, redshift, and other factors is not fully established. Moreover, our primary goal at this stage is to demonstrate lensed SN detection among all alerts, rather than to optimize classification between lensed and normal SNe. In the final pipeline, however, significant false alarms are expected to arise from cases where the system is a genuine strong lens but the transient occurs in the foreground lens galaxy rather than in the background lensed host. Mitigating this will require fine-tuning with more realistic offset distributions.} Additionally, we clip the distance at $2\arcsec$ to maintain consistency with the mock LSNe samples, where the maximum Einstein radius ($\thE$) is $2\arcsec$, and to avoid placing supernovae too far from the galactic center. Just like the LSNe Ia, we generate the time series of cutouts at different phases of the SN Ia evolution. 

\subsubsection{SNe Ia in spirals}
We also collect around 40,000 spiral galaxies from the HSC wide layer, which are used exclusively to host SN Ia. We follow the same procedure as described above for the LRGs, with a key difference: we use photometric redshifts instead of spectroscopic redshifts (as the latter is not available for all spirals). 

As explained in Section~\ref{sec:adding_blanks} above, we add blanks (primarily white noise) to the images across epochs to introduce realistic noise variation over time for both of SN in LRGs and spiral galaxies. The bottom row of Figure~\ref{fig:images} shows $i$-band time series of normal SNe Ia exploding in an LRG (left) and in a spiral galaxy (right), as illustrative examples.

Note that in the synthetic time series (for both lensed and normal SNe), we do not incorporate seeing variations for simplicity. Such variations are naturally present in the fully observed time series of HSC variables. As PSF variations are expected to have minimal impact on our method, we defer their inclusion to future work.

\subsection{Matching data quality}
\label{sec:matching_data_quality}

As outlined earlier, our dataset consists of two fundamentally different types of samples: (i) simulated time series, including mock LSNe Ia and normal SNe Ia exploding in LRGs and spiral galaxies, and (ii) real, fully observed time series of variables from the HSC Transient Survey. To ensure consistency across different components, we match the data quality in terms of depth, zero-point, and cadence. For the simulated samples, we have complete control over the temporal sampling, allowing us to freely define the cadence. However, for the observed HSC variables, the cadence is fixed by the actual observing schedule, effectively ``etched in stone", leaving no flexibility.

To ensure homogeneity between the simulated and observed time series in the dataset, it is essential that their cadences match. This means we must adjust the sampling of the simulated time series to mirror the cadence of the HSC Transient Survey. In practice, we adopt the HSC Transient Survey cadence from the 2017 observing season as our reference and enforce this distribution across all time series used in training and testing, including the simulations. This step is crucial for aligning the temporal resolution of inputs across the dataset and ensuring that the model does not inadvertently learn spurious features associated with differences in cadence. It is important to emphasize that this choice of cadence is driven by practical constraints imposed by the observed data, and it differs significantly from the cadence anticipated for LSST.

The details of how we match the cadence distributions for all components (simulated mock LSNe Ia, normal SNe Ia exploding in galaxies, and observed HSC variables) are explained in Appendix \ref{app:cadence_matching}. Overall, the cadence distributions for all components match well across bands, as seen in Figure \ref{fig:matching_cadence}, with a slight difference in the LSNe samples due to the additional criterion for capturing at least the second image.

\subsection{First detection}
In the final training, validation, and test sets, the first detection of the simulated time series (i.e., for the synthetic lensed and normal SNe Ia time series) is randomly selected within $(-20, 0)$ days relative to the peak of the first image for lensed cases or the SN light curve for the normal cases. In contrast, the first detection of the HSC variables, derived from observations in the HSC Transient Survey, is determined by the data and cannot be manually adjusted. Since the phase of the SN at a given time is unknown in real observations, we ensure that the timestamps (i.e., temporal values for each frame in the time series) do not inadvertently reveal the identity of any component. To achieve this, we set the timestamp of the first detection for each sample (i.e., the starting time of the time series) to zero, with all subsequent timestamps measured relative to this first detection.

\section{Modelling framework -- \convlstmd}
\label{sec:model}
 In this study, our primary goal is to distinguish between LSNe Ia and other transients (and bogus detections) using time series of 2D images. We aim to make predictions after each observation, with the expectation that the accuracy will improve over time as more data are processed throughout the time series.  For this purpose, we employ a \convlstmd, a specialized version of a recurrent neural network (RNN) that extends Long Short-Term Memory ({\tt LSTM}) networks to handle 2D data with both spatial and temporal correlations. For a more detailed explanation of {\tt LSTM} and \convlstmd, we refer to Appendix \ref{app:convlstm}.

In brief, \convlstmd, just like {\tt LSTM}, utilizes two types of memory: long-term memory (commonly referred to as the `cell state') and short-term memory (referred to as the `hidden state'). The key difference between \convlstmd~ and {\tt LSTM} is that \convlstmd~ applies convolutional operations to both the input images and the hidden states, as opposed to {\tt LSTM}'s use of matrix multiplications (which is analogous to a fully-connected layer). As a result, the two memory states in \convlstmd~ retain their 2D structure, preserving spatial correlations throughout the process. At each observation epoch, the \convlstmd~ produces an output based on the input image and the two memory states from the previous epoch. This output is passed forward as the hidden state for the next observation epoch, and the cell state is updated based on the current input and previous memory states.  

\begin{figure}
    \centering
    \includegraphics[width=\columnwidth]{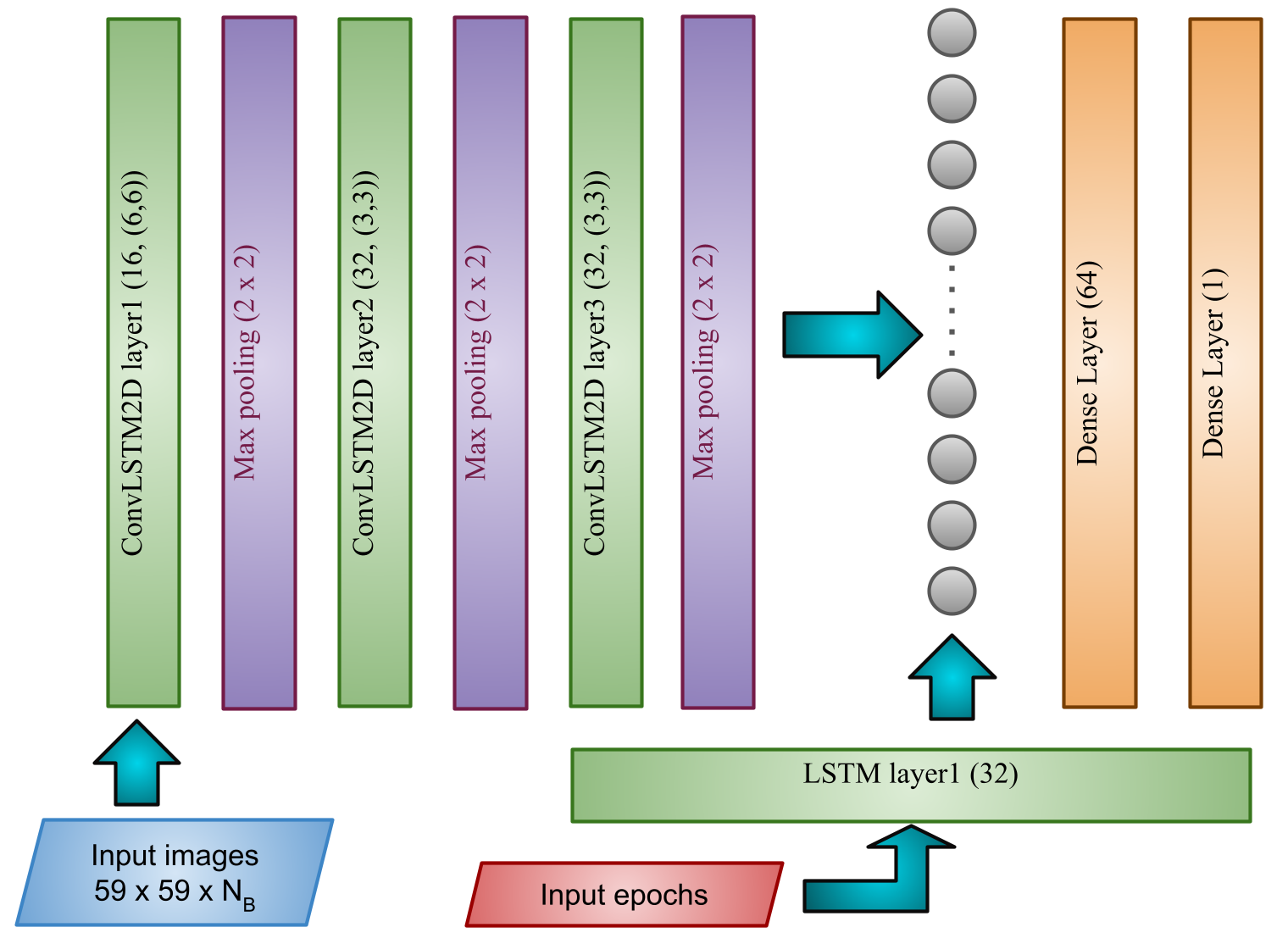}
    \caption{The figure illustrates the architecture of our deep learning model. At each observation epoch, the image cutout is processed through a \texttt{ConvLSTM2D} channel, while the corresponding time value is fed into a standard \texttt{LSTM} channel. The outputs from both channels are then concatenated and passed through a dense layer. The final layer, consisting of a single node with \texttt{sigmoid} activation, outputs a value $\P \in (0,1)$, representing the probability that the sample is lensed at that particular observation epoch. We use the \texttt{binary-cross-entropy} loss function for binary classification. Note the the number of bands is $N_{\rm B}=4$ and $1$ for the multi-band and single-band analyses, respectively.}
    \label{fig:network_architechture}
\end{figure}

Similar to convolutional layers, multiple \texttt{ConvLSTM2D} layers can be stacked together to capture spatial and temporal information more efficiently. Figure \ref{fig:network_architechture} illustrates our model architecture, which consists of two parallel branches. The first branch processes image cutouts, where the input is passed through three \texttt{ConvLSTM2D} layers with \texttt{max-pooling} and \texttt{batch-normalization} applied in between to capture both spatial features and their temporal evolution. The second branch processes the associated timestamps using an \texttt{LSTM} layer, allowing the model to learn the characteristic time scales of lensed and unlensed SNe Ia.

The outputs from both branches are concatenated and passed through a dense layer, followed by a final sigmoid-activated node that predicts a number, $\P \in (0,1)$, representing the probability that the sample is a LSN Ia (ideally $\P \to 1$) or not (ideally $\P \to 0$). At each observation epoch, the model makes decisions based on the 2D cutout and timestamp for that epoch, as well as the memories (cell and hidden states) from past observations, if any. Performance is expected to improve progressively over time as more data are processed.

We test two types of networks: separate models for each band in single-band analysis, and a single model for multi-band analysis that processes multi-band time series data.

\subsection{Arranging data for multi-band analysis}
\label{sec:data_multiband_arrangement}
\begin{figure*}
    \centering
    \includegraphics[width=\textwidth]{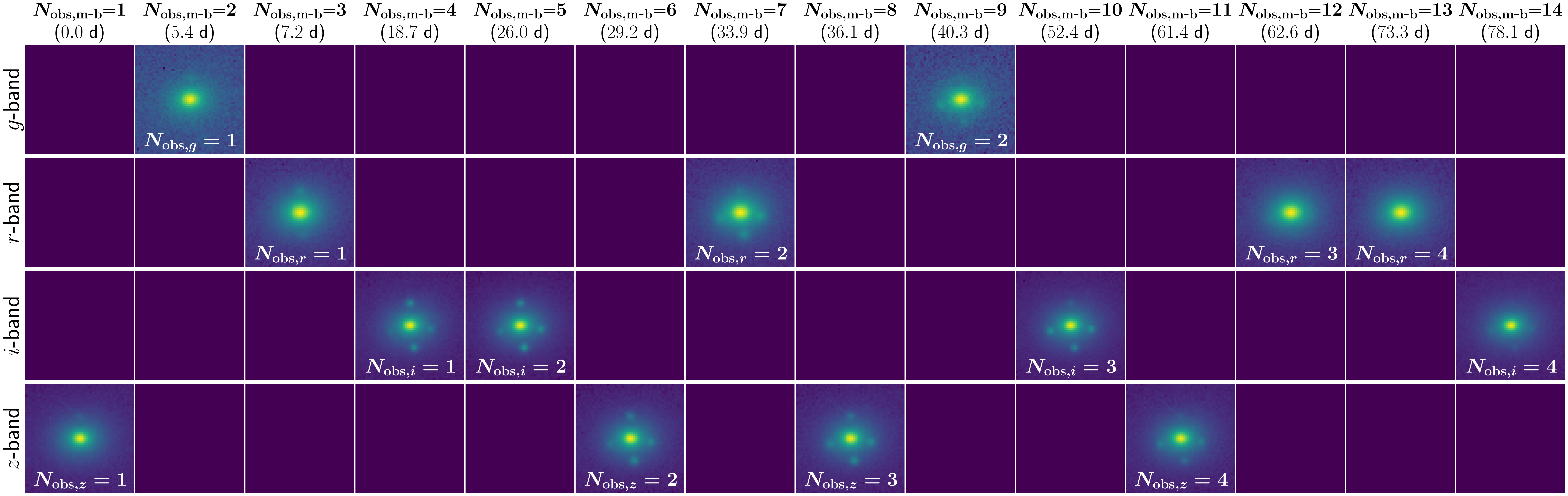}
    \caption{Example multi-band time series of a quadruply LSN Ia, the same system whose $i$-band time series is shown in the top-left panel of Figure~\ref{fig:images}. At each observation epoch ($\Ne$), data are available in only one band; missing-band frames are zero-padded to maintain a uniform input format. The timestamps corresponding to each $\Ne$ is shown above the top frames in parentheses (in days, relative to the first detection, which is always set to zero). The band-specific cumulative observation count ($N_{\text{obs},X}$, where $X \in {g, r, i, z}$) is also marked within each observed frame. In the multi-band analysis, the full time series is input as a 4-channel sequence to the multi-band model, while in the single-band case, the observation sequence in the respective band is provided to the corresponding single-band model without the need of any padding. Note that we follow the HSC Transient Survey cadence here, and the multi-band time series for other classes (HSC variables, SNe in galaxies) have statistically similar time sampling.}
    \label{fig:multi-band_ts}
\end{figure*}

Observations in different bands will generally not be synchronized, meaning that we will not have observations in all bands at the same time.\footnote{LSST aims to revisit a sky location in two bands per night, though the specific bands are not fixed. However, to build a generic pipeline, we assume completely asynchronous observations.} Since LSNe Ia are typically at higher redshifts than their unlensed counterparts, they tend to appear ``redder" in optical filters. As a result, the color information available in the multi-band scenario can be highly informative for distinguishing LSNe Ia.

The multi-band time series of a quadruple LSNe Ia is shown in Figure~\ref{fig:multi-band_ts} as an illustrative example. For a balanced analysis and without loss of generality, we ensure that each sample contains exactly 14 observation epochs in total, distributed across the $griz$ bands. We denote these observation epochs as $\Ne$, which also represents the cumulative count of multi-band observations up to a given point in time; thus, $\Ne \in {1, 2, \dots, 14}$, as marked above the top panels. The corresponding timestamps (in days, MJD to be more precise) relative to the first detection are given just below the $\Ne$ labels, enclosed in parenthesis.

At each observation epoch, i.e., in each column of the time series shown in Figure~\ref{fig:multi-band_ts}, data are available in only one band. The temporal order in which these observations occur in different bands varies from sample to sample. To enable a fair comparison with single-band analyses, we ensure that each sample contains a fixed number of observations per band: specifically, $N_{\text{obs}}^{\text{tot}} = 2$, 4, 4, and 4 for the $g$, $r$, $i$, and $z$ bands, respectively, resulting in a total of 14 observation epochs per sample. This setup ensures that the same set of samples can be used to compare model performance at every observation step, whether indexed by $\Ne$ for multi-band analysis or by $N_{\text{obs},X}$, where $X \in {g, r, i, z}$, for single-band analyses. Since our models generate predictions after each new observation, this constraint does not limit generality but instead enables a consistent comparison. For our example time series in Figure~\ref{fig:multi-band_ts}, $N_{\text{obs},X}$ in single band is also marked in each observed frame.

The asynchronous time sampling across bands can be addressed in several ways. As a first step, we adopt a simple, brute-force strategy: missing band images are zero-padded to construct artificial multi-band inputs at each observation epoch, enabling the network to exploit color information when available. This uniformly structured input, as shown in Figure~\ref{fig:multi-band_ts}, is then fed into the multi-band model, where we expect the network to learn to ignore the padded entries and focus on the real data. We experimented with different padding values and normalization schemes, and found that padding with zeros, while ensuring the range of real images in the time series remains positive after normalization, led to the fastest convergence. This approach enables the model to effectively distinguish real data from padded placeholder data, resulting in better classification performance with fewer training iterations.

\subsection{Normalization strategy}
Normalization of the input data becomes non-trivial in realistic application scenarios, particularly since predictions are made at each observation epoch starting from the first, i.e., from $\Ne=1$ to $14$. One possible approach is to avoid normalization altogether, but we found that normalization can enhance model performance. To address this, we normalize based on the first observation, which can be in any band, depending on the sample. After testing several normalization schemes, we found that the following strategy works best.

\begin{enumerate}
    \item Each image observed in the time series ($\lbrace I_t \rbrace_{t=1}^{14}$, where $I_t \in \mathbb{R}^{59\times 59}$, a real-valued $(59 \times 59)$ matrix) is first shifted to have non-negative pixel values by subtracting its minimum pixel value. We then take the square root to enhance the dynamic range:
    \begin{equation}
        I'_t = \sqrt{I_t - \min(I_t)} \in \mathbb{R}_{\geq 0}^{59\times 59} \;.
    \end{equation}
    
    \item For normalization, the first image in the sequence is min-max scaled to the $[0,1]$ range. The same maximum value (from the first image) is then used to scale all subsequent images:
    \begin{equation}
        \tilde{I}_t = \frac{I'_t}{\max(I'_1)} \;.
    \end{equation}
\end{enumerate}

After normalization, the resulting image time series spans a range from zero to order-unity, without intersecting the padding values, which are set to zero exactly. This ensures that the model can more easily disregard the padded missing band images while still retaining the color information. Moreover, this approach leverages the benefits of normalization to improve the model's ability to focus on the actual data.

\subsection{Composition of training, validation and test sets}
\begin{table}[h]
    \centering
    \begin{tabular}{|c||c|c|c|}
        \hline
        set &  Training & Validation  & Test  \\
        \hline
         LSNe Ia  &  $40,000$ & $1893$ & $5811$ \\
        \hline \hline
        HSC variables & $12,000$ & $4081$ &  $12631$  \\
        \hline
      SN in LRGs   & $20,000$ & $ 1980$ &  $6374$  \\
      \hline
      SN in spirals   & $8,000$  & $2046$  &  $6308$  \\
        \hline\hline

        Total & $80,000$ & $10,000$ & $31,124$ \\
        \hline
    \end{tabular}
    \caption{Number of samples of different components in the training, validation, and test sets.}
    \label{tab:sets}
\end{table}

Table \ref{tab:sets} summarizes the distribution of samples across different components in the training, validation, and test sets. In the training set, we maintain a balanced representation of positive and negative samples. However, initial tests revealed that normal SNe Ia in LRGs are most frequently confused with LSNe Ia, contributing the highest FPR \footnote{The lack of analogous confusion from normal SNe in spiral hosts could stem from not including spiral lenses in training for reasons described in footnote~\ref{fn:two}. Having not seen examples of SNe lensed by spiral galaxies, the model could be trivially associating a spiral galaxy with a normal SN. Although difficult to test explicitly, the model is expected to detect some SNe Ia lensed by spirals using lensing features such as image configurations, color, and temporal evolution.}. As a result, we give these samples more weight in the final training set compared to the other two negative components -- HSC variables and SNe in spirals. As a result, the negative samples in the training set are composed of $30\%$ HSC variables, $50\%$ SNe in LRGs, and $20\%$ SNe in spirals. In the validation and test sets, which share the same component ratios, we increase the number of negative samples to evaluate performance at lower FPRs (note that, in reality, negative samples will vastly outnumber positives). For these two sets, the negative samples are approximately distributed as $50\%$ HSC variables, $25\%$ SNe in LRGs, and $25\%$ SNe in spirals.

\section{Results}
\label{sec:results}

\subsection{Multi-band analysis}
\begin{figure*}
    \centering
    \includegraphics[width=\textwidth]{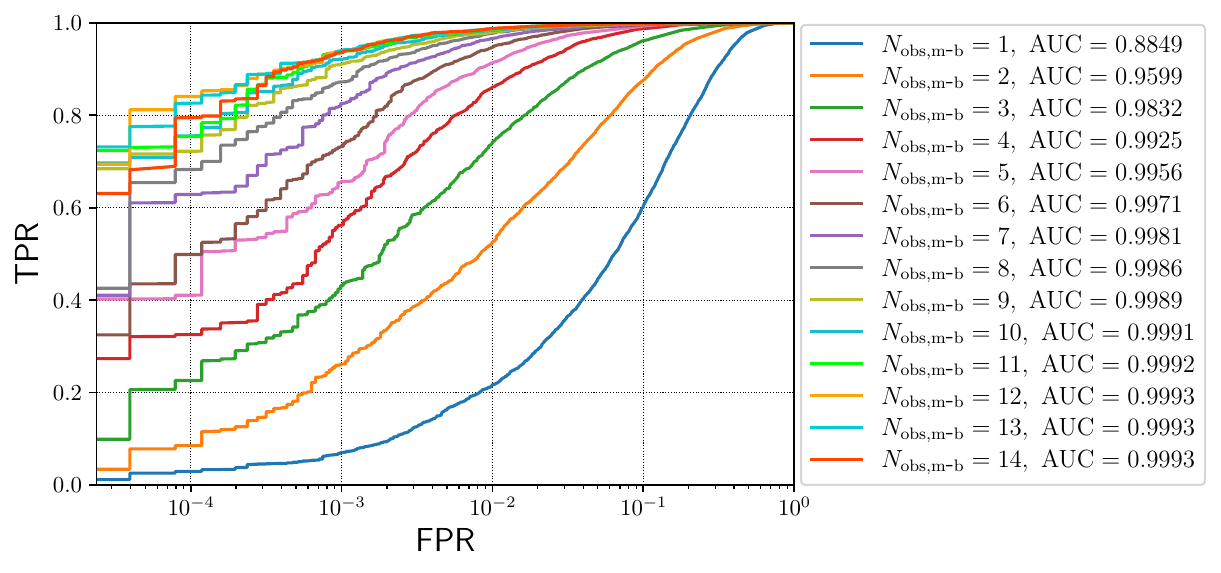}
    \caption{Receiver operating characteristic (ROC) curves for the multi-band classification results obtained from $14$ observations per sample. The corresponding area under the curve (AUC), which approaches unity for a perfect classification, is indicated in the legend. Note that the ROC curve improves rapidly over the first few observations, suggesting increased model accuracy with early observations, and begins to saturate when $\Ne \gtrsim9$. }
    \label{fig:ROC0}
\end{figure*}

\begin{figure*}
    \centering
    \includegraphics[width=\textwidth]{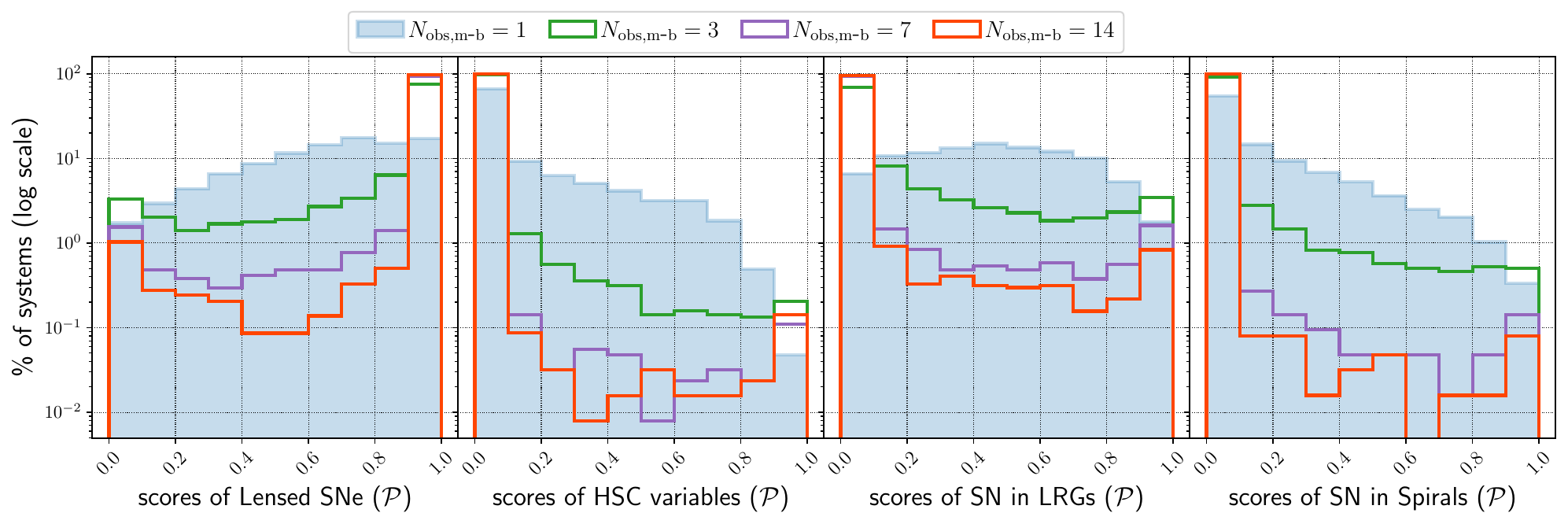}
    \caption{Distribution of model predicted scores for four different components (LSNe Ia, HSC variables, SNe Ia in LRGs and spiral galaxies) -- are plotted in four panels. Each panel shows the distributions obtained after 1st, 3rd, 7th and 14th epoch of observation ($\Ne$). Among the negative components, SNe in LRGs contributes the most to the highest score bin indicates this is responsible for most FPR as indicated explicitly in Fig.~\ref{fig:neg_fprs}.}
    \label{fig:scores}
\end{figure*}

As mentioned above in Section \ref{sec:data_multiband_arrangement}, without any loss of generality, we constructed each sample to consist of 14 observations in random order across 4 bands. To assess the classification performance and how it evolves with additional observations, Figure~\ref{fig:ROC0} shows the Receiver Operating Characteristic (ROC) curves computed after each observation, yielding 14 ROC curves in total. After each observation, i.e., at $\Ne\ \in {1, 2, \ldots, 14}$, we collect the predicted scores ($\mathcal{P}$) for all samples and compute the corresponding ROC curve, i.e., the true-positive rate (TPR) as a function of the false-positive rate (FPR). This sequence of 14 ROC curves illustrates the progressive improvement in classification performance as the model is exposed to more data over time. 

 The area under the ROC curve (AUC), shown in the legend of Figure~\ref{fig:ROC0}, serves as a useful metric for evaluating classification performance. Initially, after the first observation (i.e. at $\Ne=1$), the AUC is approximately $0.885$, reflecting suboptimal performance. However, as additional observations are incorporated, the ROC curve improves rapidly. At $\Ne=7$, the model achieves over $60\%$ TPR at the expense of FPR of $0.01\%$, and by 9th observation, the TPR exceeds $70\%$ with the AUC reaching  $\sim 0.999$. The corresponding median times for these $\Ne$, relative to the peak of the first image, are approximately 17 and 25 days, respectively. These timescales, however, are strongly influenced by our choice to follow the coarse HSC cadence and by the ad hoc prescription we used for setting the time of first detection. We expect that both the timing and the model's performance could improve under a more optimized cadence and detection strategy. If we relax FPR to $0.1\%$, TPR reaches $\sim 60\%$ even earlier at $\Ne=4$. Beyond 9th observation, the ROC curve saturates, suggesting that further observations do not lead to significant improvements in classification performance. Note that while we used diverse classes of negative examples, in a real classification setup the class proportions will inevitably differ, which will affect performance metrics such as ROC curves and their AUC values.

In Figure~\ref{fig:scores}, we show the distribution of predicted scores ($\P$) for four different components -- LSNe Ia, HSC variables, SNe Ia exploding in LRGs, and spiral galaxies -- across four panels. Each panel presents the $\P$-distribution after the 1st, 3rd, 7th, and 14th epochs of observation, illustrating how the score distribution evolves as the model processes more data. Among the negative components, SNe in LRGs most frequently receive high scores in the $0.9 < \P \leq 1.0$ bin, indicating they are most easily confused with the positives (LSNe Ia). Interestingly, as more observations are available, the model becomes slightly overconfident, increasingly assigning extreme scores. For instance, it predicts a higher number of LSNe Ia samples with scores near unity at later epoch of observation, but this also comes at the cost of a few false positives (negatives) receiving similarly extreme high (low) scores. 

\begin{figure*}
    \centering
    \includegraphics[width=\textwidth]{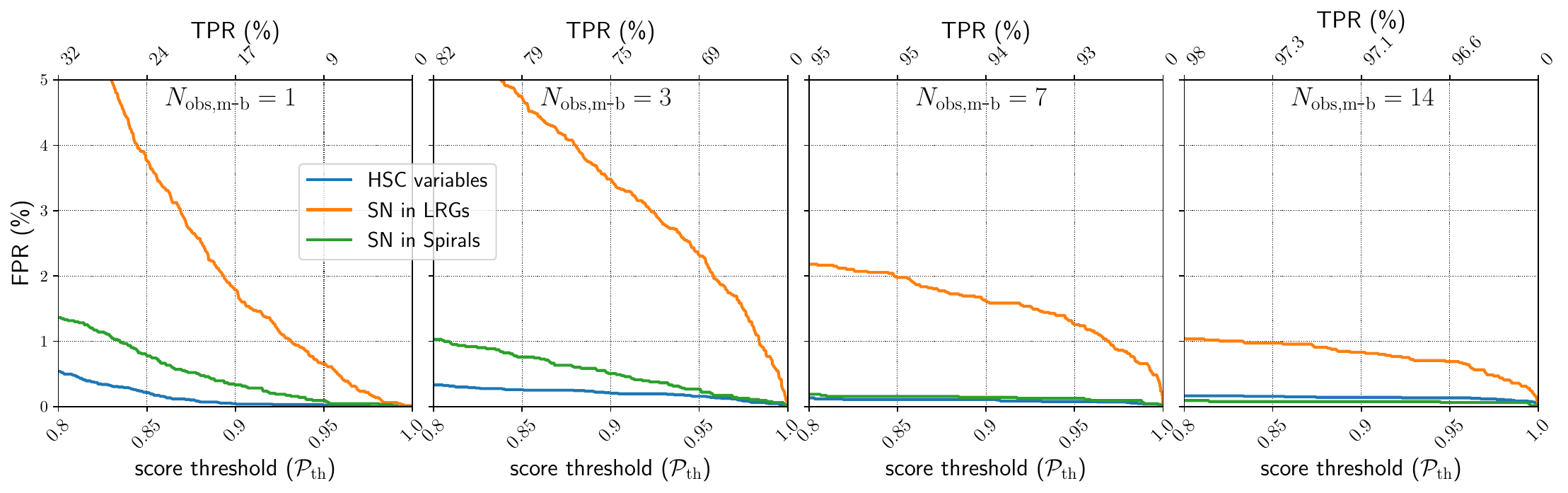}
    \caption{Four panels show the FPR as a function of score threshold, separately for different negative components, for results obtained after 1st, 3rd, 7th and 14th epochs of observation. TPR values corresponding to different score thresholds are marked on the top x-axis. The FPR for all three negative components -- HSC variables, SN exploding in LRGs and spiral galaxies -- gradually decreases with additional observations. However, it is clearly evident that SNe exploding in LRGs dominate the FPR budget, as they are more likely to be confused with the positive class.}
    \label{fig:neg_fprs}
\end{figure*}

Next, we explicitly assess the contribution of different negative components to the FPR. In Figure~\ref{fig:neg_fprs}, we show the FPR as a function of score threshold ($\mathcal{P}_{\rm th}$) for HSC variables, SNe Ia exploding in LRGs, and spiral galaxies, shown in blue, orange, and green, respectively. These curves represent the cumulative distributions, i.e., the fraction of each component with scores exceeding a given threshold, $\mathcal{P} > \mathcal{P}_{\rm th}$. The top x-axis indicates the corresponding TPR values at different score thresholds $\mathcal{P}_{\rm th}$. The four panels display results after the 1st, 3rd, 7th, and 14th observation epochs. As expected, the FPR for all components decreases as the model sees more data, i.e., with increasing $\Ne$. However, in all panels, the FPR associated with SNe Ia in LRGs consistently dominates over the other negative components, indicating that they are more frequently confused with the positive class (LSNe Ia) and are thus the primary contributors to the overall FPR. 

To better understand the sources of confusion, we investigate two extreme cases: mock lensed systems that receive very low scores ($\mathcal{P} < 0.01$) which turn out to be mostly doubles, and  SNe Ia in LRGs that receive very high scores ($\mathcal{P} > 0.99$). We identify two main scenarios in which LSNe Ia can resemble SNe Ia in LRGs:

\begin{itemize}
    \item In some occasions, when an image of a doubly LSN appears within the bright region of the lens galaxy, its light curve is heavily dominated by the lens light. As a result, the image may become difficult for the model to detect making the event resemble an unlensed SN embedded in a bright host galaxy.

    \item When the fainter image(s) of a LSN Ia are not sampled near their peaks due to poor cadence, they may fall below the background noise and go undetected in the image sequence. This again makes the system appear more like an unlensed SN Ia.
\end{itemize}

Both types of confusion are more common among doubles than quads. In principle, color information can still provide a useful cue for distinguishing between the two classes, as LSNe Ia statistically have higher source redshifts ($z_s$) compared to the unlensed SNe Ia. However, this signal is not always reliable. The first scenario could potentially be mitigated by adopting difference imaging techniques, which better isolate variable sources from static galaxy light. The second issue would be alleviated with a finer cadence, such as that expected from LSST, improving the chances of capturing the fainter images near their peaks.

Additionally the Appendix \ref{app:classification_performence_comp} compares classification performance between doubles and quads, and examines how image separation affects the recovery accuracy. Quads are generally easier to detect, and while larger separations improve the performance, the gain diminishes beyond $\sim 1 \arcsec$.

%---------------------------------------------------------------
\subsection{Multi-band vs single band results}

\begin{figure*}
    \centering
    \includegraphics[width=\textwidth]{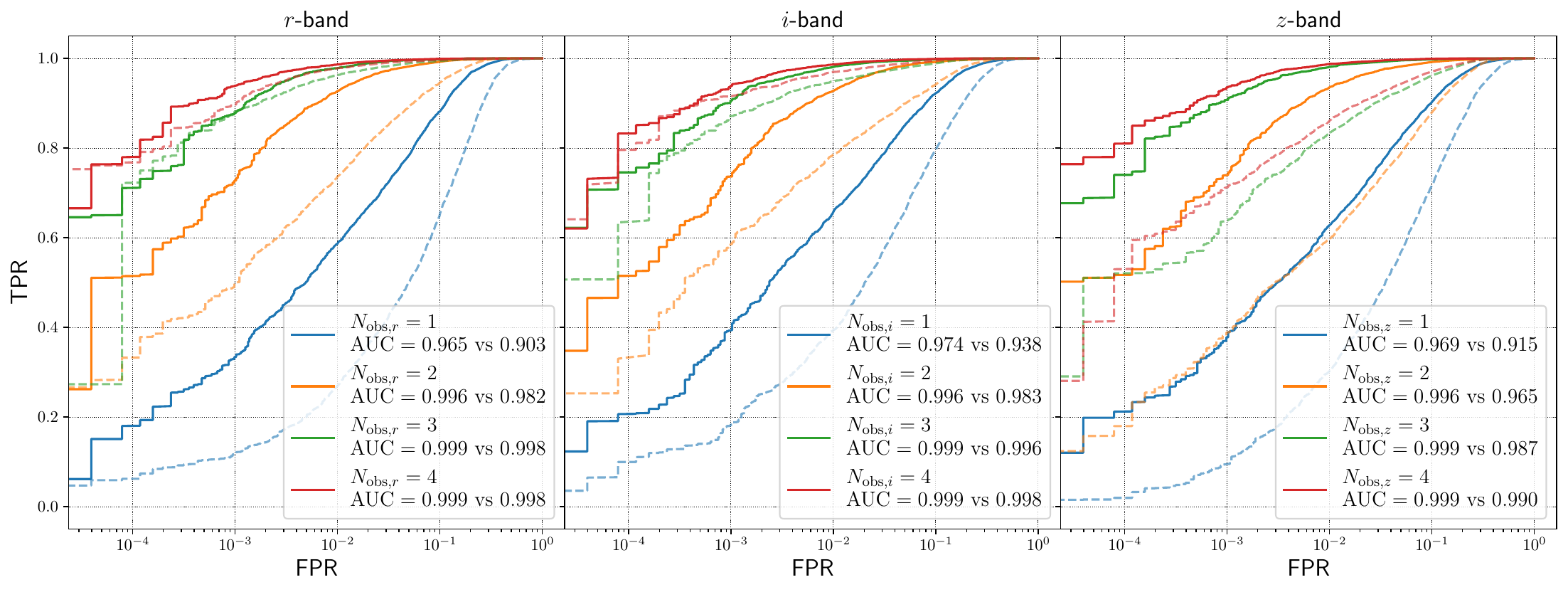}
    \caption{Comparison of classification performance between multi-band (solid) and single-band (dashed) analyses using ROC curves for the $r$, $i$, and $z$ bands, respectively shown in three panels from the left. In each panel, ROC curves corresponding to different observation epochs ($N_{\text{obs},X}$ where $X \in \lbrace r,i,z \rbrace$) are shown in different colours. Legends indicate the AUC values for each curve, with multi-band listed first, followed by single-band for the same epoch.}
    \label{fig:multi_vs_single}
\end{figure*}

Instead of using multi-band data, one can consider training separate models for each individual band, a setup we refer to as the``single-band analysis". This approach avoids the complications that arise from asynchronous time sampling across different filters in the multi-band case. Due to their higher redshifts, LSNe Ia appear redder, making color information in multi-band data particularly informative. In this section, we evaluate whether, and to what extent, the multi-band scenario offers improved classification performance over the single-band models.

We remind the reader that, for a fair comparison between single- and multi-band analyses, each sample is constrained to have exactly $N_\text{obs}^\text{tot} = 2$, 4, 4, and 4 observations in the $g$, $r$, $i$, and $z$ bands, respectively. Also, $N_{\text{obs},X}$ denotes the observation epochs in band $X$, i.e. the cumulative number of observations up to a given point in the sequence. Since the order of observations across bands varies randomly from sample to sample, a fixed value of $N_{\text{obs},X}$ corresponds to different multi-band indices $\Ne$. Nevertheless, all observations in each band are guaranteed to be present somewhere in the sequence. This setup enables a consistent comparison between single- and multi-band models by evaluating their performance after the same observations. 

To do this systematically, we track the predictions from multi-band analyses for each band separately. Consider a generic band $X \in \{r, i, z \}$, which contains four observations per sample. We exclude the $g$-band from the single-band analysis due to its limited cadence in the HSC Transient Survey, coarse cadence anticipated from LSST, and the lower signal-to-noise expected for LSNe Ia in this band. To construct the ROC curve after the $j$-th observation (where $j \in \{1,2,3,4\}$) in $X$-band, we proceed as follows: for the single-band analysis, we collect the $X$-band model's predictions after the $j$-th observation across all samples, i.e. at $N_{\text{obs},X}=j$. For the multi-band analysis, we extract the multi-band model’s predictions after it has seen the same $j$-th observation in $X$-band for each sample, regardless of where that observation occurs within the full multi-band sequence. The key difference in setup is that, unlike single-band models, the multi-band model is exposed not only to prior $X$-band observations but also to all preceding observations across other bands. This allows the multi-band model to develop a richer contextual understanding and incorporate color information, potentially enhancing its ability to learn from the data.

Figure~\ref{fig:multi_vs_single} compares the classification performance between multi-band (solid) and single-band (dashed) analyses using ROC curves for the $r$, $i$, and $z$ bands, shown in three separate panels. In each panel, ROC curves corresponding to predictions after different observation epochs in the concerned band are shown in different colours. The plot legends also indicate the AUC values of these ROC curves,  listed first for the multi-band and then for the single-band analyses at a given epoch of observation.

As expected, the multi-band ROC curves consistently outperform their single-band counterparts in all bands. Notably, the advantage of the multi-band analysis is more significant at earlier epochs of observation. For instance, in the $i$-band at a fixed FPR of $0.1\%$, the multi-band TPR reaches $\sim40\%$ and $\sim75\%$ for $N_{\text{obs},i}=1$ and $2$, respectively -- substantially higher than the corresponding TPRs of $\sim20\%$ and $\sim60\%$ from the single-band analysis.
This is intuitive, as in the earlier part of the sequence, the multi-band model benefits more from the accumulated information from all previous observations across all bands and color information. At later epochs of observation, however, the multi-band model's ROC curves from different bands tend to converge in performance, as reflected in similar AUC values. This suggests that the classifier becomes increasingly confident with more observations, making the specific band less critical to the final decision. 

One may also be tempted to compare the performance across different single-band analyses. However, we caution that the results may be affected by the cadence distributions, which differ across bands because we matched each band's cadence to that of the selected HSC variables. It can be seen that, after the first observation, the ROC curve performs best in the $i$-band, followed by the $z$- and $r$-bands, as reflected in their respective AUCs. This is expected, given that LSNe Ia typically exhibit the highest signal-to-noise ratios in the $i$ and $z$ bands. Interestingly, the single-band ROC curves for the $r$-band show the most significant improvement at later epochs of observation. This may be partly attributed to the more homogeneous cadence in the $r$-band compared to the others.

\section{Conclusions and Discussion}
\label{sec:conclusions}
We aim to develop a pipeline for detecting LSNe from other transients and bogus detections using multi-band imaging data obtained over multiple epochs, as expected from upcoming time-domain surveys like the LSST. In other words, we construct time-series of 2D image cutouts in multiple filters from these surveys' alert streams, with the goal of identifying LSNe as early as possible to enable timely follow-up observations. To achieve this, we employ the \texttt{ConvLSTM2D} model, a specialized recurrent neural network (RNN) designed to understand both spatial and temporal correlations in data, making it well-suited to handle time-series imaging data.

Given that we have to rely on simulations to train our model, we aim for as much realism as possible, ensuring that the simulations closely mirror the complexities and challenges encountered in real observational scenarios. To do so, we base our simulations on real observations, particularly from the HSC, as the LSST is expected to match HSC's data quality in terms of depths and filter throughputs. Specifically, we use LRGs from the HSC wide-layer as lenses and inject LSN images at various time points corresponding to different phases of SN evolution. Although we focus on SNe Ia in this work, the analysis can be extended to other types of SNe or transients. Negative samples are drawn from variable sources observed in the HSC Transient Survey (HSC variables) in the COSMOS field across multiple epochs. Additionally, we simulate time series of normal SNe Ia exploding in HSC LRGs and spiral galaxies to further enhance the training on these particular cases.

While multi-band data provide critical color information, temporally asynchronous observations across different bands presents a challenge in building a seamless network. As an initial approach, we pad the missing-band images to create a consistent input shape, effectively generating synthetic placeholders where data are absent. This method has already shown promising results, as outlined below.

\begin{itemize}
    \item The ROC curves for the multi-band analysis show a rapid improvement in classification performance as the model processes the first few observations across different bands. Keeping the FPR fixed at $0.01\%$, the TPR reaches $\gtrsim 60\%$ by the 7th observation ($\Ne = 7$) and exceeds $\gtrsim 70\%$ by the 9th observation ($\Ne = 9$).\footnote{The corresponding median times for these $\Ne$, measured from the peak of the first image, are approximately 17 and 25 days. These timescales are shaped by our use of the coarse HSC cadence and an ad hoc definition of the first detection. Both the timing and model performance could improve with a more optimized cadence and earlier detection. Notably, if the first detection had occurred earlier in the SN light curve, these timescales would have shifted accordingly.} If the FPR is relaxed to $0.1\%$, the TPR reaches $\sim 60\%$ as early as $\Ne=4$. All these demonstrate the model's strong potential to identify LSNe Ia early in their evolution, enabling timely follow-up observations. Beyond $\Ne = 9$, the ROC curve begins to plateau, indicating that the model has already gathered sufficient information to make confident predictions, and additional observations offer diminishing returns.

    \item Among the various negative classes, normal (unlensed) SNe Ia occurring in LRGs contribute the most to the overall FPR, as LSNe Ia can sometimes mimic these events under certain conditions, making them more prone to misclassification compared to other negative classes. For example, when one of the images from a doubly LSN falls within the bright central region of the lens galaxy, or when the dimmer counterimages are poorly sampled and missed around their peak brightness due to cadence limitations, the resulting time series may resemble that of an unlensed SN Ia in an LRG. In such cases, even though color information can help, it is not always sufficient or reliable for distinguishing between the two.

    \item As expected, quad systems are generally easier to detect than doubles. Likewise, systems with smaller image separations ($\thE < 1.0\arcsec$) tend to be more challenging to identify compared to those with larger separations ($\thE > 1.0\arcsec$).

    \item To avoid dealing with asynchronous observations across different bands, one can instead train separate models for each band and apply them individually to the corresponding single-band data, a strategy we refer to as single-band analysis. While this approach performs reasonably well, it is consistently outperformed by the full multi-band analysis. The performance gap is most pronounced at observation epochs, where the multi-band model benefits from a richer memory formed by preceding observations across multiple bands. This not only helps establish stronger temporal context but also leverages valuable color information early on. In contrast, the single-band model's memory is limited to the past observations in that specific band, lacking both broader temporal context and color cues that are crucial for distinguishing LSNe Ia.

    \item Our model tends to become overconfident in certain scenarios. As it processes more observations, it increasingly assigns very high scores to more positive samples; however, this comes at the cost of misclassifying a few negative examples, primarily normal SNe Ia occurring in LRGs, with similarly high confidence.

\end{itemize}

%\section{Discussions}

While our model is trained and tested on HSC-like data here, the method can be applied to any cadenced imaging survey. It is especially suitable for the upcoming LSST, where a cadence 5 to 10 times better than HSC is expected to further enhance detection performance.  The promising results highlight the potential of our method for real-time detection of LSNe from the LSST alert stream.

Below, we outline the assumptions and caveats of our current method to maintain clarity and avoid confusion, along with potential improvements and remaining tasks.
\begin{itemize}
    \item In the first step, this work focuses exclusively on Type Ia SNe, both lensed and unlensed (normal), as their light curves are well understood, and hence easy to implement. However, this method can easily be extended to other types of SNe and transients by simply incorporating their respective light curves. We have not yet considered microlensing, but we expect it to have minimal impact on the performance of our technique, as it primarily affects the brightness of the LSN images, not their positions. Additionally, chromatic microlensing occurs later in the SN Ia evolution, whereas our goal is to detect lensed cases early in their evolution. Therefore, not accounting for microlensing is not a critical issue at this stage. Nevertheless, we plan to include microlensing in a more comprehensive pipeline in the future.

    \item We have not included host galaxies in the mock LSNe Ia images. In many cases, the host galaxies will also be lensed and visible. However, the choice of SN–host offset distribution and host detectability threshold play a critical role in determining when lensed hosts appear in the data and how prominently. If not treated carefully, these factors can introduce biases into the model. To avoid such premature assumptions, we defer the inclusion of host galaxies to future work, where we will address these aspects systematically. Importantly, our current simulations still reflect a substantial subset of realistic scenarios (roughly half of LSST LSNe Ia) in which the host galaxy is either not strongly lensed or too faint to be detected in the surveys. Since we intend to apply this tool directly to LSST transient alert cutouts, incorporating host galaxies is important as their arc-like features will enhance the performance of our method. Additionally, although we generate synthetic time series for small-separation systems ($\thE < 0.5\arcsec$), we exclude them here to focus on cases with more easily distinguishable spatial features. These challenging systems will be considered in future work, in conjunction with the inclusion of lensed host galaxies.

    \item One limitation of the current setup is the simplified treatment of observational effects in the synthetic time series. In particular, we do not model seeing variations or Poisson noise over time—both of which are naturally present in real observations, such as those of HSC variables. While time-varying PSFs are expected to have minimal impact on our method, accurately modeling them is nontrivial and therefore deferred to future work. Similarly, adding Poisson noise to coadded layers is technically challenging, and is also left for future development. In our current setup, we add white noise at each time step, which is of the same order as Poisson noise near the bright central regions and dominates in the outskirts. Given this, we expect that the absence of explicit Poisson noise modeling is unlikely to significantly affect the results. Nonetheless, incorporating more realistic observational conditions remains an important step for improving the robustness of our approach.

    \item We use real galaxies and HSC variables to create our datasets, treating HSC as a stand-in for LSST. While this helps us address key conceptual challenges in advance, the upcoming LSST early data will offer a valuable opportunity to build a more representative training set using real LRGs and transients. 

    \item In this work, we have so far matched the cadence of the HSC Transient Survey, which, while useful, is not fully representative of LSST's cadence. With LSST's superior cadence, we expect to achieve further improved results, especially enabling earlier detection of LSNe Ia.

    \item Above, we demonstrated the inclusion of multi-band data using a brute-force approach. However, a more sophisticated alternative would involve a composite architecture, where separate \convlstm~ branches process each band in parallel. After each observation, the model could make a decision based on the aggregated outputs from all \convlstm~ branches up to that point. Furthermore, entirely different architectures, such as transformers, may be inherently better suited to handling asynchronous multi-band data, and we are exploring this possibility as well. 

\end{itemize}

\begin{acknowledgements} 
We thank Christian Vogl, Tim Meinhardt and 
Laura Leal-Taixe for many useful discussions. SB acknowledges the funding provided by the Alexander von Humboldt Foundation. This work is partially funded by the Deutsche Forschungsgemeinschaft (DFG, German Research Foundation) – SFB 1258 – 283604770. 
SHS thanks the Max Planck Society for support through the Max Planck Fellowship. 
This project has received funding from the European Research Council (ERC) under the European Union's Horizon 2020 research and innovation programme (LENSNOVA: grant agreement No 771776).
This work is supported in part by the Deutsche Forschungsgemeinschaft (DFG, German Research Foundation) under Germany's Excellence Strategy -- EXC-2094 -- 390783311.  SS has received funding from the European Union’s Horizon 2022 research and innovation programme under the Marie Skłodowska-Curie grant agreement No 101105167 — FASTIDIoUS.
\end{acknowledgements} 

\bibliographystyle{aa}
\bibliography{ref1}

@string(ApJ={Astrophys. J.})

@string(ApJL={Astrophys. J. Lett.})

@string(ApJS={Astrophys. J. Suppl.})

@string(MNRAS={Mon. Not. Roy. Astron. Soc.})

@string(apjs={Astrophys. J. Suppl.})

@ARTICLE{suyu20,
       author = {{Suyu}, S.~H. and {Huber}, S. and {Ca{\~n}ameras}, R. and {Kromer}, M. and {Schuldt}, S. and {Taubenberger}, S. and {Y{\i}ld{\i}r{\i}m}, A. and {Bonvin}, V. and {Chan}, J.~H.~H. and {Courbin}, F. and {N{\"o}bauer}, U. and {Sim}, S.~A. and {Sluse}, D.},
        title = "{HOLISMOKES. I. Highly Optimised Lensing Investigations of Supernovae, Microlensing Objects, and Kinematics of Ellipticals and Spirals}",
      journal = {\aap},
         year = 2020,
        month = dec,
       volume = {644},
          eid = {A162},
        pages = {A162},
          doi = {10.1051/0004-6361/202037757},
archivePrefix = {arXiv},
       eprint = {2002.08378},
 primaryClass = {astro-ph.CO},
       adsurl = {https://ui.adsabs.harvard.edu/abs/2020A&A...644A.162S}
}

@ARTICLE{craig24,
       author = {{Craig}, Peter and {O'Connor}, Kyle and {Chakrabarti}, Sukanya and {Rodney}, Steven A. and {Pierel}, Justin R. and {McCully}, Curtis and {Perez-Fournon}, Ismael},
        title = "{A targeted search for strongly lensed supernovae with the Las Cumbres Observatory}",
      journal = {\mnras},
         year = 2024,
        month = oct,
       volume = {534},
       number = {2},
        pages = {1077-1092},
          doi = {10.1093/mnras/stae2103},
archivePrefix = {arXiv},
       eprint = {2111.01680},
 primaryClass = {astro-ph.CO},
       adsurl = {https://ui.adsabs.harvard.edu/abs/2024MNRAS.534.1077C}
}

@ARTICLE{scoville07,
       author = {{Scoville}, N. and {Aussel}, H. and {Brusa}, M. and {Capak}, P. and {Carollo}, C.~M. and {Elvis}, M. and {Giavalisco}, M. and {Guzzo}, L. and {Hasinger}, G. and {Impey}, C. and {Kneib}, J. -P. and {LeFevre}, O. and {Lilly}, S.~J. and {Mobasher}, B. and {Renzini}, A. and {Rich}, R.~M. and {Sanders}, D.~B. and {Schinnerer}, E. and {Schminovich}, D. and {Shopbell}, P. and {Taniguchi}, Y. and {Tyson}, N.~D.},
        title = "{The Cosmic Evolution Survey (COSMOS): Overview}",
      journal = {\apjs},
         year = 2007,
        month = sep,
       volume = {172},
       number = {1},
        pages = {1-8},
          doi = {10.1086/516585},
archivePrefix = {arXiv},
       eprint = {astro-ph/0612305},
 primaryClass = {astro-ph},
       adsurl = {https://ui.adsabs.harvard.edu/abs/2007ApJS..172....1S}
}

@ARTICLE{abolfathi18,
       author = {{Abolfathi}, Bela and {Aguado}, D.~S. and {Aguilar}, Gabriela and {Allende Prieto}, Carlos and {Almeida}, Andres and {Ananna}, Tonima Tasnim and {Anders}, Friedrich and {Anderson}, Scott F. and {Andrews}, Brett H. and {Anguiano}, Borja and {Arag{\'o}n-Salamanca}, Alfonso and {Argudo-Fern{\'a}ndez}, Maria and {Armengaud}, Eric and {Ata}, Metin and {Aubourg}, Eric and {Avila-Reese}, Vladimir and {Badenes}, Carles and {Bailey}, Stephen and {Balland}, Christophe and {Barger}, Kathleen A. and {Barrera-Ballesteros}, Jorge and {Bartosz}, Curtis and {Bastien}, Fabienne and {Bates}, Dominic and {Baumgarten}, Falk and {Bautista}, Julian and {Beaton}, Rachael and {Beers}, Timothy C. and {Belfiore}, Francesco and {Bender}, Chad F. and {Bernardi}, Mariangela and {Bershady}, Matthew A. and {Beutler}, Florian and {Bird}, Jonathan C. and {Bizyaev}, Dmitry and {Blanc}, Guillermo A. and {Blanton}, Michael R. and {Blomqvist}, Michael and {Bolton}, Adam S. and {Boquien}, M{\'e}d{\'e}ric and {Borissova}, Jura and {Bovy}, Jo and {Bradna Diaz}, Christian Andres and {Brandt}, William Nielsen and {Brinkmann}, Jonathan and {Brownstein}, Joel R. and {Bundy}, Kevin and {Burgasser}, Adam J. and {Burtin}, Etienne and {Busca}, Nicol{\'a}s G. and {Ca{\~n}as}, Caleb I. and {Cano-D{\'\i}az}, Mariana and {Cappellari}, Michele and {Carrera}, Ricardo and {Casey}, Andrew R. and {Cervantes Sodi}, Bernardo and {Chen}, Yanping and {Cherinka}, Brian and {Chiappini}, Cristina and {Choi}, Peter Doohyun and {Chojnowski}, Drew and {Chuang}, Chia-Hsun and {Chung}, Haeun and {Clerc}, Nicolas and {Cohen}, Roger E. and {Comerford}, Julia M. and {Comparat}, Johan and {Correa do Nascimento}, Janaina and {da Costa}, Luiz and {Cousinou}, Marie-Claude and {Covey}, Kevin and {Crane}, Jeffrey D. and {Cruz-Gonzalez}, Irene and {Cunha}, Katia and {da Silva Ilha}, Gabriele and {Damke}, Guillermo J. and {Darling}, Jeremy and {Davidson}, Jr., James W. and {Dawson}, Kyle and {de Icaza Lizaola}, Miguel Angel C. and {de la Macorra}, Axel and {de la Torre}, Sylvain and {De Lee}, Nathan and {de Sainte Agathe}, Victoria and {Deconto Machado}, Alice and {Dell'Agli}, Flavia and {Delubac}, Timoth{\'e}e and {Diamond-Stanic}, Aleksandar M. and {Donor}, John and {Downes}, Juan Jos{\'e} and {Drory}, Niv and {du Mas des Bourboux}, H{\'e}lion and {Duckworth}, Christopher J. and {Dwelly}, Tom and {Dyer}, Jamie and {Ebelke}, Garrett and {Davis Eigenbrot}, Arthur and {Eisenstein}, Daniel J. and {Elsworth}, Yvonne P. and {Emsellem}, Eric and {Eracleous}, Michael and {Erfanianfar}, Ghazaleh and {Escoffier}, Stephanie and {Fan}, Xiaohui and {Fern{\'a}ndez Alvar}, Emma and {Fernandez-Trincado}, J.~G. and {Fernando Cirolini}, Rafael and {Feuillet}, Diane and {Finoguenov}, Alexis and {Fleming}, Scott W. and {Font-Ribera}, Andreu and {Freischlad}, Gordon and {Frinchaboy}, Peter and {Fu}, Hai and {G{\'o}mez Maqueo Chew}, Yilen and {Galbany}, Llu{\'\i}s and {Garc{\'\i}a P{\'e}rez}, Ana E. and {Garcia-Dias}, R. and {Garc{\'\i}a-Hern{\'a}ndez}, D.~A. and {Garma Oehmichen}, Luis Alberto and {Gaulme}, Patrick and {Gelfand}, Joseph and {Gil-Mar{\'\i}n}, H{\'e}ctor and {Gillespie}, Bruce A. and {Goddard}, Daniel and {Gonz{\'a}lez Hern{\'a}ndez}, Jonay I. and {Gonzalez-Perez}, Violeta and {Grabowski}, Kathleen and {Green}, Paul J. and {Grier}, Catherine J. and {Gueguen}, Alain and {Guo}, Hong and {Guy}, Julien and {Hagen}, Alex and {Hall}, Patrick and {Harding}, Paul and {Hasselquist}, Sten and {Hawley}, Suzanne and {Hayes}, Christian R. and {Hearty}, Fred and {Hekker}, Saskia and {Hernandez}, Jesus and {Hernandez Toledo}, Hector and {Hogg}, David W. and {Holley-Bockelmann}, Kelly and {Holtzman}, Jon A. and {Hou}, Jiamin and {Hsieh}, Bau-Ching and {Hunt}, Jason A.~S. and {Hutchinson}, Timothy A. and {Hwang}, Ho Seong and {Jimenez Angel}, Camilo Eduardo and {Johnson}, Jennifer A. and {Jones}, Amy and {J{\"o}nsson}, Henrik and {Jullo}, Eric and {Khan}, Fahim Sakil and {Kinemuchi}, Karen and {Kirkby}, David and {Kirkpatrick}, IV, Charles C. and {Kitaura}, Francisco-Shu and {Knapp}, Gillian R. and {Kneib}, Jean-Paul and {Kollmeier}, Juna A. and {Lacerna}, Ivan and {Lane}, Richard R. and {Lang}, Dustin and {Law}, David R. and {Le Goff}, Jean-Marc and {Lee}, Young-Bae and {Li}, Hongyu and {Li}, Cheng and {Lian}, Jianhui and {Liang}, Yu and {Lima}, Marcos and {Lin}, Lihwai and {Long}, Dan and {Lucatello}, Sara and {Lundgren}, Britt and {Mackereth}, J. Ted and {MacLeod}, Chelsea L. and {Mahadevan}, Suvrath and {Maia}, Marcio Antonio Geimba and {Majewski}, Steven and {Manchado}, Arturo and {Maraston}, Claudia and {Mariappan}, Vivek and {Marques-Chaves}, Rui and {Masseron}, Thomas and {Masters}, Karen L. and {McDermid}, Richard M. and {McGreer}, Ian D. and {Melendez}, Matthew and {Meneses-Goytia}, Sofia and {Merloni}, Andrea and {Merrifield}, Michael R. and {Meszaros}, Szabolcs and {Meza}, Andres and {Minchev}, Ivan and {Minniti}, Dante},
        title = "{The Fourteenth Data Release of the Sloan Digital Sky Survey: First Spectroscopic Data from the Extended Baryon Oscillation Spectroscopic Survey and from the Second Phase of the Apache Point Observatory Galactic Evolution Experiment}",
      journal = {\apjs},
         year = 2018,
        month = apr,
       volume = {235},
       number = {2},
          eid = {42},
        pages = {42},
          doi = {10.3847/1538-4365/aa9e8a},
archivePrefix = {arXiv},
       eprint = {1707.09322},
 primaryClass = {astro-ph.GA},
       adsurl = {https://ui.adsabs.harvard.edu/abs/2018ApJS..235...42A}
}

@ARTICLE{bautista18,
       author = {{Bautista}, Julian E. and {Vargas-Maga{\~n}a}, Mariana and {Dawson}, Kyle S. and {Percival}, Will J. and {Brinkmann}, Jonathan and {Brownstein}, Joel and {Camacho}, Benjamin and {Comparat}, Johan and {Gil-Mar{\'\i}n}, Hector and {Mueller}, Eva-Maria and {Newman}, Jeffrey A. and {Prakash}, Abhishek and {Ross}, Ashley J. and {Schneider}, Donald P. and {Seo}, Hee-Jong and {Tinker}, Jeremy and {Tojeiro}, Rita and {Zhai}, Zhongxu and {Zhao}, Gong-Bo},
        title = "{The SDSS-IV Extended Baryon Oscillation Spectroscopic Survey: Baryon Acoustic Oscillations at Redshift of 0.72 with the DR14 Luminous Red Galaxy Sample}",
      journal = {\apj},
         year = 2018,
        month = aug,
       volume = {863},
       number = {1},
          eid = {110},
        pages = {110},
          doi = {10.3847/1538-4357/aacea5},
archivePrefix = {arXiv},
       eprint = {1712.08064},
 primaryClass = {astro-ph.CO},
       adsurl = {https://ui.adsabs.harvard.edu/abs/2018ApJ...863..110B}
}

@ARTICLE{chao20,
       author = {{Chao}, Dani C. -Y. and {Chan}, James H. -H. and {Suyu}, Sherry H. and {Yasuda}, Naoki and {More}, Anupreeta and {Oguri}, Masamune and {Morokuma}, Tomoki and {Jaelani}, Anton T.},
        title = "{Lensed quasar search via time variability with the HSC transient survey}",
      journal = {\aap},
         year = 2020,
        month = aug,
       volume = {640},
          eid = {A88},
        pages = {A88},
          doi = {10.1051/0004-6361/201936806},
archivePrefix = {arXiv},
       eprint = {1910.01140},
 primaryClass = {astro-ph.GA},
       adsurl = {https://ui.adsabs.harvard.edu/abs/2020A&A...640A..88C}
}

@ARTICLE{chao21,
       author = {{Chao}, Dani C. -Y. and {Chan}, James H. -H. and {Suyu}, Sherry H. and {Yasuda}, Naoki and {Morokuma}, Tomoki and {Jaelani}, Anton T. and {Nagao}, Tohru and {Rusu}, Cristian E.},
        title = "{Strongly lensed candidates from the HSC transient survey}",
      journal = {\aap},
         year = 2021,
        month = nov,
       volume = {655},
          eid = {A114},
        pages = {A114},
          doi = {10.1051/0004-6361/202039376},
archivePrefix = {arXiv},
       eprint = {2009.07854},
 primaryClass = {astro-ph.GA},
       adsurl = {https://ui.adsabs.harvard.edu/abs/2021A&A...655A.114C}
}

@ARTICLE{vandokkum01,
       author = {{van Dokkum}, Pieter G.},
        title = "{Cosmic-Ray Rejection by Laplacian Edge Detection}",
      journal = {\pasp},
         year = 2001,
        month = nov,
       volume = {113},
       number = {789},
        pages = {1420-1427},
          doi = {10.1086/323894},
archivePrefix = {arXiv},
       eprint = {astro-ph/0108003},
 primaryClass = {astro-ph},
       adsurl = {https://ui.adsabs.harvard.edu/abs/2001PASP..113.1420V}
}

@ARTICLE{aihara19,
       author = {{Aihara}, Hiroaki and {AlSayyad}, Yusra and {Ando}, Makoto and {Armstrong}, Robert and {Bosch}, James and {Egami}, Eiichi and {Furusawa}, Hisanori and {Furusawa}, Junko and {Goulding}, Andy and {Harikane}, Yuichi and {Hikage}, Chiaki and {Ho}, Paul T.~P. and {Hsieh}, Bau-Ching and {Huang}, Song and {Ikeda}, Hiroyuki and {Imanishi}, Masatoshi and {Ito}, Kei and {Iwata}, Ikuru and {Jaelani}, Anton T. and {Kakuma}, Ryota and {Kawana}, Kojiro and {Kikuta}, Satoshi and {Kobayashi}, Umi and {Koike}, Michitaro and {Komiyama}, Yutaka and {Li}, Xiangchong and {Liang}, Yongming and {Lin}, Yen-Ting and {Luo}, Wentao and {Lupton}, Robert and {Lust}, Nate B. and {MacArthur}, Lauren A. and {Matsuoka}, Yoshiki and {Mineo}, Sogo and {Miyatake}, Hironao and {Miyazaki}, Satoshi and {More}, Surhud and {Murata}, Ryoma and {Namiki}, Shigeru V. and {Nishizawa}, Atsushi J. and {Oguri}, Masamune and {Okabe}, Nobuhiro and {Okamoto}, Sakurako and {Okura}, Yuki and {Ono}, Yoshiaki and {Onodera}, Masato and {Onoue}, Masafusa and {Osato}, Ken and {Ouchi}, Masami and {Shibuya}, Takatoshi and {Strauss}, Michael A. and {Sugiyama}, Naoshi and {Suto}, Yasushi and {Takada}, Masahiro and {Takagi}, Yuhei and {Takata}, Tadafumi and {Takita}, Satoshi and {Tanaka}, Masayuki and {Terai}, Tsuyoshi and {Toba}, Yoshiki and {Uchiyama}, Hisakazu and {Utsumi}, Yousuke and {Wang}, Shiang-Yu and {Wang}, Wenting and {Yamada}, Yoshihiko},
        title = "{Second data release of the Hyper Suprime-Cam Subaru Strategic Program}",
      journal = {\pasj},
         year = 2019,
        month = dec,
       volume = {71},
       number = {6},
          eid = {114},
        pages = {114},
          doi = {10.1093/pasj/psz103},
archivePrefix = {arXiv},
       eprint = {1905.12221},
 primaryClass = {astro-ph.IM},
       adsurl = {https://ui.adsabs.harvard.edu/abs/2019PASJ...71..114A}
}

@ARTICLE{maoz14,
       author = {{Maoz}, Dan and {Mannucci}, Filippo and {Nelemans}, Gijs},
        title = "{Observational Clues to the Progenitors of Type Ia Supernovae}",
      journal = {\araa},
         year = 2014,
        month = aug,
       volume = {52},
        pages = {107-170},
          doi = {10.1146/annurev-astro-082812-141031},
archivePrefix = {arXiv},
       eprint = {1312.0628},
 primaryClass = {astro-ph.CO},
       adsurl = {https://ui.adsabs.harvard.edu/abs/2014ARA&A..52..107M}
}

@ARTICLE{aihara18,
       author = {{Aihara}, Hiroaki and {Arimoto}, Nobuo and {Armstrong}, Robert and {Arnouts}, St{\'e}phane and {Bahcall}, Neta A. and {Bickerton}, Steven and {Bosch}, James and {Bundy}, Kevin and {Capak}, Peter L. and {Chan}, James H.~H. and {Chiba}, Masashi and {Coupon}, Jean and {Egami}, Eiichi and {Enoki}, Motohiro and {Finet}, Francois and {Fujimori}, Hiroki and {Fujimoto}, Seiji and {Furusawa}, Hisanori and {Furusawa}, Junko and {Goto}, Tomotsugu and {Goulding}, Andy and {Greco}, Johnny P. and {Greene}, Jenny E. and {Gunn}, James E. and {Hamana}, Takashi and {Harikane}, Yuichi and {Hashimoto}, Yasuhiro and {Hattori}, Takashi and {Hayashi}, Masao and {Hayashi}, Yusuke and {He{\l}miniak}, Krzysztof G. and {Higuchi}, Ryo and {Hikage}, Chiaki and {Ho}, Paul T.~P. and {Hsieh}, Bau-Ching and {Huang}, Kuiyun and {Huang}, Song and {Ikeda}, Hiroyuki and {Imanishi}, Masatoshi and {Inoue}, Akio K. and {Iwasawa}, Kazushi and {Iwata}, Ikuru and {Jaelani}, Anton T. and {Jian}, Hung-Yu and {Kamata}, Yukiko and {Karoji}, Hiroshi and {Kashikawa}, Nobunari and {Katayama}, Nobuhiko and {Kawanomoto}, Satoshi and {Kayo}, Issha and {Koda}, Jin and {Koike}, Michitaro and {Kojima}, Takashi and {Komiyama}, Yutaka and {Konno}, Akira and {Koshida}, Shintaro and {Koyama}, Yusei and {Kusakabe}, Haruka and {Leauthaud}, Alexie and {Lee}, Chien-Hsiu and {Lin}, Lihwai and {Lin}, Yen-Ting and {Lupton}, Robert H. and {Mandelbaum}, Rachel and {Matsuoka}, Yoshiki and {Medezinski}, Elinor and {Mineo}, Sogo and {Miyama}, Shoken and {Miyatake}, Hironao and {Miyazaki}, Satoshi and {Momose}, Rieko and {More}, Anupreeta and {More}, Surhud and {Moritani}, Yuki and {Moriya}, Takashi J. and {Morokuma}, Tomoki and {Mukae}, Shiro and {Murata}, Ryoma and {Murayama}, Hitoshi and {Nagao}, Tohru and {Nakata}, Fumiaki and {Niida}, Mana and {Niikura}, Hiroko and {Nishizawa}, Atsushi J. and {Obuchi}, Yoshiyuki and {Oguri}, Masamune and {Oishi}, Yukie and {Okabe}, Nobuhiro and {Okamoto}, Sakurako and {Okura}, Yuki and {Ono}, Yoshiaki and {Onodera}, Masato and {Onoue}, Masafusa and {Osato}, Ken and {Ouchi}, Masami and {Price}, Paul A. and {Pyo}, Tae-Soo and {Sako}, Masao and {Sawicki}, Marcin and {Shibuya}, Takatoshi and {Shimasaku}, Kazuhiro and {Shimono}, Atsushi and {Shirasaki}, Masato and {Silverman}, John D. and {Simet}, Melanie and {Speagle}, Joshua and {Spergel}, David N. and {Strauss}, Michael A. and {Sugahara}, Yuma and {Sugiyama}, Naoshi and {Suto}, Yasushi and {Suyu}, Sherry H. and {Suzuki}, Nao and {Tait}, Philip J. and {Takada}, Masahiro and {Takata}, Tadafumi and {Tamura}, Naoyuki and {Tanaka}, Manobu M. and {Tanaka}, Masaomi and {Tanaka}, Masayuki and {Tanaka}, Yoko and {Terai}, Tsuyoshi and {Terashima}, Yuichi and {Toba}, Yoshiki and {Tominaga}, Nozomu and {Toshikawa}, Jun and {Turner}, Edwin L. and {Uchida}, Tomohisa and {Uchiyama}, Hisakazu and {Umetsu}, Keiichi and {Uraguchi}, Fumihiro and {Urata}, Yuji and {Usuda}, Tomonori and {Utsumi}, Yousuke and {Wang}, Shiang-Yu and {Wang}, Wei-Hao and {Wong}, Kenneth C. and {Yabe}, Kiyoto and {Yamada}, Yoshihiko and {Yamanoi}, Hitomi and {Yasuda}, Naoki and {Yeh}, Sherry and {Yonehara}, Atsunori and {Yuma}, Suraphong},
        title = "{The Hyper Suprime-Cam SSP Survey: Overview and survey design}",
      journal = {\pasj},
         year = 2018,
        month = jan,
       volume = {70},
          eid = {S4},
        pages = {S4},
          doi = {10.1093/pasj/psx066},
archivePrefix = {arXiv},
       eprint = {1704.05858},
 primaryClass = {astro-ph.IM},
       adsurl = {https://ui.adsabs.harvard.edu/abs/2018PASJ...70S...4A}
}

@ARTICLE{furusawa08,
       author = {{Furusawa}, Hisanori and {Kosugi}, George and {Akiyama}, Masayuki and {Takata}, Tadafumi and {Sekiguchi}, Kazuhiro and {Tanaka}, Ichi and {Iwata}, Ikuru and {Kajisawa}, Masaru and {Yasuda}, Naoki and {Doi}, Mamoru and {Ouchi}, Masami and {Simpson}, Chris and {Shimasaku}, Kazuhiro and {Yamada}, Toru and {Furusawa}, Junko and {Morokuma}, Tomoki and {Ishida}, Catherine M. and {Aoki}, Kentaro and {Fuse}, Tetsuharu and {Imanishi}, Masatoshi and {Iye}, Masanori and {Karoji}, Hiroshi and {Kobayashi}, Naoto and {Kodama}, Tadayuki and {Komiyama}, Yutaka and {Maeda}, Yoshitomo and {Miyazaki}, Satoshi and {Mizumoto}, Yoshihiko and {Nakata}, Fumiaki and {Noumaru}, Jun'ichi and {Ogasawara}, Ryusuke and {Okamura}, Sadanori and {Saito}, Tomoki and {Sasaki}, Toshiyuki and {Ueda}, Yoshihiro and {Yoshida}, Michitoshi},
        title = "{The Subaru/XMM-Newton Deep Survey (SXDS). II. Optical Imaging and Photometric Catalogs}",
      journal = {\apjs},
         year = 2008,
        month = may,
       volume = {176},
       number = {1},
        pages = {1-18},
          doi = {10.1086/527321},
archivePrefix = {arXiv},
       eprint = {0801.4017},
 primaryClass = {astro-ph},
       adsurl = {https://ui.adsabs.harvard.edu/abs/2008ApJS..176....1F}
}

@ARTICLE{goobar17,
       author = {{Goobar}, A. and {Amanullah}, R. and {Kulkarni}, S.~R. and {Nugent}, P.~E. and {Johansson}, J. and {Steidel}, C. and {Law}, D. and {M{\"o}rtsell}, E. and {Quimby}, R. and {Blagorodnova}, N. and {Brandeker}, A. and {Cao}, Y. and {Cooray}, A. and {Ferretti}, R. and {Fremling}, C. and {Hangard}, L. and {Kasliwal}, M. and {Kupfer}, T. and {Lunnan}, R. and {Masci}, F. and {Miller}, A.~A. and {Nayyeri}, H. and {Neill}, J.~D. and {Ofek}, E.~O. and {Papadogiannakis}, S. and {Petrushevska}, T. and {Ravi}, V. and {Sollerman}, J. and {Sullivan}, M. and {Taddia}, F. and {Walters}, R. and {Wilson}, D. and {Yan}, L. and {Yaron}, O.},
        title = "{iPTF16geu: A multiply imaged, gravitationally lensed type Ia supernova}",
      journal = {Science},
         year = 2017,
        month = apr,
       volume = {356},
       number = {6335},
        pages = {291-295},
          doi = {10.1126/science.aal2729},
archivePrefix = {arXiv},
       eprint = {1611.00014},
 primaryClass = {astro-ph.CO},
       adsurl = {https://ui.adsabs.harvard.edu/abs/2017Sci...356..291G}
}

@ARTICLE{planck16,
   author = {{Planck Collaboration XIII}},
    title = "{Planck 2015 results. XIII. Cosmological parameters}",
  journal = {\aap},
archivePrefix = "arXiv",
   eprint = {1502.01589},
     year = 2016,
    month = sep,
   volume = 594,
      eid = {A13},
    pages = {A13},
      doi = {10.1051/0004-6361/201525830},
   adsurl = {http://adsabs.harvard.edu/abs/2016A%26A...594A..13P}
}

@ARTICLE{schuldt25,
       author = {{Schuldt}, S. and {Ca{\~n}ameras}, R. and {Andika}, I.~T. and {Bag}, S. and {Melo}, A. and {Shu}, Y. and {Suyu}, S.~H. and {Taubenberger}, S. and {Grillo}, C.},
        title = "{HOLISMOKES: XIII. Strong-lens candidates at all mass scales and their environments from the Hyper-Suprime Cam and deep learning}",
      journal = {\aap},
         year = 2025,
        month = jan,
       volume = {693},
          eid = {A291},
        pages = {A291},
          doi = {10.1051/0004-6361/202450927},
archivePrefix = {arXiv},
       eprint = {2405.20383},
 primaryClass = {astro-ph.GA},
       adsurl = {https://ui.adsabs.harvard.edu/abs/2025A&A...693A.291S}
}

@ARTICLE{yasuda19,
       author = {{Yasuda}, Naoki and {Tanaka}, Masaomi and {Tominaga}, Nozomu and {Jiang}, Ji-an and {Moriya}, Takashi J. and {Morokuma}, Tomoki and {Suzuki}, Nao and {Takahashi}, Ichiro and {Yamaguchi}, Masaki S. and {Maeda}, Keiichi and {Sako}, Masao and {Ikeda}, Shiro and {Kimura}, Akisato and {Morii}, Mikio and {Ueda}, Naonori and {Yoshida}, Naoki and {Lee}, Chien-Hsiu and {Suyu}, Sherry H. and {Komiyama}, Yutaka and {Regnault}, Nicolas and {Rubin}, David},
        title = "{The Hyper Suprime-Cam SSP transient survey in COSMOS: Overview}",
      journal = {\pasj},
         year = 2019,
        month = aug,
       volume = {71},
       number = {4},
          eid = {74},
        pages = {74},
          doi = {10.1093/pasj/psz050},
archivePrefix = {arXiv},
       eprint = {1904.09697},
 primaryClass = {astro-ph.GA},
       adsurl = {https://ui.adsabs.harvard.edu/abs/2019PASJ...71...74Y}
}

@ARTICLE{canameras24,
       author = {{Ca{\~n}ameras}, R. and {Schuldt}, S. and {Shu}, Y. and {Suyu}, S.~H. and {Taubenberger}, S. and {Andika}, I.~T. and {Bag}, S. and {Inoue}, K.~T. and {Jaelani}, A.~T. and {Leal-Taix{\'e}}, L. and {Meinhardt}, T. and {Melo}, A. and {More}, A.},
        title = "{HOLISMOKES: XI. Evaluation of supervised neural networks for strong-lens searches in ground-based imaging surveys}",
      journal = {\aap},
         year = 2024,
        month = dec,
       volume = {692},
          eid = {A72},
        pages = {A72},
          doi = {10.1051/0004-6361/202347072},
archivePrefix = {arXiv},
       eprint = {2306.03136},
 primaryClass = {astro-ph.GA},
       adsurl = {https://ui.adsabs.harvard.edu/abs/2024A&A...692A..72C}
}

@ARTICLE{canameras20,
       author = {{Ca{\~n}ameras}, R. and {Schuldt}, S. and {Suyu}, S.~H. and {Taubenberger}, S. and {Meinhardt}, T. and {Leal-Taix{\'e}}, L. and {Lemon}, C. and {Rojas}, K. and {Savary}, E.},
        title = "{HOLISMOKES. II. Identifying galaxy-scale strong gravitational lenses in Pan-STARRS using convolutional neural networks}",
      journal = {\aap},
         year = 2020,
        month = dec,
       volume = {644},
          eid = {A163},
        pages = {A163},
          doi = {10.1051/0004-6361/202038219},
archivePrefix = {arXiv},
       eprint = {2004.13048},
 primaryClass = {astro-ph.GA},
       adsurl = {https://ui.adsabs.harvard.edu/abs/2020A&A...644A.163C}
}

@ARTICLE{metcalf19,
       author = {{Metcalf}, R.~B. and {Meneghetti}, M. and {Avestruz}, C. and {Bellagamba}, F. and {Bom}, C.~R. and {Bertin}, E. and {Cabanac}, R. and {Courbin}, F. and {Davies}, A. and {Decenci{\`e}re}, E. and {Flamary}, R. and {Gavazzi}, R. and {Geiger}, M. and {Hartley}, P. and {Huertas-Company}, M. and {Jackson}, N. and {Jacobs}, C. and {Jullo}, E. and {Kneib}, J. -P. and {Koopmans}, L.~V.~E. and {Lanusse}, F. and {Li}, C. -L. and {Ma}, Q. and {Makler}, M. and {Li}, N. and {Lightman}, M. and {Petrillo}, C.~E. and {Serjeant}, S. and {Sch{\"a}fer}, C. and {Sonnenfeld}, A. and {Tagore}, A. and {Tortora}, C. and {Tuccillo}, D. and {Valent{\'\i}n}, M.~B. and {Velasco-Forero}, S. and {Verdoes Kleijn}, G.~A. and {Vernardos}, G.},
        title = "{The strong gravitational lens finding challenge}",
      journal = {\aap},
         year = 2019,
        month = may,
       volume = {625},
          eid = {A119},
        pages = {A119},
          doi = {10.1051/0004-6361/201832797},
archivePrefix = {arXiv},
       eprint = {1802.03609},
 primaryClass = {astro-ph.GA},
       adsurl = {https://ui.adsabs.harvard.edu/abs/2019A&A...625A.119M}
}

@ARTICLE{jacobs19,
       author = {{Jacobs}, C. and {Collett}, T. and {Glazebrook}, K. and {Buckley-Geer}, E. and {Diehl}, H.~T. and {Lin}, H. and {McCarthy}, C. and {Qin}, A.~K. and {Odden}, C. and {Caso Escudero}, M. and {Dial}, P. and {Yung}, V.~J. and {Gaitsch}, S. and {Pellico}, A. and {Lindgren}, K.~A. and {Abbott}, T.~M.~C. and {Annis}, J. and {Avila}, S. and {Brooks}, D. and {Burke}, D.~L. and {Carnero Rosell}, A. and {Carrasco Kind}, M. and {Carretero}, J. and {da Costa}, L.~N. and {De Vicente}, J. and {Fosalba}, P. and {Frieman}, J. and {Garc{\'\i}a-Bellido}, J. and {Gaztanaga}, E. and {Goldstein}, D.~A. and {Gruen}, D. and {Gruendl}, R.~A. and {Gschwend}, J. and {Hollowood}, D.~L. and {Honscheid}, K. and {Hoyle}, B. and {James}, D.~J. and {Krause}, E. and {Kuropatkin}, N. and {Lahav}, O. and {Lima}, M. and {Maia}, M.~A.~G. and {Marshall}, J.~L. and {Miquel}, R. and {Plazas}, A.~A. and {Roodman}, A. and {Sanchez}, E. and {Scarpine}, V. and {Serrano}, S. and {Sevilla-Noarbe}, I. and {Smith}, M. and {Sobreira}, F. and {Suchyta}, E. and {Swanson}, M.~E.~C. and {Tarle}, G. and {Vikram}, V. and {Walker}, A.~R. and {Zhang}, Y. and {DES Collaboration}},
        title = "{An Extended Catalog of Galaxy-Galaxy Strong Gravitational Lenses Discovered in DES Using Convolutional Neural Networks}",
      journal = {\apjs},
         year = 2019,
        month = jul,
       volume = {243},
       number = {1},
          eid = {17},
        pages = {17},
          doi = {10.3847/1538-4365/ab26b6},
archivePrefix = {arXiv},
       eprint = {1905.10522},
 primaryClass = {astro-ph.GA},
       adsurl = {https://ui.adsabs.harvard.edu/abs/2019ApJS..243...17J}
}

@ARTICLE{petrillo17,
       author = {{Petrillo}, C.~E. and {Tortora}, C. and {Chatterjee}, S. and {Vernardos}, G. and {Koopmans}, L.~V.~E. and {Verdoes Kleijn}, G. and {Napolitano}, N.~R. and {Covone}, G. and {Schneider}, P. and {Grado}, A. and {McFarland}, J.},
        title = "{Finding strong gravitational lenses in the Kilo Degree Survey with Convolutional Neural Networks}",
      journal = {\mnras},
         year = 2017,
        month = nov,
       volume = {472},
       number = {1},
        pages = {1129-1150},
          doi = {10.1093/mnras/stx2052},
archivePrefix = {arXiv},
       eprint = {1702.07675},
 primaryClass = {astro-ph.GA},
       adsurl = {https://ui.adsabs.harvard.edu/abs/2017MNRAS.472.1129P}
}

@ARTICLE{shu18,
       author = {{Shu}, Yiping and {Bolton}, Adam S. and {Mao}, Shude and {Kang}, Xi and {Li}, Guoliang and {Soraisam}, Monika},
        title = "{Prediction of Supernova Rates in Known Galaxy-Galaxy Strong-lens Systems}",
      journal = {\apj},
         year = 2018,
        month = sep,
       volume = {864},
       number = {1},
          eid = {91},
        pages = {91},
          doi = {10.3847/1538-4357/aad5ea},
archivePrefix = {arXiv},
       eprint = {1803.07569},
 primaryClass = {astro-ph.GA},
       adsurl = {https://ui.adsabs.harvard.edu/abs/2018ApJ...864...91S}
}

@ARTICLE{sagues24,
       author = {{Sagu{\'e}s Carracedo}, A. and {Goobar}, A. and {M{\"o}rtsell}, E. and {Arendse}, N. and {Johansson}, J. and {Townsend}, A. and {Dhawan}, S. and {Nordin}, J. and {Sollerman}, J. and {Schulze}, S.},
        title = "{Detectability and Characterisation of Strongly Lensed Supernova Lightcurves in the Zwicky Transient Facility}",
      journal = {arXiv e-prints},
         year = 2024,
        month = may,
          eid = {arXiv:2406.00052},
        pages = {arXiv:2406.00052},
          doi = {10.48550/arXiv.2406.00052},
archivePrefix = {arXiv},
       eprint = {2406.00052},
 primaryClass = {astro-ph.HE},
       adsurl = {https://ui.adsabs.harvard.edu/abs/2024arXiv240600052S}
}

@ARTICLE{goldstein19,
       author = {{Goldstein}, Daniel A. and {Nugent}, Peter E. and {Goobar}, Ariel},
        title = "{Rates and Properties of Supernovae Strongly Gravitationally Lensed by Elliptical Galaxies in Time-domain Imaging Surveys}",
      journal = {\apjs},
         year = 2019,
        month = jul,
       volume = {243},
       number = {1},
          eid = {6},
        pages = {6},
          doi = {10.3847/1538-4365/ab1fe0},
archivePrefix = {arXiv},
       eprint = {1809.10147},
 primaryClass = {astro-ph.GA},
       adsurl = {https://ui.adsabs.harvard.edu/abs/2019ApJS..243....6G}
}

@ARTICLE{ivezic19,
       author = {{Ivezi{\'c}}, {\v{Z}}eljko and {Kahn}, Steven M. and {Tyson}, J. Anthony and {Abel}, Bob and {Acosta}, Emily and {Allsman}, Robyn and {Alonso}, David and {AlSayyad}, Yusra and {Anderson}, Scott F. and {Andrew}, John and {Angel}, James Roger P. and {Angeli}, George Z. and {Ansari}, Reza and {Antilogus}, Pierre and {Araujo}, Constanza and {Armstrong}, Robert and {Arndt}, Kirk T. and {Astier}, Pierre and {Aubourg}, {\'E}ric and {Auza}, Nicole and {Axelrod}, Tim S. and {Bard}, Deborah J. and {Barr}, Jeff D. and {Barrau}, Aurelian and {Bartlett}, James G. and {Bauer}, Amanda E. and {Bauman}, Brian J. and {Baumont}, Sylvain and {Bechtol}, Ellen and {Bechtol}, Keith and {Becker}, Andrew C. and {Becla}, Jacek and {Beldica}, Cristina and {Bellavia}, Steve and {Bianco}, Federica B. and {Biswas}, Rahul and {Blanc}, Guillaume and {Blazek}, Jonathan and {Blandford}, Roger D. and {Bloom}, Josh S. and {Bogart}, Joanne and {Bond}, Tim W. and {Booth}, Michael T. and {Borgland}, Anders W. and {Borne}, Kirk and {Bosch}, James F. and {Boutigny}, Dominique and {Brackett}, Craig A. and {Bradshaw}, Andrew and {Brandt}, William Nielsen and {Brown}, Michael E. and {Bullock}, James S. and {Burchat}, Patricia and {Burke}, David L. and {Cagnoli}, Gianpietro and {Calabrese}, Daniel and {Callahan}, Shawn and {Callen}, Alice L. and {Carlin}, Jeffrey L. and {Carlson}, Erin L. and {Chandrasekharan}, Srinivasan and {Charles-Emerson}, Glenaver and {Chesley}, Steve and {Cheu}, Elliott C. and {Chiang}, Hsin-Fang and {Chiang}, James and {Chirino}, Carol and {Chow}, Derek and {Ciardi}, David R. and {Claver}, Charles F. and {Cohen-Tanugi}, Johann and {Cockrum}, Joseph J. and {Coles}, Rebecca and {Connolly}, Andrew J. and {Cook}, Kem H. and {Cooray}, Asantha and {Covey}, Kevin R. and {Cribbs}, Chris and {Cui}, Wei and {Cutri}, Roc and {Daly}, Philip N. and {Daniel}, Scott F. and {Daruich}, Felipe and {Daubard}, Guillaume and {Daues}, Greg and {Dawson}, William and {Delgado}, Francisco and {Dellapenna}, Alfred and {de Peyster}, Robert and {de Val-Borro}, Miguel and {Digel}, Seth W. and {Doherty}, Peter and {Dubois}, Richard and {Dubois-Felsmann}, Gregory P. and {Durech}, Josef and {Economou}, Frossie and {Eifler}, Tim and {Eracleous}, Michael and {Emmons}, Benjamin L. and {Fausti Neto}, Angelo and {Ferguson}, Henry and {Figueroa}, Enrique and {Fisher-Levine}, Merlin and {Focke}, Warren and {Foss}, Michael D. and {Frank}, James and {Freemon}, Michael D. and {Gangler}, Emmanuel and {Gawiser}, Eric and {Geary}, John C. and {Gee}, Perry and {Geha}, Marla and {Gessner}, Charles J.~B. and {Gibson}, Robert R. and {Gilmore}, D. Kirk and {Glanzman}, Thomas and {Glick}, William and {Goldina}, Tatiana and {Goldstein}, Daniel A. and {Goodenow}, Iain and {Graham}, Melissa L. and {Gressler}, William J. and {Gris}, Philippe and {Guy}, Leanne P. and {Guyonnet}, Augustin and {Haller}, Gunther and {Harris}, Ron and {Hascall}, Patrick A. and {Haupt}, Justine and {Hernandez}, Fabio and {Herrmann}, Sven and {Hileman}, Edward and {Hoblitt}, Joshua and {Hodgson}, John A. and {Hogan}, Craig and {Howard}, James D. and {Huang}, Dajun and {Huffer}, Michael E. and {Ingraham}, Patrick and {Innes}, Walter R. and {Jacoby}, Suzanne H. and {Jain}, Bhuvnesh and {Jammes}, Fabrice and {Jee}, M. James and {Jenness}, Tim and {Jernigan}, Garrett and {Jevremovi{\'c}}, Darko and {Johns}, Kenneth and {Johnson}, Anthony S. and {Johnson}, Margaret W.~G. and {Jones}, R. Lynne and {Juramy-Gilles}, Claire and {Juri{\'c}}, Mario and {Kalirai}, Jason S. and {Kallivayalil}, Nitya J. and {Kalmbach}, Bryce and {Kantor}, Jeffrey P. and {Karst}, Pierre and {Kasliwal}, Mansi M. and {Kelly}, Heather and {Kessler}, Richard and {Kinnison}, Veronica and {Kirkby}, David and {Knox}, Lloyd and {Kotov}, Ivan V. and {Krabbendam}, Victor L. and {Krughoff}, K. Simon and {Kub{\'a}nek}, Petr and {Kuczewski}, John and {Kulkarni}, Shri and {Ku}, John and {Kurita}, Nadine R. and {Lage}, Craig S. and {Lambert}, Ron and {Lange}, Travis and {Langton}, J. Brian and {Le Guillou}, Laurent and {Levine}, Deborah and {Liang}, Ming and {Lim}, Kian-Tat and {Lintott}, Chris J. and {Long}, Kevin E. and {Lopez}, Margaux and {Lotz}, Paul J. and {Lupton}, Robert H. and {Lust}, Nate B. and {MacArthur}, Lauren A. and {Mahabal}, Ashish and {Mandelbaum}, Rachel and {Markiewicz}, Thomas W. and {Marsh}, Darren S. and {Marshall}, Philip J. and {Marshall}, Stuart and {May}, Morgan and {McKercher}, Robert and {McQueen}, Michelle and {Meyers}, Joshua and {Migliore}, Myriam and {Miller}, Michelle and {Mills}, David J.},
        title = "{LSST: From Science Drivers to Reference Design and Anticipated Data Products}",
      journal = {\apj},
         year = 2019,
        month = mar,
       volume = {873},
       number = {2},
          eid = {111},
        pages = {111},
          doi = {10.3847/1538-4357/ab042c},
archivePrefix = {arXiv},
       eprint = {0805.2366},
 primaryClass = {astro-ph},
       adsurl = {https://ui.adsabs.harvard.edu/abs/2019ApJ...873..111I}
}

@ARTICLE{Millon2020,
       author = {{Millon}, M. and {Galan}, A. and {Courbin}, F. and {Treu}, T. and {Suyu}, S.~H. and {Ding}, X. and {Birrer}, S. and {Chen}, G.~C. -F. and {Shajib}, A.~J. and {Sluse}, D. and {Wong}, K.~C. and {Agnello}, A. and {Auger}, M.~W. and {Buckley-Geer}, E.~J. and {Chan}, J.~H.~H. and {Collett}, T. and {Fassnacht}, C.~D. and {Hilbert}, S. and {Koopmans}, L.~V.~E. and {Motta}, V. and {Mukherjee}, S. and {Rusu}, C.~E. and {Sonnenfeld}, A. and {Spiniello}, C. and {Van de Vyvere}, L.},
        title = "{TDCOSMO. I. An exploration of systematic uncertainties in the inference of H$_{0}$ from time-delay cosmography}",
      journal = {\aap},
         year = 2020,
        month = jul,
       volume = {639},
          eid = {A101},
        pages = {A101},
          doi = {10.1051/0004-6361/201937351},
archivePrefix = {arXiv},
       eprint = {1912.08027},
 primaryClass = {astro-ph.CO},
       adsurl = {https://ui.adsabs.harvard.edu/abs/2020A&A...639A.101M}
}

@ARTICLE{2020A&A...643A.165B,
       author = {{Birrer}, S. and {Shajib}, A.~J. and {Galan}, A. and {Millon}, M. and {Treu}, T. and {Agnello}, A. and {Auger}, M. and {Chen}, G.~C. -F. and {Christensen}, L. and {Collett}, T. and {Courbin}, F. and {Fassnacht}, C.~D. and {Koopmans}, L.~V.~E. and {Marshall}, P.~J. and {Park}, J. -W. and {Rusu}, C.~E. and {Sluse}, D. and {Spiniello}, C. and {Suyu}, S.~H. and {Wagner-Carena}, S. and {Wong}, K.~C. and {Barnab{\`e}}, M. and {Bolton}, A.~S. and {Czoske}, O. and {Ding}, X. and {Frieman}, J.~A. and {Van de Vyvere}, L.},
        title = "{TDCOSMO. IV. Hierarchical time-delay cosmography - joint inference of the Hubble constant and galaxy density profiles}",
      journal = {\aap},
         year = 2020,
        month = nov,
       volume = {643},
          eid = {A165},
        pages = {A165},
          doi = {10.1051/0004-6361/202038861},
archivePrefix = {arXiv},
       eprint = {2007.02941},
 primaryClass = {astro-ph.CO},
       adsurl = {https://ui.adsabs.harvard.edu/abs/2020A&A...643A.165B}
}

@ARTICLE{om10,
       author = {{Oguri}, Masamune and {Marshall}, Philip J.},
        title = "{Gravitationally lensed quasars and supernovae in future wide-field optical imaging surveys}",
      journal = {\mnras},
         year = 2010,
        month = jul,
       volume = {405},
       number = {4},
        pages = {2579-2593},
          doi = {10.1111/j.1365-2966.2010.16639.x},
archivePrefix = {arXiv},
       eprint = {1001.2037},
 primaryClass = {astro-ph.CO},
       adsurl = {https://ui.adsabs.harvard.edu/abs/2010MNRAS.405.2579O}
}

@article{Riess2022,
    author = "Riess, Adam G. and others",
    title = "{A Comprehensive Measurement of the Local Value of the Hubble Constant with 1 km s$^{−1}$ Mpc$^{−1}$ Uncertainty from the Hubble Space Telescope and the SH0ES Team}",
    eprint = "2112.04510",
    archivePrefix = "arXiv",
    primaryClass = "astro-ph.CO",
    doi = "10.3847/2041-8213/ac5c5b",
    journal = "Astrophys. J. Lett.",
    volume = "934",
    number = "1",
    pages = "L7",
    year = "2022"
}

@ARTICLE{2016A&ARv..24...11T,
       author = {{Treu}, Tommaso and {Marshall}, Philip J.},
        title = "{Time delay cosmography}",
      journal = {\aapr},
         year = 2016,
        month = jul,
       volume = {24},
       number = {1},
          eid = {11},
        pages = {11},
          doi = {10.1007/s00159-016-0096-8},
archivePrefix = {arXiv},
       eprint = {1605.05333},
 primaryClass = {astro-ph.CO},
       adsurl = {https://ui.adsabs.harvard.edu/abs/2016A&ARv..24...11T}
}

@ARTICLE{Treu2010,
       author = {{Treu}, Tommaso},
        title = "{Strong Lensing by Galaxies}",
      journal = {\araa},
         year = 2010,
        month = sep,
       volume = {48},
        pages = {87-125},
          doi = {10.1146/annurev-astro-081309-130924},
archivePrefix = {arXiv},
       eprint = {1003.5567},
 primaryClass = {astro-ph.CO},
       adsurl = {https://ui.adsabs.harvard.edu/abs/2010ARA&A..48...87T}
}

@article{Bag:2020pbg,
    author = "Bag, Satadru and Kim, Alex G. and Linder, Eric V. and Shafieloo, Arman",
    title = "{Be It Unresolved: Measuring Time Delays from Lensed Supernovae}",
    eprint = "2010.03774",
    archivePrefix = "arXiv",
    primaryClass = "astro-ph.CO",
    doi = "10.3847/1538-4357/abe238",
    journal = "Astrophys. J.",
    volume = "910",
    number = "1",
    pages = "65",
    year = "2021"
}

@article{Planck:2018vyg,
    author = "Aghanim, N. and others",
    collaboration = "Planck",
    title = "{Planck 2018 results. VI. Cosmological parameters}",
    eprint = "1807.06209",
    archivePrefix = "arXiv",
    primaryClass = "astro-ph.CO",
    doi = "10.1051/0004-6361/201833910",
    journal = "Astron. Astrophys.",
    volume = "641",
    pages = "A6",
    year = "2020"
}

@article{Wong:2019kwg,
    author = "Wong, Kenneth C. and others",
    title = "{H0LiCOW \textendash{} XIII. A 2.4 per cent measurement of H0 from lensed quasars: 5.3\ensuremath{\sigma} tension between early- and late-Universe probes}",
    eprint = "1907.04869",
    archivePrefix = "arXiv",
    primaryClass = "astro-ph.CO",
    doi = "10.1093/mnras/stz3094",
    journal = "Mon. Not. Roy. Astron. Soc.",
    volume = "498",
    number = "1",
    pages = "1420--1439",
    year = "2020"
}

@article{Refsdal1964_1,
    author = {Refsdal, Sjur and Bondi, H.},
    title = "{ The Gravitational Lens Effect ⋆}",
    journal = {Monthly Notices of the Royal Astronomical Society},
    volume = {128},
    number = {4},
    pages = {295-306},
    year = {1964},
    month = {09},
    issn = {0035-8711},
    doi = {10.1093/mnras/128.4.295},
    url = {https://doi.org/10.1093/mnras/128.4.295},
    eprint = {https://academic.oup.com/mnras/article-pdf/128/4/295/8073511/mnras128-0295.pdf},
    }

@article{Refsdal1964_2,
    author = {Refsdal, Sjur},
    title = "{ On the Possibility of Determining Hubble's Parameter and the Masses of Galaxies from the Gravitational Lens Effect ⋆}",
    journal = {Monthly Notices of the Royal Astronomical Society},
    volume = {128},
    number = {4},
    pages = {307-310},
    year = {1964},
    month = {09},
    issn = {0035-8711},
    doi = {10.1093/mnras/128.4.307},
    url = {https://doi.org/10.1093/mnras/128.4.307},
    eprint = {https://academic.oup.com/mnras/article-pdf/128/4/307/8073517/mnras128-0307.pdf},
}

@ARTICLE{Saha,
       author = {{Saha}, Prasenjit and {Coles}, Jonathan and {Macci{\`o}}, Andrea V. and {Williams}, Liliya L.~R.},
        title = "{The Hubble Time Inferred from 10 Time Delay Lenses}",
      journal = {\apjl},
         year = 2006,
        month = oct,
       volume = {650},
       number = {1},
        pages = {L17-L20},
          doi = {10.1086/507583},
archivePrefix = {arXiv},
       eprint = {astro-ph/0607240},
 primaryClass = {astro-ph},
       adsurl = {https://ui.adsabs.harvard.edu/abs/2006ApJ...650L..17S}
}

@article{Oguri_2007,
	doi = {10.1086/513093},
	url = {https://doi.org/10.1086/513093},
	year = 2007,
	month = {may},
	publisher = {American Astronomical Society},
	volume = {660},
	number = {1},
	pages = {1--15},
	author = {Masamune Oguri},
	title = {Gravitational Lens Time Delays: A Statistical Assessment of Lens Model Dependences and Implications for the Global Hubble Constant},
	journal = {The Astrophysical Journal}
	}

@ARTICLE{Bonvin2017,
       author = {{Bonvin}, V. and {Courbin}, F. and {Suyu}, S.~H. and {Marshall}, P.~J. and {Rusu}, C.~E. and {Sluse}, D. and {Tewes}, M. and {Wong}, K.~C. and {Collett}, T. and {Fassnacht}, C.~D. and {Treu}, T. and {Auger}, M.~W. and {Hilbert}, S. and {Koopmans}, L.~V.~E. and {Meylan}, G. and {Rumbaugh}, N. and {Sonnenfeld}, A. and {Spiniello}, C.},
        title = "{H0LiCOW - V. New COSMOGRAIL time delays of HE 0435-1223: H$_{0}$ to 3.8 per cent precision from strong lensing in a flat {\ensuremath{\Lambda}}CDM model}",
      journal = {\mnras},
         year = 2017,
        month = mar,
       volume = {465},
       number = {4},
        pages = {4914-4930},
          doi = {10.1093/mnras/stw3006},
archivePrefix = {arXiv},
       eprint = {1607.01790},
 primaryClass = {astro-ph.CO},
       adsurl = {https://ui.adsabs.harvard.edu/abs/2017MNRAS.465.4914B}
}

@ARTICLE{Birrer:2020jyr,
       author = {{Birrer}, Simon and {Treu}, Tommaso},
        title = "{TDCOSMO. V. Strategies for precise and accurate measurements of the Hubble constant with strong lensing}",
      journal = {\aap},
         year = 2021,
        month = may,
       volume = {649},
          eid = {A61},
        pages = {A61},
          doi = {10.1051/0004-6361/202039179},
archivePrefix = {arXiv},
       eprint = {2008.06157},
 primaryClass = {astro-ph.CO},
       adsurl = {https://ui.adsabs.harvard.edu/abs/2021A&A...649A..61B}
}

@article{Mao:1997ek,
    author = "Mao, Shu-de and Schneider, Peter",
    title = "{Evidence for substructure in lens galaxies?}",
    eprint = "astro-ph/9707187",
    archivePrefix = "arXiv",
    doi = "10.1046/j.1365-8711.1998.01319.x",
    journal = "Mon. Not. Roy. Astron. Soc.",
    volume = "295",
    pages = "587--594",
    year = "1998"
}

@article{Metcalf:2001ap,
    author = "Metcalf, R. Benton and Madau, Piero",
    title = "{Compound gravitational lensing as a probe of dark matter substructure within galaxy halos}",
    eprint = "astro-ph/0108224",
    archivePrefix = "arXiv",
    doi = "10.1086/323695",
    journal = "Astrophys. J.",
    volume = "563",
    pages = "9",
    year = "2001"
}

@ARTICLE{Dalal2002,
       author = {{Dalal}, N. and {Kochanek}, C.~S.},
        title = "{Direct Detection of Cold Dark Matter Substructure}",
      journal = {\apj},
         year = 2002,
        month = jun,
       volume = {572},
       number = {1},
        pages = {25-33},
          doi = {10.1086/340303},
archivePrefix = {arXiv},
       eprint = {astro-ph/0111456},
 primaryClass = {astro-ph},
       adsurl = {https://ui.adsabs.harvard.edu/abs/2002ApJ...572...25D}
}

@article{Pooley_2009,
	doi = {10.1088/0004-637x/697/2/1892},
	url = {https://doi.org/10.1088/0004-637x/697/2/1892},
	year = 2009,
	month = {may},
	publisher = {American Astronomical Society},
	volume = {697},
	number = {2},
	pages = {1892--1900},
	author = {D. Pooley and S. Rappaport and J. Blackburne and P. L. Schechter and J. Schwab and J. Wambsganss},
	title = {{THE} {DARK}-{MATTER} {FRACTION} {IN} {THE} {ELLIPTICAL} {GALAXY} {LENSING} {THE} {QUASAR} {PG} 1115$\mathplus$080},
	journal = {The Astrophysical Journal}
}

@article{Oguri2014,
    author = {Oguri, Masamune and Rusu, Cristian E. and Falco, Emilio E.},
    title = "{The stellar and dark matter distributions in elliptical galaxies from the ensemble of strong gravitational lenses}",
    journal = {Monthly Notices of the Royal Astronomical Society},
    volume = {439},
    number = {3},
    pages = {2494-2504},
    year = {2014},
    month = {02},
    issn = {0035-8711},
    doi = {10.1093/mnras/stu106},
    url = {https://doi.org/10.1093/mnras/stu106},
    eprint = {https://academic.oup.com/mnras/article-pdf/439/3/2494/3809833/stu106.pdf},
}

@article{Jim_nez_Vicente_2019,
	doi = {10.3847/1538-4357/ab46b8},
	url = {https://doi.org/10.3847/1538-4357/ab46b8},
	year = 2019,
	month = {nov},
	publisher = {American Astronomical Society},
	volume = {885},
	number = {1},
	pages = {75},
	author = {J. Jim{\'{e}}nez-Vicente and E. Mediavilla},
	title = {The Initial Mass Function of Lens Galaxies from Quasar Microlensing},
	journal = {The Astrophysical Journal}
}

@article{Kelly:2014mwa,
    author = "Kelly, Patrick L. and others",
    title = "{Multiple Images of a Highly Magnified Supernova Formed by an Early-Type Cluster Galaxy Lens}",
    eprint = "1411.6009",
    archivePrefix = "arXiv",
    primaryClass = "astro-ph.CO",
    doi = "10.1126/science.aaa3350",
    journal = "Science",
    volume = "347",
    pages = "1123",
    year = "2015"
}

@article{Goobar:2016uuf,
    author = "Goobar, A. and others",
    title = "{iPTF16geu: A multiply imaged, gravitationally lensed type Ia supernova}",
    eprint = "1611.00014",
    archivePrefix = "arXiv",
    primaryClass = "astro-ph.CO",
    doi = "10.1126/science.aal2729",
    journal = "Science",
    volume = "356",
    pages = "291--295",
    year = "2017"
}

@ARTICLE{lsst1,
       author = {{LSST Science Collaboration} and {Abell}, Paul A. and {Allison}, Julius and {Anderson}, Scott F. and {Andrew}, John R. and {Angel}, J. Roger P. and {Armus}, Lee and {Arnett}, David and {Asztalos}, S.~J. and {Axelrod}, Tim S. and {Bailey}, Stephen and {Ballantyne}, D.~R. and {Bankert}, Justin R. and {Barkhouse}, Wayne A. and {Barr}, Jeffrey D. and {Barrientos}, L. Felipe and {Barth}, Aaron J. and {Bartlett}, James G. and {Becker}, Andrew C. and {Becla}, Jacek and {Beers}, Timothy C. and {Bernstein}, Joseph P. and {Biswas}, Rahul and {Blanton}, Michael R. and {Bloom}, Joshua S. and {Bochanski}, John J. and {Boeshaar}, Pat and {Borne}, Kirk D. and {Bradac}, Marusa and {Brandt}, W.~N. and {Bridge}, Carrie R. and {Brown}, Michael E. and {Brunner}, Robert J. and {Bullock}, James S. and {Burgasser}, Adam J. and {Burge}, James H. and {Burke}, David L. and {Cargile}, Phillip A. and {Chandrasekharan}, Srinivasan and {Chartas}, George and {Chesley}, Steven R. and {Chu}, You-Hua and {Cinabro}, David and {Claire}, Mark W. and {Claver}, Charles F. and {Clowe}, Douglas and {Connolly}, A.~J. and {Cook}, Kem H. and {Cooke}, Jeff and {Cooray}, Asantha and {Covey}, Kevin R. and {Culliton}, Christopher S. and {de Jong}, Roelof and {de Vries}, Willem H. and {Debattista}, Victor P. and {Delgado}, Francisco and {Dell'Antonio}, Ian P. and {Dhital}, Saurav and {Di Stefano}, Rosanne and {Dickinson}, Mark and {Dilday}, Benjamin and {Djorgovski}, S.~G. and {Dobler}, Gregory and {Donalek}, Ciro and {Dubois-Felsmann}, Gregory and {Durech}, Josef and {Eliasdottir}, Ardis and {Eracleous}, Michael and {Eyer}, Laurent and {Falco}, Emilio E. and {Fan}, Xiaohui and {Fassnacht}, Christopher D. and {Ferguson}, Harry C. and {Fernandez}, Yanga R. and {Fields}, Brian D. and {Finkbeiner}, Douglas and {Figueroa}, Eduardo E. and {Fox}, Derek B. and {Francke}, Harold and {Frank}, James S. and {Frieman}, Josh and {Fromenteau}, Sebastien and {Furqan}, Muhammad and {Galaz}, Gaspar and {Gal-Yam}, A. and {Garnavich}, Peter and {Gawiser}, Eric and {Geary}, John and {Gee}, Perry and {Gibson}, Robert R. and {Gilmore}, Kirk and {Grace}, Emily A. and {Green}, Richard F. and {Gressler}, William J. and {Grillmair}, Carl J. and {Habib}, Salman and {Haggerty}, J.~S. and {Hamuy}, Mario and {Harris}, Alan W. and {Hawley}, Suzanne L. and {Heavens}, Alan F. and {Hebb}, Leslie and {Henry}, Todd J. and {Hileman}, Edward and {Hilton}, Eric J. and {Hoadley}, Keri and {Holberg}, J.~B. and {Holman}, Matt J. and {Howell}, Steve B. and {Infante}, Leopoldo and {Ivezic}, Zeljko and {Jacoby}, Suzanne H. and {Jain}, Bhuvnesh and {R} and {Jedicke} and {Jee}, M. James and {Garrett Jernigan}, J. and {Jha}, Saurabh W. and {Johnston}, Kathryn V. and {Jones}, R. Lynne and {Juric}, Mario and {Kaasalainen}, Mikko and {Styliani} and {Kafka} and {Kahn}, Steven M. and {Kaib}, Nathan A. and {Kalirai}, Jason and {Kantor}, Jeff and {Kasliwal}, Mansi M. and {Keeton}, Charles R. and {Kessler}, Richard and {Knezevic}, Zoran and {Kowalski}, Adam and {Krabbendam}, Victor L. and {Krughoff}, K. Simon and {Kulkarni}, Shrinivas and {Kuhlman}, Stephen and {Lacy}, Mark and {Lepine}, Sebastien and {Liang}, Ming and {Lien}, Amy and {Lira}, Paulina and {Long}, Knox S. and {Lorenz}, Suzanne and {Lotz}, Jennifer M. and {Lupton}, R.~H. and {Lutz}, Julie and {Macri}, Lucas M. and {Mahabal}, Ashish A. and {Mandelbaum}, Rachel and {Marshall}, Phil and {May}, Morgan and {McGehee}, Peregrine M. and {Meadows}, Brian T. and {Meert}, Alan and {Milani}, Andrea and {Miller}, Christopher J. and {Miller}, Michelle and {Mills}, David and {Minniti}, Dante and {Monet}, David and {Mukadam}, Anjum S. and {Nakar}, Ehud and {Neill}, Douglas R. and {Newman}, Jeffrey A. and {Nikolaev}, Sergei and {Nordby}, Martin and {O'Connor}, Paul and {Oguri}, Masamune and {Oliver}, John and {Olivier}, Scot S. and {Olsen}, Julia K. and {Olsen}, Knut and {Olszewski}, Edward W. and {Oluseyi}, Hakeem and {Padilla}, Nelson D. and {Parker}, Alex and {Pepper}, Joshua and {Peterson}, John R. and {Petry}, Catherine and {Pinto}, Philip A. and {Pizagno}, James L. and {Popescu}, Bogdan and {Prsa}, Andrej and {Radcka}, Veljko and {Raddick}, M. Jordan and {Rasmussen}, Andrew and {Rau}, Arne and {Rho}, Jeonghee and {Rhoads}, James E. and {Richards}, Gordon T. and {Ridgway}, Stephen T. and {Robertson}, Brant E. and {Roskar}, Rok and {Saha}, Abhijit and {Sarajedini}, Ata and {Scannapieco}, Evan and {Schalk}, Terry and {Schindler}, Rafe and {Schmidt}, Samuel and {Schmidt}, Sarah and {Schneider}, Donald P. and {Schumacher}, German and {Scranton}, Ryan and {Sebag}, Jacques and {Seppala}, Lynn G. and {Shemmer}, Ohad and {Simon}, Joshua D. and {Sivertz}, M. and {Smith}, Howard A. and {Allyn Smith}, J. and {Smith}, Nathan and {Spitz}, Anna H. and {Stanford}, Adam and {Stassun}, Keivan G. and {Strader}, Jay and {Strauss}, Michael A. and {Stubbs}, Christopher W. and {Sweeney}, Donald W. and {Szalay}, Alex and {Szkody}, Paula and {Takada}, Masahiro and {Thorman}, Paul and {Trilling}, David E. and {Trimble}, Virginia and {Tyson}, Anthony and {Van Berg}, Richard and {Vanden Berk}, Daniel and {VanderPlas}, Jake and {Verde}, Licia and {Vrsnak}, Bojan and {Walkowicz}, Lucianne M. and {Wandelt}, Benjamin D. and {Wang}, Sheng and {Wang}, Yun and {Warner}, Michael and {Wechsler}, Risa H. and {West}, Andrew A. and {Wiecha}, Oliver and {Williams}, Benjamin F. and {Willman}, Beth and {Wittman}, David and {Wolff}, Sidney C. and {Wood-Vasey}, W. Michael and {Wozniak}, Przemek and {Young}, Patrick and {Zentner}, Andrew and {Zhan}, Hu},
        title = "{LSST Science Book, Version 2.0}",
      journal = {arXiv e-prints},
         year = 2009,
        month = dec,
          eid = {arXiv:0912.0201},
        pages = {arXiv:0912.0201},
archivePrefix = {arXiv},
       eprint = {0912.0201},
 primaryClass = {astro-ph.IM},
       adsurl = {https://ui.adsabs.harvard.edu/abs/2009arXiv0912.0201L}
}

@article{misha2021,
       author = {{Denissenya}, Mikhail and {Bag}, Satadru and {Kim}, Alex G. and {Linder}, Eric V. and {Shafieloo}, Arman},
        title = "{Out of one, many: distinguishing time delays from lensed supernovae}",
      journal = {\mnras},
         year = 2022,
        month = mar,
       volume = {511},
       number = {1},
        pages = {1210-1217},
          doi = {10.1093/mnras/stac143},
archivePrefix = {arXiv},
       eprint = {2109.13282},
 primaryClass = {astro-ph.CO},
       adsurl = {https://ui.adsabs.harvard.edu/abs/2022MNRAS.511.1210D}
}

@ARTICLE{2021NatAs.tmp..164R,
       author = {{Rodney}, Steven A. and {Brammer}, Gabriel B. and {Pierel}, Justin D.~R. and {Richard}, Johan and {Toft}, Sune and {O'Connor}, Kyle F. and {Akhshik}, Mohammad and {Whitaker}, Katherine E.},
        title = "{A gravitationally lensed supernova with an observable two-decade time delay}",
      journal = {Nature Astronomy},
         year = 2021,
        month = sep,
          doi = {10.1038/s41550-021-01450-9},
archivePrefix = {arXiv},
       eprint = {2106.08935},
 primaryClass = {astro-ph.CO},
       adsurl = {https://ui.adsabs.harvard.edu/abs/2021NatAs.tmp..164R}
}

@ARTICLE{misha2022,
       author = {{Denissenya}, Mikhail and {Linder}, Eric V.},
        title = "{Deep learning unresolved lensed light curves}",
      journal = {\mnras},
         year = 2022,
        month = sep,
       volume = {515},
       number = {1},
        pages = {977-983},
          doi = {10.1093/mnras/stac1726},
archivePrefix = {arXiv},
       eprint = {2202.11903},
 primaryClass = {astro-ph.IM},
       adsurl = {https://ui.adsabs.harvard.edu/abs/2022MNRAS.515..977D}
}

@ARTICLE{1996astro.ph..6001N,
       author = {{Narayan}, Ramesh and {Bartelmann}, Matthias},
        title = "{Lectures on Gravitational Lensing}",
      journal = {arXiv e-prints},
         year = 1996,
        month = jun,
          eid = {astro-ph/9606001},
        pages = {astro-ph/9606001},
archivePrefix = {arXiv},
       eprint = {astro-ph/9606001},
 primaryClass = {astro-ph},
       adsurl = {https://ui.adsabs.harvard.edu/abs/1996astro.ph..6001N}
}

@ARTICLE{Oguri2018,
       author = {{Oguri}, Masamune},
        title = "{Effect of gravitational lensing on the distribution of gravitational waves from distant binary black hole mergers}",
      journal = {\mnras},
         year = 2018,
        month = nov,
       volume = {480},
       number = {3},
        pages = {3842-3855},
          doi = {10.1093/mnras/sty2145},
archivePrefix = {arXiv},
       eprint = {1807.02584},
 primaryClass = {astro-ph.CO},
       adsurl = {https://ui.adsabs.harvard.edu/abs/2018MNRAS.480.3842O}
}

@ARTICLE{Quimby2014,
       author = {{Quimby}, Robert M. and {Oguri}, Masamune and {More}, Anupreeta and {More}, Surhud and {Moriya}, Takashi J. and {Werner}, Marcus C. and {Tanaka}, Masayuki and {Folatelli}, Gaston and {Bersten}, Melina C. and {Maeda}, Keiichi and {Nomoto}, Ken'ichi},
        title = "{Detection of the Gravitational Lens Magnifying a Type Ia Supernova}",
      journal = {Science},
         year = 2014,
        month = apr,
       volume = {344},
       number = {6182},
        pages = {396-399},
          doi = {10.1126/science.1250903},
archivePrefix = {arXiv},
       eprint = {1404.6014},
 primaryClass = {astro-ph.CO},
       adsurl = {https://ui.adsabs.harvard.edu/abs/2014Sci...344..396Q}
}

@ARTICLE{Suyu_2024,
       author = {{Suyu}, Sherry H. and {Goobar}, Ariel and {Collett}, Thomas and {More}, Anupreeta and {Vernardos}, Giorgos},
        title = "{Strong Gravitational Lensing and Microlensing of Supernovae}",
      journal = {\ssr},
         year = 2024,
        month = feb,
       volume = {220},
       number = {1},
          eid = {13},
        pages = {13},
          doi = {10.1007/s11214-024-01044-7},
archivePrefix = {arXiv},
       eprint = {2301.07729},
 primaryClass = {astro-ph.CO},
       adsurl = {https://ui.adsabs.harvard.edu/abs/2024SSRv..220...13S}
}

@ARTICLE{Nikki_2023,
       author = {{Arendse}, Nikki and {Dhawan}, Suhail and {Sagu{\'e}s Carracedo}, Ana and {Peiris}, Hiranya V. and {Goobar}, Ariel and {Wojtak}, Radek and {Alves}, Catarina and {Biswas}, Rahul and {Huber}, Simon and {Birrer}, Simon and {The LSST Dark Energy Science Collaboration}},
        title = "{Detecting strongly lensed type Ia supernovae with LSST}",
      journal = {\mnras},
         year = 2024,
        month = jul,
       volume = {531},
       number = {3},
        pages = {3509-3523},
          doi = {10.1093/mnras/stae1356},
archivePrefix = {arXiv},
       eprint = {2312.04621},
 primaryClass = {astro-ph.CO},
       adsurl = {https://ui.adsabs.harvard.edu/abs/2024MNRAS.531.3509A}
}

@ARTICLE{Wojtak2019,
       author = {{Wojtak}, Rados{\l}aw and {Hjorth}, Jens and {Gall}, Christa},
        title = "{Magnified or multiply imaged? - Search strategies for gravitationally lensed supernovae in wide-field surveys}",
      journal = {\mnras},
         year = 2019,
        month = aug,
       volume = {487},
       number = {3},
        pages = {3342-3355},
          doi = {10.1093/mnras/stz1516},
archivePrefix = {arXiv},
       eprint = {1903.07687},
 primaryClass = {astro-ph.CO},
       adsurl = {https://ui.adsabs.harvard.edu/abs/2019MNRAS.487.3342W}
}

@ARTICLE{Frye2023,
       author = {{Frye}, B. and {Pascale}, M. and {Cohen}, S. and {Summers}, J. and {Foo}, N. and {Kamieneski}, P. and {Carleton}, T. and {Jansen}, R.~A. and {Pierel}, J. and {Engesser}, M. and {Chen}, W. and {Austin}, D. and {Marshall}, M. and {Trussler}, J. and {Meena}, A. and {Leimbach}, R. and {Garuda}, N. and {Honor}, R. and {Furtak}, L.~J. and {Strolger}, L. and {Windhorst}, R.~A. and {Koekemoer}, A. and {Zitrin}, A. and {Diego}, J. and {Kelly}, P. and {Coe}, D. and {Conselice}, C. and {Dai}, L. and {D{\^a}Silva}, J. and {Dole}, H. and {Driver}, S. and {Grogin}, N. and {Nonino}, M. and {Pirzkal}, N. and {Polletta}, M. and {Robotham}, A. and {Rutkowski}, M. and {Ryan}, R. and {Tompkins}, S. and {Willmer}, C. and {Willner}, S. and {Yan}, H. and {Yun}, M.},
        title = "{SN H0pe: three images of a SN detected near the central region of the galaxy cluster field PLCK G165.7+67.0}",
      journal = {Transient Name Server AstroNote},
         year = 2023,
        month = apr,
       volume = {96},
        pages = {1},
       adsurl = {https://ui.adsabs.harvard.edu/abs/2023TNSAN..96....1F}
}

@ARTICLE{Goobar2023,
       author = {{Goobar}, Ariel and {Johansson}, Joel and {Schulze}, Steve and {Arendse}, Nikki and {Carracedo}, Ana Sagu{\'e}s and {Dhawan}, Suhail and {M{\"o}rtsell}, Edvard and {Fremling}, Christoffer and {Yan}, Lin and {Perley}, Daniel and {Sollerman}, Jesper and {Joseph}, R{\'e}my and {Hinds}, K. -Ryan and {Meynardie}, William and {Andreoni}, Igor and {Bellm}, Eric and {Bloom}, Josh and {Collett}, Thomas E. and {Drake}, Andrew and {Graham}, Matthew and {Kasliwal}, Mansi and {Kulkarni}, Shri R. and {Lemon}, Cameron and {Miller}, Adam A. and {Neill}, James D. and {Nordin}, Jakob and {Pierel}, Justin and {Richard}, Johan and {Riddle}, Reed and {Rigault}, Mickael and {Rusholme}, Ben and {Sharma}, Yashvi and {Stein}, Robert and {Stewart}, Gabrielle and {Townsend}, Alice and {Vinko}, Jozsef and {Wheeler}, J. Craig and {Wold}, Avery},
        title = "{Uncovering a population of gravitational lens galaxies with magnified standard candle SN Zwicky}",
      journal = {Nature Astronomy},
         year = 2023,
        month = jun,
       volume = {7},
        pages = {1098-1107},
          doi = {10.1038/s41550-023-01981-3},
archivePrefix = {arXiv},
       eprint = {2211.00656},
 primaryClass = {astro-ph.CO},
       adsurl = {https://ui.adsabs.harvard.edu/abs/2023NatAs...7.1098G}
}

@ARTICLE{chen2022,
       author = {{Chen}, Wenlei and {Kelly}, Patrick L. and {Oguri}, Masamune and {Broadhurst}, Thomas J. and {Diego}, Jose M. and {Emami}, Najmeh and {Filippenko}, Alexei V. and {Treu}, Tommaso L. and {Zitrin}, Adi},
        title = "{Shock cooling of a red-supergiant supernova at redshift 3 in lensed images}",
      journal = {\nat},
         year = 2022,
        month = nov,
       volume = {611},
       number = {7935},
        pages = {256-259},
          doi = {10.1038/s41586-022-05252-5},
archivePrefix = {arXiv},
       eprint = {2306.12985},
 primaryClass = {astro-ph.GA},
       adsurl = {https://ui.adsabs.harvard.edu/abs/2022Natur.611..256C}
}

@ARTICLE{sn_encore,
        author = {{Pierel}, J.~D.~R. and {Newman}, A.~B. and {Dhawan}, S. and {Gu}, M. and {Joshi}, B.~A. and {Li}, T. and {Schuldt}, S. and {Strolger}, L.~G. and {Suyu}, S.~H. and {Caminha}, G.~B. and {Cohen}, S.~H. and {Diego}, J.~M. and {D{\'S}ilva}, J.~C.~J. and {Ertl}, S. and {Frye}, B.~L. and {Granata}, G. and {Grillo}, C. and {Koekemoer}, A.~M. and {Li}, J. and {Robotham}, A. and {Summers}, J. and {Treu}, T. and {Windhorst}, R.~A. and {Zitrin}, A. and {Agarwal}, S. and {Agrawal}, A. and {Arendse}, N. and {Belli}, S. and {Burns}, C. and {Ca{\~n}ameras}, R. and {Chakrabarti}, S. and {Chen}, W. and {Collett}, T.~E. and {Coulter}, D.~A. and {Ellis}, R.~S. and {Engesser}, M. and {Foo}, N. and {Fox}, O.~D. and {Gall}, C. and {Garuda}, N. and {Gezari}, S. and {Gomez}, S. and {Glazebrook}, K. and {Hjorth}, J. and {Huang}, X. and {Jha}, S.~W. and {Kamieneski}, P.~S. and {Kelly}, P. and {Larison}, C. and {Moustakas}, L.~A. and {Pascale}, M. and {P{\'e}rez-Fournon}, I. and {Petrushevska}, T. and {Poidevin}, F. and {Rest}, A. and {Shahbandeh}, M. and {Shajib}, A.~J. and {Siebert}, M. and {Storfer}, C. and {Talbot}, M. and {Wang}, Q. and {Wevers}, T. and {Zenati}, Y.},
        title = "{Lensed Type Ia Supernova ``Encore'' at z = 2: The First Instance of Two Multiply Imaged Supernovae in the Same Host Galaxy}",
      journal = {\apjl},
         year = 2024,
        month = jun,
       volume = {967},
       number = {2},
          eid = {L37},
        pages = {L37},
          doi = {10.3847/2041-8213/ad4648},
archivePrefix = {arXiv},
       eprint = {2404.02139},
 primaryClass = {astro-ph.CO},
       adsurl = {https://ui.adsabs.harvard.edu/abs/2024ApJ...967L..37P}
}

@ARTICLE{Ramanah2022,
       author = {{Kodi Ramanah}, Doogesh and {Arendse}, Nikki and {Wojtak}, Rados{\l}aw},
        title = "{AI-driven spatio-temporal engine for finding gravitationally lensed type Ia supernovae}",
      journal = {\mnras},
         year = 2022,
        month = jun,
       volume = {512},
       number = {4},
        pages = {5404-5417},
          doi = {10.1093/mnras/stac838},
archivePrefix = {arXiv},
       eprint = {2107.12399},
 primaryClass = {astro-ph.IM},
       adsurl = {https://ui.adsabs.harvard.edu/abs/2022MNRAS.512.5404K}
}

@article{Treu:2022aqp,
    author = "Treu, Tommaso and Suyu, Sherry H. and Marshall, Philip J.",
    title = "{Strong lensing time-delay cosmography in the 2020s}",
    eprint = "2210.15794",
    archivePrefix = "arXiv",
    primaryClass = "astro-ph.CO",
    doi = "10.1007/s00159-022-00145-y",
    journal = "Astron. Astrophys. Rev.",
    volume = "30",
    number = "1",
    pages = "8",
    year = "2022"
}

@ARTICLE{Grillo2024,
       author = {{Grillo}, C. and {Pagano}, L. and {Rosati}, P. and {Suyu}, S.~H.},
        title = "{Cosmography with supernova Refsdal through time-delay cluster lensing: Independent measurements of the Hubble constant and geometry of the Universe}",
      journal = {\aap},
         year = 2024,
        month = apr,
       volume = {684},
          eid = {L23},
        pages = {L23},
          doi = {10.1051/0004-6361/202449278},
archivePrefix = {arXiv},
       eprint = {2401.10980},
 primaryClass = {astro-ph.CO},
       adsurl = {https://ui.adsabs.harvard.edu/abs/2024A&A...684L..23G}
}

@ARTICLE{convlstm,
       author = {{Shi}, Xingjian and {Chen}, Zhourong and {Wang}, Hao and {Yeung}, Dit-Yan and {Wong}, Wai-kin and {Woo}, Wang-chun},
        title = "{Convolutional LSTM Network: A Machine Learning Approach for Precipitation Nowcasting}",
      journal = {arXiv e-prints},
         year = 2015,
        month = jun,
          eid = {arXiv:1506.04214},
        pages = {arXiv:1506.04214},
          doi = {10.48550/arXiv.1506.04214},
archivePrefix = {arXiv},
       eprint = {1506.04214},
 primaryClass = {cs.CV},
       adsurl = {https://ui.adsabs.harvard.edu/abs/2015arXiv150604214S}
}

@INPROCEEDINGS{bertin02,
       author = {{Bertin}, Emmanuel and {Mellier}, Yannick and {Radovich}, Mario and {Missonnier}, Gilles and {Didelon}, Pierre and {Morin}, Bertrand},
        title = "{The TERAPIX Pipeline}",
    booktitle = {Astronomical Data Analysis Software and Systems XI},
         year = 2002,
       editor = {{Bohlender}, David A. and {Durand}, Daniel and {Handley}, Thomas H.},
       series = {Astronomical Society of the Pacific Conference Series},
       volume = {281},
        month = jan,
        pages = {228},
       adsurl = {https://ui.adsabs.harvard.edu/abs/2002ASPC..281..228B}
}

@ARTICLE{bosch18,
       author = {{Bosch}, James and {Armstrong}, Robert and {Bickerton}, Steven and {Furusawa}, Hisanori and {Ikeda}, Hiroyuki and {Koike}, Michitaro and {Lupton}, Robert and {Mineo}, Sogo and {Price}, Paul and {Takata}, Tadafumi and {Tanaka}, Masayuki and {Yasuda}, Naoki and {AlSayyad}, Yusra and {Becker}, Andrew C. and {Coulton}, William and {Coupon}, Jean and {Garmilla}, Jose and {Huang}, Song and {Krughoff}, K. Simon and {Lang}, Dustin and {Leauthaud}, Alexie and {Lim}, Kian-Tat and {Lust}, Nate B. and {MacArthur}, Lauren A. and {Mandelbaum}, Rachel and {Miyatake}, Hironao and {Miyazaki}, Satoshi and {Murata}, Ryoma and {More}, Surhud and {Okura}, Yuki and {Owen}, Russell and {Swinbank}, John D. and {Strauss}, Michael A. and {Yamada}, Yoshihiko and {Yamanoi}, Hitomi},
        title = "{The Hyper Suprime-Cam software pipeline}",
      journal = {\pasj},
         year = 2018,
        month = jan,
       volume = {70},
          eid = {S5},
        pages = {S5},
          doi = {10.1093/pasj/psx080},
archivePrefix = {arXiv},
       eprint = {1705.06766},
 primaryClass = {astro-ph.IM},
       adsurl = {https://ui.adsabs.harvard.edu/abs/2018PASJ...70S...5B}
}

@ARTICLE{cosmograil1,
       author = {{Millon}, M. and {Courbin}, F. and {Bonvin}, V. and {Paic}, E. and {Meylan}, G. and {Tewes}, M. and {Sluse}, D. and {Magain}, P. and {Chan}, J.~H.~H. and {Galan}, A. and {Joseph}, R. and {Lemon}, C. and {Tihhonova}, O. and {Anderson}, R.~I. and {Marmier}, M. and {Chazelas}, B. and {Lendl}, M. and {Triaud}, A.~H.~M.~J. and {Wyttenbach}, A.},
        title = "{COSMOGRAIL. XIX. Time delays in 18 strongly lensed quasars from 15 years of optical monitoring}",
      journal = {\aap},
         year = 2020,
        month = aug,
       volume = {640},
          eid = {A105},
        pages = {A105},
          doi = {10.1051/0004-6361/202037740},
archivePrefix = {arXiv},
       eprint = {2002.05736},
 primaryClass = {astro-ph.CO},
       adsurl = {https://ui.adsabs.harvard.edu/abs/2020A&A...640A.105M}
}

@ARTICLE{glee1,
       author = {{Suyu}, S.~H. and {Halkola}, A.},
        title = "{The halos of satellite galaxies: the companion of the massive elliptical lens SL2S J08544-0121}",
      journal = {\aap},
         year = 2010,
        month = dec,
       volume = {524},
          eid = {A94},
        pages = {A94},
          doi = {10.1051/0004-6361/201015481},
archivePrefix = {arXiv},
       eprint = {1007.4815},
 primaryClass = {astro-ph.CO},
       adsurl = {https://ui.adsabs.harvard.edu/abs/2010A&A...524A..94S}
}

@ARTICLE{glee2,
       author = {{Suyu}, S.~H. and {Hensel}, S.~W. and {McKean}, J.~P. and {Fassnacht}, C.~D. and {Treu}, T. and {Halkola}, A. and {Norbury}, M. and {Jackson}, N. and {Schneider}, P. and {Thompson}, D. and {Auger}, M.~W. and {Koopmans}, L.~V.~E. and {Matthews}, K.},
        title = "{Disentangling Baryons and Dark Matter in the Spiral Gravitational Lens B1933+503}",
      journal = {\apj},
         year = 2012,
        month = may,
       volume = {750},
       number = {1},
          eid = {10},
        pages = {10},
          doi = {10.1088/0004-637X/750/1/10},
archivePrefix = {arXiv},
       eprint = {1110.2536},
 primaryClass = {astro-ph.CO},
       adsurl = {https://ui.adsabs.harvard.edu/abs/2012ApJ...750...10S}
}

@ARTICLE{Stefan21,
       author = {{Schuldt}, S. and {Suyu}, S.~H. and {Meinhardt}, T. and {Leal-Taix{\'e}}, L. and {Ca{\~n}ameras}, R. and {Taubenberger}, S. and {Halkola}, A.},
        title = "{HOLISMOKES. IV. Efficient mass modeling of strong lenses through deep learning}",
      journal = {\aap},
         year = 2021,
        month = feb,
       volume = {646},
          eid = {A126},
        pages = {A126},
          doi = {10.1051/0004-6361/202039574},
archivePrefix = {arXiv},
       eprint = {2010.00602},
 primaryClass = {astro-ph.GA},
       adsurl = {https://ui.adsabs.harvard.edu/abs/2021A&A...646A.126S}
}

@article{Bag:2024kbk,
    author = "Bag, Satadru and others",
    title = "{Detecting unresolved lensed SNe Ia in LSST using blended light curves}",
    eprint = "2404.15389",
    archivePrefix = "arXiv",
    primaryClass = "astro-ph.IM",
    doi = "10.1051/0004-6361/202450485",
    journal = "Astron. Astrophys.",
    volume = "691",
    pages = "A100",
    year = "2024"
}

@ARTICLE{Ana_2024,
       author = {{Sainz de Murieta}, Ana and {Collett}, Thomas E. and {Magee}, Mark R. and {Pierel}, Justin D.~R. and {Enzi}, Wolfgang J.~R. and {Lokken}, Martine and {Gagliano}, Alex and {Ryczanowski}, Dan},
        title = "{Find the haystacks, then look for needles: the rate of strongly lensed supernovae in galaxy-galaxy strong gravitational lenses}",
      journal = {\mnras},
         year = 2024,
        month = dec,
       volume = {535},
       number = {3},
        pages = {2523-2537},
          doi = {10.1093/mnras/stae2486},
archivePrefix = {arXiv},
       eprint = {2407.04080},
 primaryClass = {astro-ph.CO},
       adsurl = {https://ui.adsabs.harvard.edu/abs/2024MNRAS.535.2523S}
}

@ARTICLE{Ana_2023,
       author = {{Sainz de Murieta}, Ana and {Collett}, Thomas E. and {Magee}, Mark R. and {Weisenbach}, Luke and {Krawczyk}, Coleman M. and {Enzi}, Wolfgang},
        title = "{Lensed Type Ia supernovae in light of SN Zwicky and iPTF16geu}",
      journal = {\mnras},
         year = 2023,
        month = dec,
       volume = {526},
       number = {3},
        pages = {4296-4307},
          doi = {10.1093/mnras/stad3031},
archivePrefix = {arXiv},
       eprint = {2307.12881},
 primaryClass = {astro-ph.CO},
       adsurl = {https://ui.adsabs.harvard.edu/abs/2023MNRAS.526.4296S}
}

@ARTICLE{Morgan2022,
       author = {{Morgan}, R. and {Nord}, B. and {Bechtol}, K. and {Gonz{\'a}lez}, S.~J. and {Buckley-Geer}, E. and {M{\"o}ller}, A. and {Park}, J.~W. and {Kim}, A.~G. and {Birrer}, S. and {Aguena}, M. and {Annis}, J. and {Bocquet}, S. and {Brooks}, D. and {Carnero Rosell}, A. and {Carrasco Kind}, M. and {Carretero}, J. and {Cawthon}, R. and {da Costa}, L.~N. and {Davis}, T.~M. and {De Vicente}, J. and {Doel}, P. and {Ferrero}, I. and {Friedel}, D. and {Frieman}, J. and {Garc{\'\i}a-Bellido}, J. and {Gatti}, M. and {Gaztanaga}, E. and {Giannini}, G. and {Gruen}, D. and {Gruendl}, R.~A. and {Gutierrez}, G. and {Hollowood}, D.~L. and {Honscheid}, K. and {James}, D.~J. and {Kuehn}, K. and {Kuropatkin}, N. and {Maia}, M.~A.~G. and {Miquel}, R. and {Palmese}, A. and {Paz-Chinch{\'o}n}, F. and {Pereira}, M.~E.~S. and {Pieres}, A. and {Plazas Malag{\'o}n}, A.~A. and {Reil}, K. and {Roodman}, A. and {Sanchez}, E. and {Smith}, M. and {Suchyta}, E. and {Swanson}, M.~E.~C. and {Tarle}, G. and {To}, C.},
        title = "{DeepZipper: A Novel Deep-learning Architecture for Lensed Supernovae Identification}",
      journal = {\apj},
         year = 2022,
        month = mar,
       volume = {927},
       number = {1},
          eid = {109},
        pages = {109},
          doi = {10.3847/1538-4357/ac5178},
archivePrefix = {arXiv},
       eprint = {2112.01541},
 primaryClass = {astro-ph.CO},
       adsurl = {https://ui.adsabs.harvard.edu/abs/2022ApJ...927..109M}
}

@ARTICLE{Morgan2023,
       author = {{Morgan}, R. and {Nord}, B. and {Bechtol}, K. and {M{\"o}ller}, A. and {Hartley}, W.~G. and {Birrer}, S. and {Gonz{\'a}lez}, S.~J. and {Martinez}, M. and {Gruendl}, R.~A. and {Buckley-Geer}, E.~J. and {Shajib}, A.~J. and {Carnero Rosell}, A. and {Lidman}, C. and {Collett}, T. and {Abbott}, T.~M.~C. and {Aguena}, M. and {Andrade-Oliveira}, F. and {Annis}, J. and {Bacon}, D. and {Bocquet}, S. and {Brooks}, D. and {Burke}, D.~L. and {Carrasco Kind}, M. and {Carretero}, J. and {Castander}, F.~J. and {Conselice}, C. and {da Costa}, L.~N. and {Costanzi}, M. and {De Vicente}, J. and {Desai}, S. and {Doel}, P. and {Everett}, S. and {Ferrero}, I. and {Flaugher}, B. and {Friedel}, D. and {Frieman}, J. and {Garc{\'\i}a-Bellido}, J. and {Gaztanaga}, E. and {Gruen}, D. and {Gutierrez}, G. and {Hinton}, S.~R. and {Hollowood}, D.~L. and {Honscheid}, K. and {Kuehn}, K. and {Kuropatkin}, N. and {Lahav}, O. and {Lima}, M. and {Menanteau}, F. and {Miquel}, R. and {Palmese}, A. and {Paz-Chinch{\'o}n}, F. and {Pereira}, M.~E.~S. and {Pieres}, A. and {Plazas Malag{\'o}n}, A.~A. and {Prat}, J. and {Rodriguez-Monroy}, M. and {Romer}, A.~K. and {Roodman}, A. and {Sanchez}, E. and {Scarpine}, V. and {Sevilla-Noarbe}, I. and {Smith}, M. and {Suchyta}, E. and {Swanson}, M.~E.~C. and {Tarle}, G. and {Thomas}, D. and {Varga}, T.~N.},
        title = "{DeepZipper. II. Searching for Lensed Supernovae in Dark Energy Survey Data with Deep Learning}",
      journal = {\apj},
         year = 2023,
        month = jan,
       volume = {943},
       number = {1},
          eid = {19},
        pages = {19},
          doi = {10.3847/1538-4357/ac721b},
archivePrefix = {arXiv},
       eprint = {2204.05924},
 primaryClass = {astro-ph.CO},
       adsurl = {https://ui.adsabs.harvard.edu/abs/2023ApJ...943...19M}
}

@ARTICLE{Tripp1998,
       author = {{Tripp}, Robert},
        title = "{A two-parameter luminosity correction for Type IA supernovae}",
      journal = {\aap},
         year = 1998,
        month = mar,
       volume = {331},
        pages = {815-820},
       adsurl = {https://ui.adsabs.harvard.edu/abs/1998A&A...331..815T}
}

@software{sncosmo,
       author = {{Barbary}, Kyle and {Barclay}, Tom and {Biswas}, Rahul and {Craig}, Matt and {Feindt}, Ulrich and {Friesen}, Brian and {Goldstein}, Danny and {Jha}, Saurabh and {Rodney}, Steve and {Sofiatti}, Caroline and {Thomas}, Rollin C. and {Wood-Vasey}, Michael},
        title = "{SNCosmo: Python library for supernova cosmology}",
 howpublished = {Astrophysics Source Code Library, record ascl:1611.017},
         year = 2016,
        month = nov,
          eid = {ascl:1611.017},
       adsurl = {https://ui.adsabs.harvard.edu/abs/2016ascl.soft11017B}
}

@ARTICLE{Hsiao2007,
       author = {{Hsiao}, E.~Y. and {Conley}, A. and {Howell}, D.~A. and {Sullivan}, M. and {Pritchet}, C.~J. and {Carlberg}, R.~G. and {Nugent}, P.~E. and {Phillips}, M.~M.},
        title = "{K-Corrections and Spectral Templates of Type Ia Supernovae}",
      journal = {\apj},
         year = 2007,
        month = jul,
       volume = {663},
       number = {2},
        pages = {1187-1200},
          doi = {10.1086/518232},
archivePrefix = {arXiv},
       eprint = {astro-ph/0703529},
 primaryClass = {astro-ph},
       adsurl = {https://ui.adsabs.harvard.edu/abs/2007ApJ...663.1187H}
}

@Article{convlstm2,
AUTHOR = {Naz, Farah and She, Lei and Sinan, Muhammad and Shao, Jie},
TITLE = {Enhancing Radar Echo Extrapolation by ConvLSTM2D for Precipitation Nowcasting},
JOURNAL = {Sensors},
VOLUME = {24},
YEAR = {2024},
NUMBER = {2},
ARTICLE-NUMBER = {459},
URL = {https://www.mdpi.com/1424-8220/24/2/459},
PubMedID = {38257552},
ISSN = {1424-8220},
}

@ARTICLE{TDCOSMO2025,
       author = {{TDCOSMO Collaboration} and {Birrer}, Simon and {Buckley-Geer}, Elizabeth J. and {Cappellari}, Michele and {Courbin}, Fr{\'e}d{\'e}ric and {Dux}, Fr{\'e}d{\'e}ric and {Fassnacht}, Christopher D. and {Frieman}, Joshua A. and {Galan}, Aymeric and {Gilman}, Daniel and {Huang}, Xiang-Yu and {Knabel}, Shawn and {Langeroodi}, Danial and {Lin}, Huan and {Millon}, Martin and {Morishita}, Takahiro and {Motta}, Veronica and {Mozumdar}, Pritom and {Paic}, Eric and {Shajib}, Anowar J. and {Sheu}, William and {Sluse}, Dominique and {Sonnenfeld}, Alessandro and {Spiniello}, Chiara and {Stiavelli}, Massimo and {Suyu}, Sherry H. and {Tan}, Chin Yi and {Treu}, Tommaso and {Van de Vyvere}, Lyne and {Wang}, Han and {Wells}, Patrick and {Williams}, Devon M. and {Wong}, Kenneth C.},
        title = "{TDCOSMO 2025: Cosmological constraints from strong lensing time delays}",
      journal = {arXiv e-prints},
         year = 2025,
        month = jun,
          eid = {arXiv:2506.03023},
        pages = {arXiv:2506.03023},
          doi = {10.48550/arXiv.2506.03023},
archivePrefix = {arXiv},
       eprint = {2506.03023},
 primaryClass = {astro-ph.CO},
       adsurl = {https://ui.adsabs.harvard.edu/abs/2025arXiv250603023T}
}

@ARTICLE{Euclid2025,
       author = {{Euclid Collaboration} and {Mellier}, Y. and {Abdurro'uf} and {Acevedo Barroso}, J.~A. and {Ach{\'u}carro}, A. and {Adamek}, J. and {Adam}, R. and {Addison}, G.~E. and {Aghanim}, N. and {Aguena}, M. and {Ajani}, V. and {Akrami}, Y. and {Al-Bahlawan}, A. and {Alavi}, A. and {Albuquerque}, I.~S. and {Alestas}, G. and {Alguero}, G. and {Allaoui}, A. and {Allen}, S.~W. and {Allevato}, V. and {Alonso-Tetilla}, A.~V. and {Altieri}, B. and {Alvarez-Candal}, A. and {Alvi}, S. and {Amara}, A. and {Amendola}, L. and {Amiaux}, J. and {Andika}, I.~T. and {Andreon}, S. and {Andrews}, A. and {Angora}, G. and {Angulo}, R.~E. and {Annibali}, F. and {Anselmi}, A. and {Anselmi}, S. and {Arcari}, S. and {Archidiacono}, M. and {Aric{\`o}}, G. and {Arnaud}, M. and {Arnouts}, S. and {Asgari}, M. and {Asorey}, J. and {Atayde}, L. and {Atek}, H. and {Atrio-Barandela}, F. and {Aubert}, M. and {Aubourg}, E. and {Auphan}, T. and {Auricchio}, N. and {Aussel}, B. and {Aussel}, H. and {Avelino}, P.~P. and {Avgoustidis}, A. and {Avila}, S. and {Awan}, S. and {Azzollini}, R. and {Baccigalupi}, C. and {Bachelet}, E. and {Bacon}, D. and {Baes}, M. and {Bagley}, M.~B. and {Bahr-Kalus}, B. and {Balaguera-Antolinez}, A. and {Balbinot}, E. and {Balcells}, M. and {Baldi}, M. and {Baldry}, I. and {Balestra}, A. and {Ballardini}, M. and {Ballester}, O. and {Balogh}, M. and {Ba{\~n}ados}, E. and {Barbier}, R. and {Bardelli}, S. and {Baron}, M. and {Barreiro}, T. and {Barrena}, R. and {Barriere}, J. -C. and {Barros}, B.~J. and {Barthelemy}, A. and {Bartolo}, N. and {Basset}, A. and {Battaglia}, P. and {Battisti}, A.~J. and {Baugh}, C.~M. and {Baumont}, L. and {Bazzanini}, L. and {Beaulieu}, J. -P. and {Beckmann}, V. and {Belikov}, A.~N. and {Bel}, J. and {Bellagamba}, F. and {Bella}, M. and {Bellini}, E. and {Benabed}, K. and {Bender}, R. and {Benevento}, G. and {Bennett}, C.~L. and {Benson}, K. and {Bergamini}, P. and {Bermejo-Climent}, J.~R. and {Bernardeau}, F. and {Bertacca}, D. and {Berthe}, M. and {Berthier}, J. and {Bethermin}, M. and {Beutler}, F. and {Bevillon}, C. and {Bhargava}, S. and {Bhatawdekar}, R. and {Bianchi}, D. and {Bisigello}, L. and {Biviano}, A. and {Blake}, R.~P. and {Blanchard}, A. and {Blazek}, J. and {Blot}, L. and {Bosco}, A. and {Bodendorf}, C. and {Boenke}, T. and {B{\"o}hringer}, H. and {Boldrini}, P. and {Bolzonella}, M. and {Bonchi}, A. and {Bonici}, M. and {Bonino}, D. and {Bonino}, L. and {Bonvin}, C. and {Bon}, W. and {Booth}, J.~T. and {Borgani}, S. and {Borlaff}, A.~S. and {Borsato}, E. and {Bose}, B. and {Botticella}, M.~T. and {Boucaud}, A. and {Bouche}, F. and {Boucher}, J.~S. and {Boutigny}, D. and {Bouvard}, T. and {Bouwens}, R. and {Bouy}, H. and {Bowler}, R.~A.~A. and {Bozza}, V. and {Bozzo}, E. and {Branchini}, E. and {Brando}, G. and {Brau-Nogue}, S. and {Brekke}, P. and {Bremer}, M.~N. and {Brescia}, M. and {Breton}, M. -A. and {Brinchmann}, J. and {Brinckmann}, T. and {Brockley-Blatt}, C. and {Brodwin}, M. and {Brouard}, L. and {Brown}, M.~L. and {Bruton}, S. and {Bucko}, J. and {Buddelmeijer}, H. and {Buenadicha}, G. and {Buitrago}, F. and {Burger}, P. and {Burigana}, C. and {Busillo}, V. and {Busonero}, D. and {Cabanac}, R. and {Cabayol-Garcia}, L. and {Cagliari}, M.~S. and {Caillat}, A. and {Caillat}, L. and {Calabrese}, M. and {Calabro}, A. and {Calderone}, G. and {Calura}, F. and {Camacho Quevedo}, B. and {Camera}, S. and {Campos}, L. and {Ca{\~n}as-Herrera}, G. and {Candini}, G.~P. and {Cantiello}, M. and {Capobianco}, V. and {Cappellaro}, E. and {Cappelluti}, N. and {Cappi}, A. and {Caputi}, K.~I. and {Cara}, C. and {Carbone}, C. and {Cardone}, V.~F. and {Carella}, E. and {Carlberg}, R.~G. and {Carle}, M. and {Carminati}, L. and {Caro}, F. and {Carrasco}, J.~M. and {Carretero}, J. and {Carrilho}, P. and {Carron Duque}, J. and {Carry}, B.},
        title = "{Euclid: I. Overview of the Euclid mission}",
      journal = {\aap},
         year = 2025,
        month = may,
       volume = {697},
          eid = {A1},
        pages = {A1},
          doi = {10.1051/0004-6361/202450810},
archivePrefix = {arXiv},
       eprint = {2405.13491},
 primaryClass = {astro-ph.CO},
       adsurl = {https://ui.adsabs.harvard.edu/abs/2025A&A...697A...1E}
}

@ARTICLE{shajib2024,
       author = {{Shajib}, A.~J. and {Vernardos}, G. and {Collett}, T.~E. and {Motta}, V. and {Sluse}, D. and {Williams}, L.~L.~R. and {Saha}, P. and {Birrer}, S. and {Spiniello}, C. and {Treu}, T.},
        title = "{Strong Lensing by Galaxies}",
      journal = {\ssr},
         year = 2024,
        month = dec,
       volume = {220},
       number = {8},
          eid = {87},
        pages = {87},
          doi = {10.1007/s11214-024-01105-x},
archivePrefix = {arXiv},
       eprint = {2210.10790},
 primaryClass = {astro-ph.GA},
       adsurl = {https://ui.adsabs.harvard.edu/abs/2024SSRv..220...87S}
}

@ARTICLE{vegetti2023,
author = {{Vegetti}, S. and {Birrer}, S. and {Despali}, G. and {Fassnacht}, C.~D. and {Gilman}, D. and {Hezaveh}, Y. and {Perreault Levasseur}, L. and {McKean}, J.~P. and {Powell}, D.~M. and {O'Riordan}, C.~M. and {Vernardos}, G.},
        title = "{Strong gravitational lensing as a probe of dark matter}",
      journal = {arXiv e-prints},
         year = 2023,
        month = jun,
          eid = {arXiv:2306.11781},
        pages = {arXiv:2306.11781},
          doi = {10.48550/arXiv.2306.11781},
archivePrefix = {arXiv},
       eprint = {2306.11781},
 primaryClass = {astro-ph.CO},
       adsurl = {https://ui.adsabs.harvard.edu/abs/2023arXiv230611781V}
}

@ARTICLE{roman2015,
       author = {{Spergel}, D. and {Gehrels}, N. and {Baltay}, C. and {Bennett}, D. and {Breckinridge}, J. and {Donahue}, M. and {Dressler}, A. and {Gaudi}, B.~S. and {Greene}, T. and {Guyon}, O. and {Hirata}, C. and {Kalirai}, J. and {Kasdin}, N.~J. and {Macintosh}, B. and {Moos}, W. and {Perlmutter}, S. and {Postman}, M. and {Rauscher}, B. and {Rhodes}, J. and {Wang}, Y. and {Weinberg}, D. and {Benford}, D. and {Hudson}, M. and {Jeong}, W. -S. and {Mellier}, Y. and {Traub}, W. and {Yamada}, T. and {Capak}, P. and {Colbert}, J. and {Masters}, D. and {Penny}, M. and {Savransky}, D. and {Stern}, D. and {Zimmerman}, N. and {Barry}, R. and {Bartusek}, L. and {Carpenter}, K. and {Cheng}, E. and {Content}, D. and {Dekens}, F. and {Demers}, R. and {Grady}, K. and {Jackson}, C. and {Kuan}, G. and {Kruk}, J. and {Melton}, M. and {Nemati}, B. and {Parvin}, B. and {Poberezhskiy}, I. and {Peddie}, C. and {Ruffa}, J. and {Wallace}, J.~K. and {Whipple}, A. and {Wollack}, E. and {Zhao}, F.},
        title = "{Wide-Field InfrarRed Survey Telescope-Astrophysics Focused Telescope Assets WFIRST-AFTA 2015 Report}",
      journal = {arXiv e-prints},
         year = 2015,
        month = mar,
          eid = {arXiv:1503.03757},
        pages = {arXiv:1503.03757},
          doi = {10.48550/arXiv.1503.03757},
archivePrefix = {arXiv},
       eprint = {1503.03757},
 primaryClass = {astro-ph.IM},
       adsurl = {https://ui.adsabs.harvard.edu/abs/2015arXiv150303757S}
}

@ARTICLE{valentino2021,
       author = {{Di Valentino}, Eleonora and {Mena}, Olga and {Pan}, Supriya and {Visinelli}, Luca and {Yang}, Weiqiang and {Melchiorri}, Alessandro and {Mota}, David F. and {Riess}, Adam G. and {Silk}, Joseph},
        title = "{In the realm of the Hubble tension-a review of solutions}",
      journal = {Classical and Quantum Gravity},
         year = 2021,
        month = jul,
       volume = {38},
       number = {15},
          eid = {153001},
        pages = {153001},
          doi = {10.1088/1361-6382/ac086d},
archivePrefix = {arXiv},
       eprint = {2103.01183},
 primaryClass = {astro-ph.CO},
       adsurl = {https://ui.adsabs.harvard.edu/abs/2021CQGra..38o3001D}
}

@ARTICLE{verde2024,
       author = {{Verde}, Licia and {Sch{\"o}neberg}, Nils and {Gil-Mar{\'\i}n}, H{\'e}ctor},
        title = "{A Tale of Many H $_{0}$}",
      journal = {\araa},
         year = 2024,
        month = sep,
       volume = {62},
       number = {1},
        pages = {287-331},
          doi = {10.1146/annurev-astro-052622-033813},
archivePrefix = {arXiv},
       eprint = {2311.13305},
 primaryClass = {astro-ph.CO},
       adsurl = {https://ui.adsabs.harvard.edu/abs/2024ARA&A..62..287V}
}

@article{Linder2011,
  title = {Lensing time delays and cosmological complementarity},
  author = {Linder, Eric V.},
  journal = {Phys. Rev. D},
  volume = {84},
  issue = {12},
  pages = {123529},
  numpages = {6},
  year = {2011},
  month = {Dec},
  publisher = {American Physical Society},
  doi = {10.1103/PhysRevD.84.123529},
  url = {https://link.aps.org/doi/10.1103/PhysRevD.84.123529}
}

@ARTICLE{Jee2016,
       author = {{Jee}, I. and {Komatsu}, E. and {Suyu}, S.~H. and {Huterer}, D.},
        title = "{Time-delay cosmography: increased leverage with angular diameter distances}",
      journal = {\jcap},
         year = 2016,
        month = apr,
       volume = {2016},
       number = {4},
          eid = {031},
        pages = {031},
          doi = {10.1088/1475-7516/2016/04/031},
archivePrefix = {arXiv},
       eprint = {1509.03310},
 primaryClass = {astro-ph.CO},
       adsurl = {https://ui.adsabs.harvard.edu/abs/2016JCAP...04..031J}
}

@ARTICLE{Lokken_2023,
       author = {{Lokken}, Martine and {Gagliano}, Alexander and {Narayan}, Gautham and {Hlo{\v{z}}ek}, Ren{\'e}e and {Kessler}, Richard and {Crenshaw}, John Franklin and {Salo}, Laura and {Alves}, Catarina S. and {Chatterjee}, Deep and {Vincenzi}, Maria and {Malz}, Alex I. and {LSST Dark Energy Science Collaboration}},
        title = "{The simulated catalogue of optical transients and correlated hosts (SCOTCH)}",
      journal = {\mnras},
         year = 2023,
        month = apr,
       volume = {520},
       number = {2},
        pages = {2887-2912},
          doi = {10.1093/mnras/stad302},
archivePrefix = {arXiv},
       eprint = {2206.02815},
 primaryClass = {astro-ph.IM},
       adsurl = {https://ui.adsabs.harvard.edu/abs/2023MNRAS.520.2887L}
}

@ARTICLE{Suyu2012,
       author = {{Suyu}, S.~H. and {Treu}, T. and {Blandford}, R.~D. and {Freedman}, W.~L. and {Hilbert}, S. and {Blake}, C. and {Braatz}, J. and {Courbin}, F. and {Dunkley}, J. and {Greenhill}, L. and {Humphreys}, E. and {Jha}, S. and {Kirshner}, R. and {Lo}, K.~Y. and {Macri}, L. and {Madore}, B.~F. and {Marshall}, P.~J. and {Meylan}, G. and {Mould}, J. and {Reid}, B. and {Reid}, M. and {Riess}, A. and {Schlegel}, D. and {Scowcroft}, V. and {Verde}, L.},
        title = "{The Hubble constant and new discoveries in cosmology}",
      journal = {arXiv e-prints},
         year = 2012,
        month = feb,
          eid = {arXiv:1202.4459},
        pages = {arXiv:1202.4459},
          doi = {10.48550/arXiv.1202.4459},
archivePrefix = {arXiv},
       eprint = {1202.4459},
 primaryClass = {astro-ph.CO},
       adsurl = {https://ui.adsabs.harvard.edu/abs/2012arXiv1202.4459S}
}
\appendix

\section{Simulating mock LSNe time series}
\label{app:mock_sim}
\begin{figure*}
    \centering
    \includegraphics[width=\textwidth]{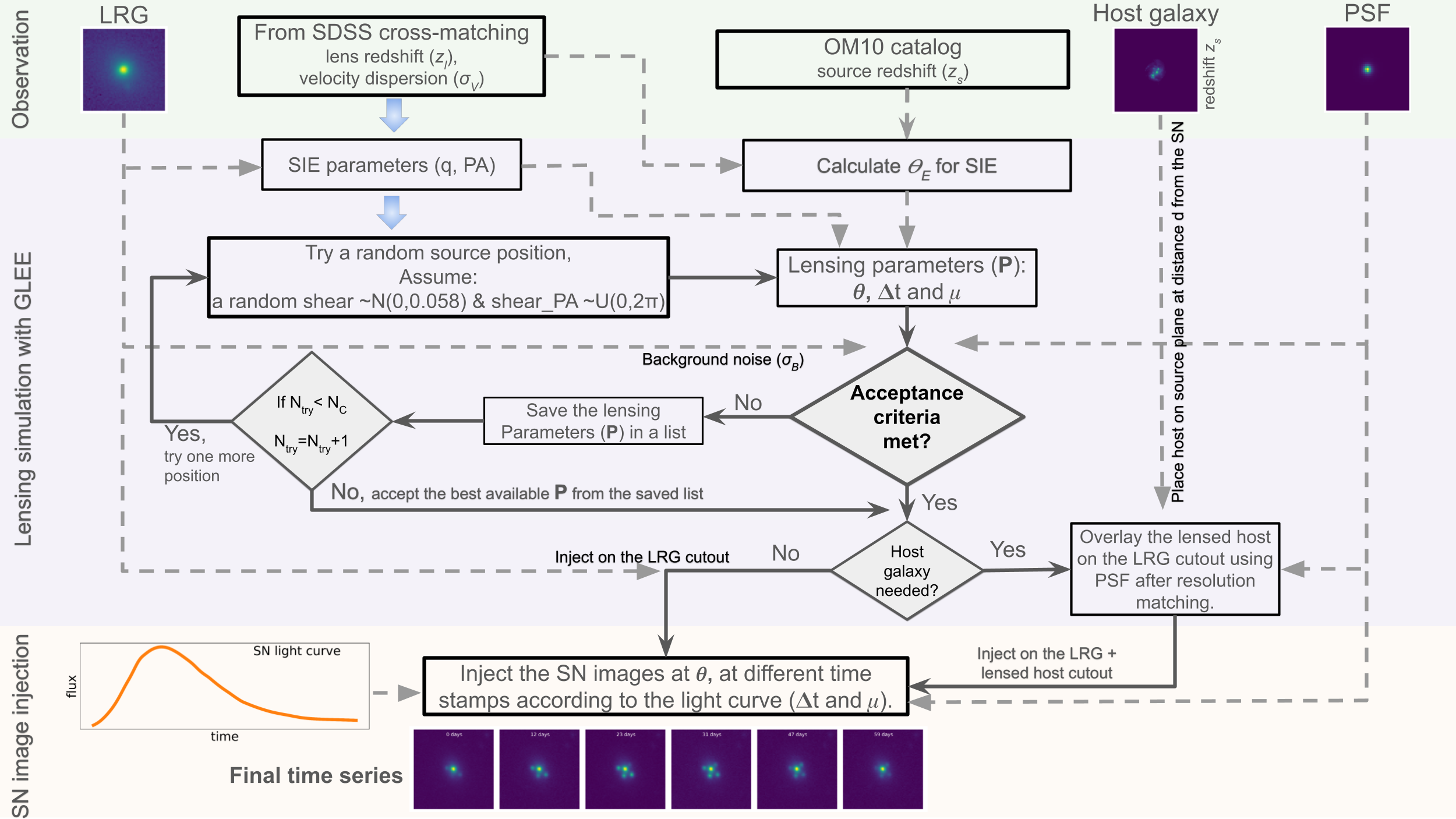}
    \caption{Schematic illustration of the simulation pipeline used to create time series of LSNe Ia. Starting from HSC Wide layer observations, LRG samples are cross-matched with SDSS, and lensed SN images are injected onto the LRG cutouts at different epochs, with or without the host galaxy as needed. While the host-inclusion feature is demonstrated for completeness, host galaxies are not included in the simulations used in this work.}
    \label{fig:sim}
\end{figure*}

\subsection{Preparing LRG catalog}\label{app:lrg_cat}

Galaxy-scale LSNe Ia are simulated by injecting point-like LSNe onto cutouts of real HSC galaxy images and embedding them into empty regions of the HSC Transient Survey. As population of foreground deflectors, we use samples of luminous red galaxies (LRGs) which have the highest lensing cross-section. Given the dearth of massive galaxies present in the COSMOS field, the sample of LRGs is obtained from a larger footprint, namely the Wide layer of the HSC-SSP second public data release \citep[PDR2,][]{aihara19} that covers nearly 800 deg$^2$ with depth of 26.6, 26.2, 26.2, and 25.3 mag in $griz$ filters. We select LRGs with spectroscopic redshift and velocity dispersion measurements available from DR14 of the SDSS-IV survey \citep{abolfathi18,bautista18}, in order to get robust proxy of the total galaxy masses for computing realistic light deflection angles. An upper limit on the velocity dispersion uncertainty, $\delta v_{\rm disp} < 100$ km s$^{-1}$, is set at this stage.

To focus on clean LRG cutouts without artefacts, we use the HSC tables to discard galaxies and their neighbour within 1\arcsec\ that were flagged by the pipeline. We request that the following flags are set to {\tt False} in all $g$, $r$, $i$, and $z$-band: {\tt cmodel\_flux\_flag}, {\tt pixelflags\_edge}, {\tt pixelflags\_interpolatedcenter}, {\tt pixelflags\_saturatedcenter}, {\tt pixelflags\_crcenter}, {\tt pixelflags\_bad}, {\tt sdsscentroid\_flag}. The cutouts passing these standardised HSC image quality flags are further analysed to exclude remaining artefacts. We identify cutouts with repeated pixel values indicating partial coverage in PDR2, and cutouts with few pixel wide artefacts detected with LACosmic due to their sharp edges \citep{vandokkum01}. This results in a sample of 36138 unique LRGs. Their cutouts were imported with the Data Archive System service\footnote{https://hsc-release.mtk.nao.ac.jp/das\_cutout/pdr2/} together with the relevant PSF frames\footnote{https://hsc-release.mtk.nao.ac.jp/psf/pdr2/} in order to produce the LSN simulations. However, to ensure sufficient sample size, we augment the LRG cutouts with up to four rotations.

\subsection{The simulation pipeline}\label{app:sim_pipe}

Here, we detail the key steps for simulating time series data for mock LSNe. Our focus is on type Ia SN, which are standardizable candles with well-constrained intrinsic luminosities. However, this pipeline can be easily extended to other types of SNe or transient sources.

We base the redshift distributions of both lenses and sources on the Oguri \& Marshall (2010, hereafter OM10) simulation \citep{om10, Oguri2018}. We select our final LRG sample such that its redshift distribution broadly matches that of the lens population for LSNe Ia in OM10. Rather than adopting the realistic Einstein radius distribution from OM10, we use a flat distribution between $0.1\arcsec$ and $2.0\arcsec$ to ensure a more balanced training set. With these choices, we construct a matched lens-source catalog by assigning each lens a source redshift, at which the SN Ia will be injected.

We assume a Singular Isothermal Ellipsoid (SIE) + shear model to represent lens galaxies and create the mock lensed images using a pipeline \citep{canameras20, Stefan21} based on the \texttt{GLEE} software \citep{glee1, glee2}. The complete pipeline is schematically illustrated in Figure~\ref{fig:sim}, and its key components are summarized below.
\begin{itemize}
   \item For each LRG, we have its spectroscopic redshift, velocity dispersion, and the PSF of its field. We first compute the SIE parameters. Using the source redshift given in the lens-source matched catalog, we obtain the Einstein radius ($\thE$) under a \lcdm background cosmology. A point source is then randomly placed on the source plane within $1.5 \times \thE$ to maximize the likelihood of strong lensing. At each random position, we assume a random shear value. Together with the SIE parameters, we compute the lensing configuration and derive the image positions ($\boldsymbol{\theta}$), magnifications ($\mu$), and time delays ($\dt$) using the \texttt{GLEE} package.

\item \textbf{Lens configuration acceptance criteria:}

\begin{itemize}

    \item The major criterion is whether the faintest image is detectable against the background noise. We validate this in the $i$-band, where the brightness of most LSNe Ia is expected to peak. The background noise level ($\sigma_{\rm B}$) for each LRG is estimated from the standard deviation of pixels in the outskirts.
    
    \item We focus on LSNe Ia, which are standardizable candles with well-constrained intrinsic luminosities. For simplicity, we assume a fixed peak absolute magnitude in the $B$-band, $M_B = -19.3$.\footnote{In reality, $M_B$ varies and can be standardized using Tripp's relation \citep{Tripp1998}. However, for computational efficiency, we assume a fixed value while allowing for a potential boost in $M_B$ if required for detectability. The effect of allowing Tripp's relation-based variations is expected to be minimal on our exercise.} 
    
    \item Placing the SN Ia at the source redshift, we compute the peak brightness of all images in the $i$-band, accounting for the PSF associated with the LRG and using the Hsiao template \citep{Hsiao2007} from the \texttt{SNCosmo} package \cite{sncosmo}. We check whether the peak flux of the faintest image (\( F_{0, \rm faint} \)) satisfies the detectability condition \( F_{0, \rm faint} > 5\sigma_{\rm B} \). If the condition is not met, we calculate the necessary boost in absolute magnitude to achieve a $5\sigma$ detection as:  
    \begin{equation}
       \Delta M_{\rm faint}  \equiv 2.5 \times\log_{10} \left(\frac{F_{0, \rm faint}}{5\sigma_{\rm B}} \right)|_{i\rm -band}
    \end{equation}
    Note that, only when $\Delta M_{\rm faint} <0$ we need a magnitude boost in the source, since $\Delta M_{\rm faint} > 0$ implies the faintest image is already brighter than five times the background noise. 
    We then impose a threshold, ${\Delta M}_{\rm th} \leq 0$, such that only if the source position satisfies $\Delta M_{\rm faint} > {\Delta M}_{\rm th}$ is accepted. Otherwise, we try another random source position.

    \item Condition on the time delays: Between the arrival of the first and second images, the frames may resemble samples of normal SNe exploding in LRGs, where the only differentiating factor could be the color information. The model may confuse these two cases, as color information alone may not be sufficient to distinguish them. Therefore, we require that the second image does not arrive too late, ensuring it can be accommodated within the time series of reasonable length. Thus, we impose an upper limit on the time delay between the first and second images.

    \item Lastly, if the SN is placed too close to the caustic curves due to random positioning, the resulting magnification can become extremely high ($\mathcal{O}(10^3)$ or even greater). This can cause one of the LSN images to become excessively bright, potentially dominating the entire mock frames and creating issues during training. Such extreme magnifications are not observed in real lensed systems. To prevent unrealistic scenarios with extreme magnifications, we impose a limit on the maximum magnification, requiring $|\mu|_{\rm max} < \mu_{\rm th}$, where $\mu_{\rm th}$ is set to 100.

    \item In summary, we accept a random position of the SN Ia on the source plane if it satisfies the above three conditions that are recast more generically below,
        \begin{align}
        {\Delta M}_{\rm faint} > {\Delta M}_{\rm th},~ &\text{and} ~\dt_{12}<\dt_{\rm th}, \nonumber \\ ~ &\text{and}~ |\mu|_{\rm max}< \mu_{\rm th} =100 \;,
    \end{align}
    where ${\Delta M}_{\rm th}\leq 0$ by construction. 
\end{itemize}

\item If the lensing parameters derived from a trial random source position fail to satisfy the acceptance criteria outlined above, we attempt another random source position with new random values for the shear parameters. During this process, we record the source position and corresponding lensing parameters for each trail.  We continue this procedure until the number of trials reaches a maximum threshold, $N_C$, which we set to $400$ to balance computational efficiency with scientific objectives.  If no acceptable solution is found by then, we systematically relax the magnitude boost (${\Delta M}_{\rm faint}$) and time-delay ($\dt_{\rm th}$) thresholds to identify the most suitable lensing parameters from the pool of recorded attempts, as long as the resulting configurations remain physically plausible. In rare occasions when no realistic set of lensing parameters could be found within the finite ($N_C$) number of trails we discard that lens-source pair. This ensures that we obtain a reasonable lensing system for almost all lens-source pairs.

We note that in this initial work, we only consider systems with $\Delta t_{\rm th} = 50$ days and $\Delta M_{\rm th} = 0$, meaning only cases that require no magnitude boost are included. Notably, the majority of lens-source pairs (approximately 60\%) yield acceptable lensing configurations without any magnitude boost and with $\Delta t_{12} < 50$ days. This subset forms the primary focus of our study. Nevertheless, we present the full pipeline here, which can accommodate fainter sources requiring a magnitude boost and allow larger time delays if necessary, attempting to minimize the required magnitude boost and large $\Delta t_{12}$ whenever possible.

\item Once the lensing parameters ($\boldsymbol{\theta}$, $\Delta t$, $\mu$) are determined for each lens-source pair based on the $i$-band data alone, we use these parameters to generate SN Ia images as point sources across all bands. Specifically, we place the multiple SN images at the corresponding positions on the LRG cutouts using the respective PSFs, applying the appropriate magnitude boost ($\Delta M$) if necessary. The time delays ($\Delta t$) and magnification factors ($\mu$) are incorporated consistently when placing the images in all bands. To build the multi-band time series, we inject SN images at various epochs representing different phases of the SN evolution, modeling the SN light curves using the Hsiao template from \texttt{SNCosmo}. Additionally, a random positional shift of up to $0.85\arcsec$ (5 pixels) is applied in any direction for each sample (or time series) to mimic observational variations.
\end{itemize}

The final cutouts have a size of $\sim (10\arcsec \times 10\arcsec)$, corresponding to $(59 \times 59)$ pixels at a given time step across all bands. Consequently, the resulting time series have a shape of $(N_t \times 59 \times 59 \times 4)$ where $N_t$ is the number of time steps and 4 is the number of bands. 

The pipeline described above enables the creation of multi-band time series of mock LSNe Ia at arbitrary numbers of timestamps ($N_t$) with freely chosen values. We emphasize that, in our training, we generate asynchronous time sampling across multiple bands, i.e., observations occur in only one band at each given time, using this pipeline. 

Some examples of time series of LSNe Ia are shown in the top row of Figure~\ref{fig:images} and Figure~\ref{fig:mock_lensed_examples} for $i$-band and, in Figure~\ref{fig:multi-band_ts} for multi-band.

\begin{figure*}
    \centering
    \includegraphics[width=0.485\textwidth]{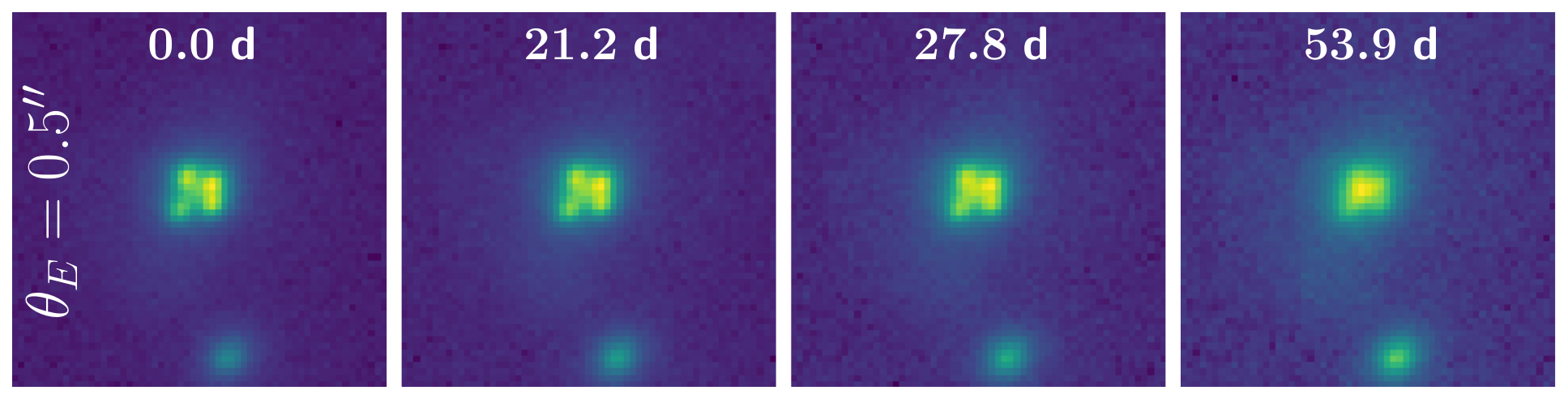}\hspace*{1 mm} 
    \includegraphics[width=0.485\textwidth]{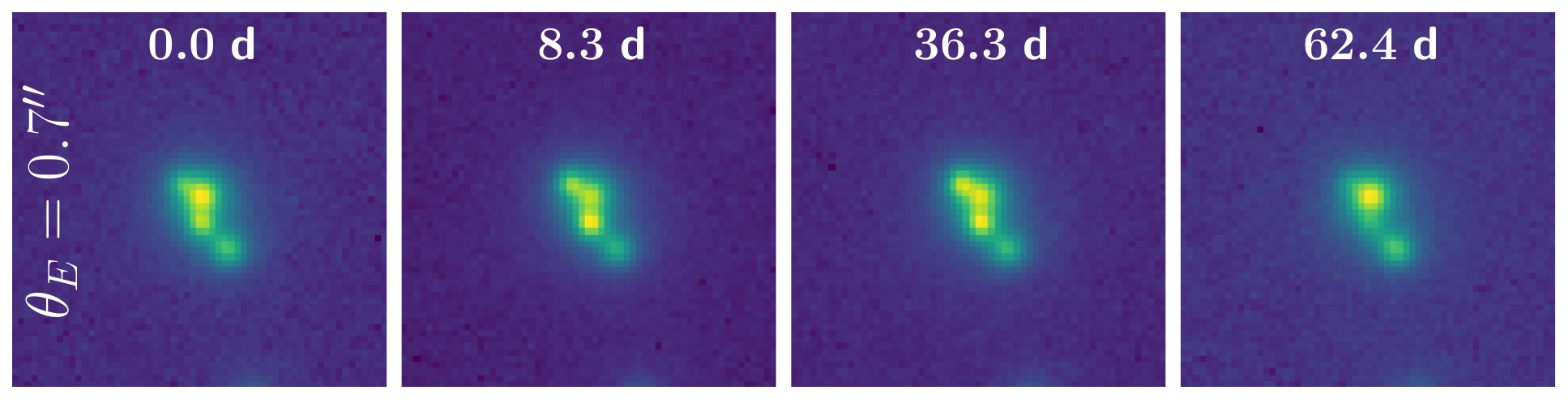} \\
    \includegraphics[width=0.485\textwidth]{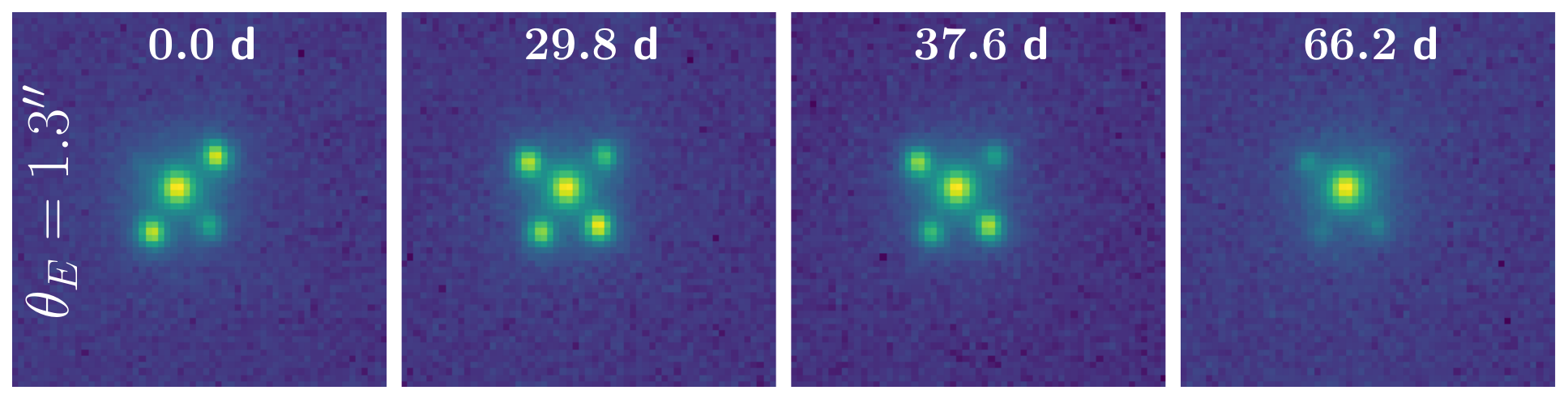}\hspace*{1 mm} 
    \includegraphics[width=0.485\textwidth]{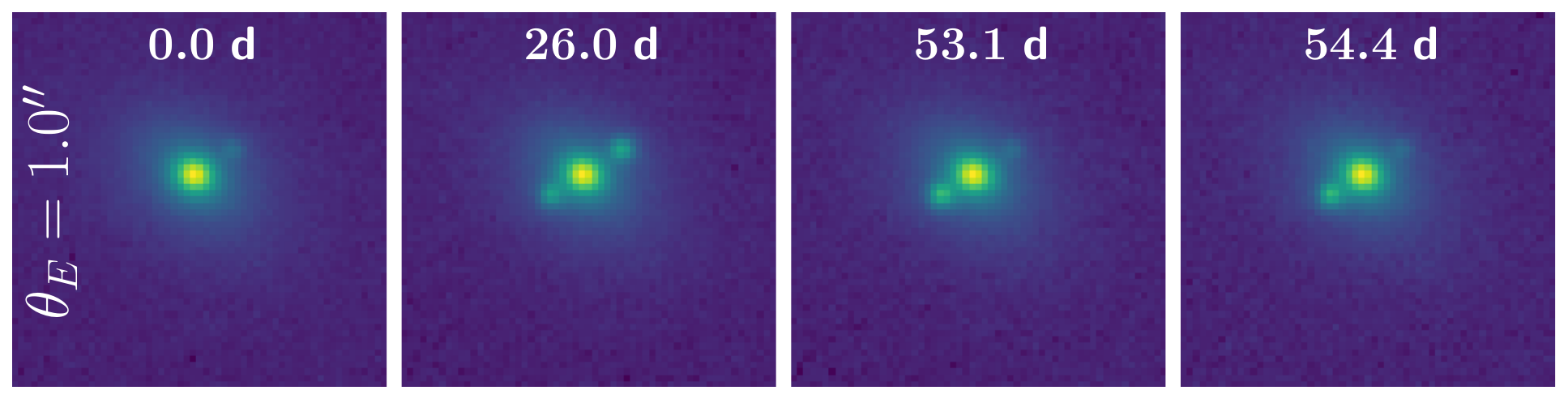} \\
    \includegraphics[width=0.485\textwidth]{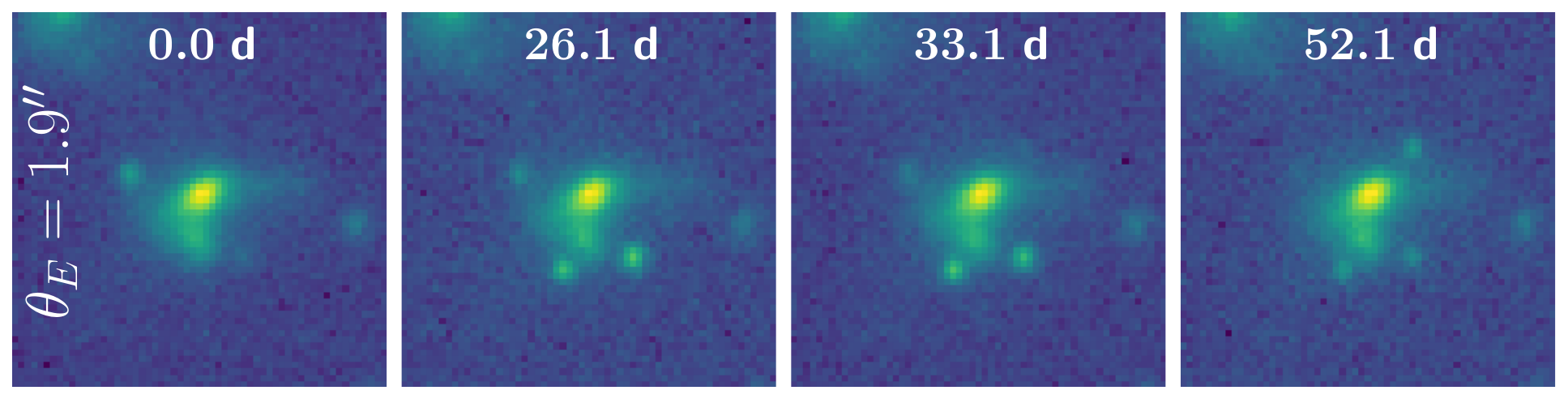}\hspace*{1 mm} 
    \includegraphics[width=0.485\textwidth]{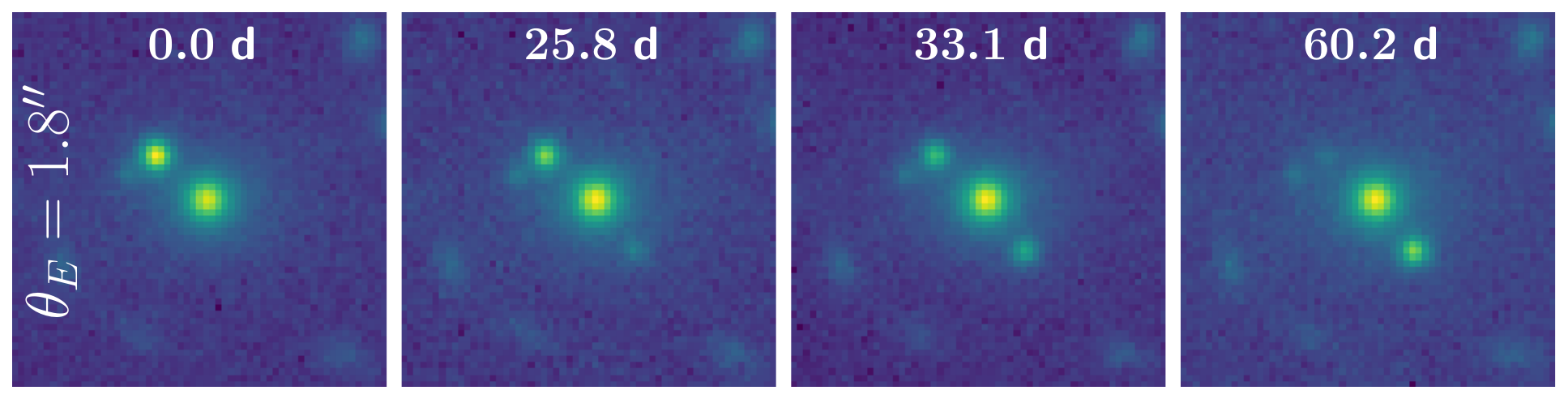}
    \caption{Examples of time series of mock LSNe Ia systems in $i$-band. The three rows correspond to systems with smaller (top), intermediate (middle), and larger (bottom) Einstein radii, with the specific values indicated at the first time step of each series. Time stamps, relative to the first observation, are displayed at the top of each frame. The time sampling follows the HSC Transient Survey cadence, as discussed in Section~\ref{sec:matching_data_quality} and detailed further in Appendix~\ref{app:cadence_matching}. In each row, two systems are presented: a quad on the left and a double on the right. For visual clarity, we show in this figure systems with prominent lensing features; however, many samples in the training set exhibit more subtle and less easily identifiable signals. Still, some examples shown here include foreground or background contaminants, reflecting the observational complexities incorporated in the training data.}
    \label{fig:mock_lensed_examples}
\end{figure*}

\section{Cadence matching}
\label{app:cadence_matching}

\begin{figure*}
    \centering
    \includegraphics[width=\textwidth]{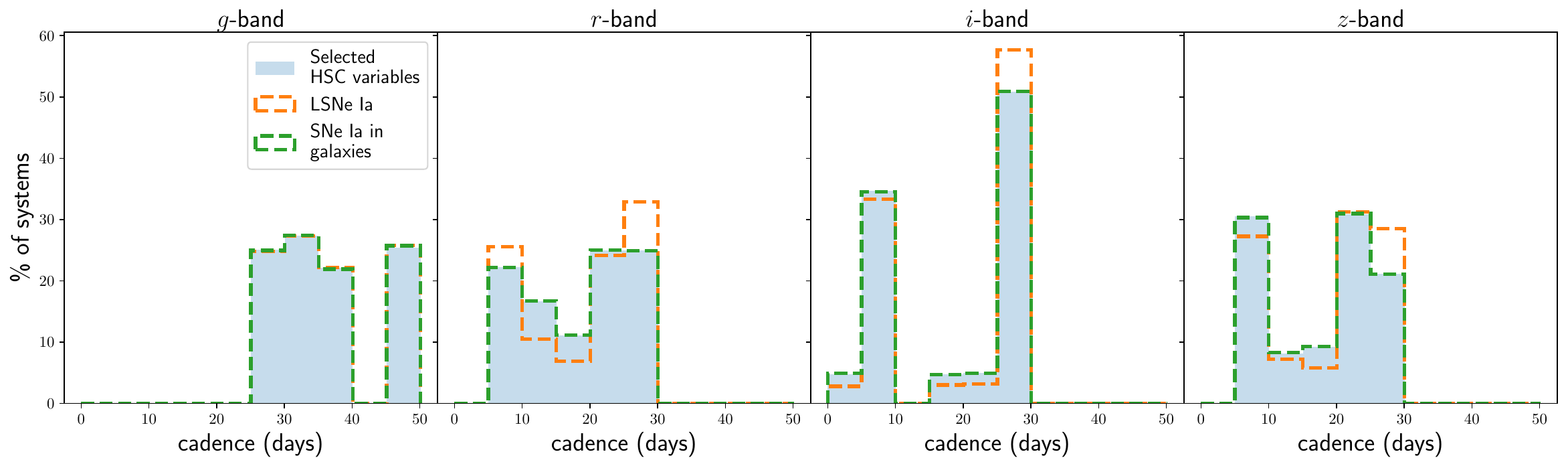}
    \caption{The cadence distributions for different components across four bands are shown in four panels. HSC variables are selected with ``good" cadences according to the criteria described in Appendix \ref{app:cadence_matching}. We then match the cadence distributions of our simulated time series to those of the selected observed HSC variables, separately for each band.  Due to the additional requirement of capturing at least the second trailing image for the LSNe Ia, their cadence distribution tends to slightly favor coarse cadence, i.e. larger gaps between successive observations.}
    \label{fig:matching_cadence}
\end{figure*}

In Section \ref{sec:matching_data_quality}, we explain the importance of matching the cadence of the synthetic time series (for LSNe Ia and normal SNe Ia) to that of the HSC variable time series to ensure consistent time sampling. Random sampling from the available observation epochs of the HSC variables, as shown in Figure~\ref{fig:cosmos_epochs}, often results in very coarse cadences and consequently long, sparsely sampled light curves. Such cadences are typically not well-suited for the efficient detection of LSNe. To mitigate this, we define a set of selection criteria to enforce more suitable cadences for the HSC variables. As mentioned earlier, we fix the number of observations per band across all samples: specifically, $N_\text{obs}^\text{tot} = 2$, 4, 4, and 4 for the $g$, $r$, $i$, and $z$ bands, respectively. The $g$-band is excluded when defining our cadence selection criteria due to its very sparse observations in the HSC Transient Survey. Consequently, the criteria are based on the $r$, $i$, and $z$ bands, each having 4 observations.

\begin{itemize}
\item For each HSC variable, and for each of the bands, we randomly draw a combination of $N_\text{obs}^\text{tot}$ observing epochs and check whether the following two conditions are satisfied.
    \begin{enumerate}
        \item The gap between any two successive observations in a given band must not exceed a predefined threshold: ${\delta t} \leq {\delta t}_{\rm th}$, chosen here as 30 days for $riz$ bands.
        \item The total duration of the light curve in that band must remain below a specified limit: $T_{\rm tot} \leq T_{\rm th}$, chosen here as 80 days for $riz$ bands and 50 days for the $g$-band. This prevents the light curves from becoming excessively long, which would otherwise reduce their relevance for SN-like variability timescales and hinder early detection.       
    \end{enumerate}\label{item:cadence_criteria}

    \item For each HSC variable and each band, we store all combinations of observing epochs that satisfy the above criteria. If no such valid combination is found in any of the bands, we discard that HSC variable from further consideration.

    \item While constructing the training set, we randomly select a ``good" combination of observing epochs from the saved cadence database for each band for every HSC variable.

    \item Similarly, while simulating the time series for SNe exploding in galaxies (either LRGs or spirals), we again randomly select a ``good'' combination of observing epochs from our saved database for each band.

    \item We apply the same procedure for the LSNe Ia, but with one additional criterion: by the last observation in any band (except for $g$), the peak of the second arriving image must be included. This ensures that even for long time delay lensed systems (most of which are doubles), the model sees the second image somewhere in the time series. For each band, we randomly select a ``good" combination of observing epochs from our saved cadence database until this additional condition is met. 
\end{itemize}

The resulting cadence distribution for the different components in our training and test sets is shown in Figure \ref{fig:matching_cadence} for each band. The cadence distributions for HSC variables and SNe exploding in galaxies match exactly in all bands due to the formulation. However, because of the additional criterion for the LSNe Ia, their cadence distribution slightly favors slightly larger gaps between successive observations (i.e. somewhat coarse cadence). Despite this, the difference is minimal, and the distributions across all components generally match well, as intended.

\section{Details of \convlstmd}
\label{app:convlstm}

\begin{figure*}[hbt]
    \centering
    \includegraphics[width=0.48\textwidth]{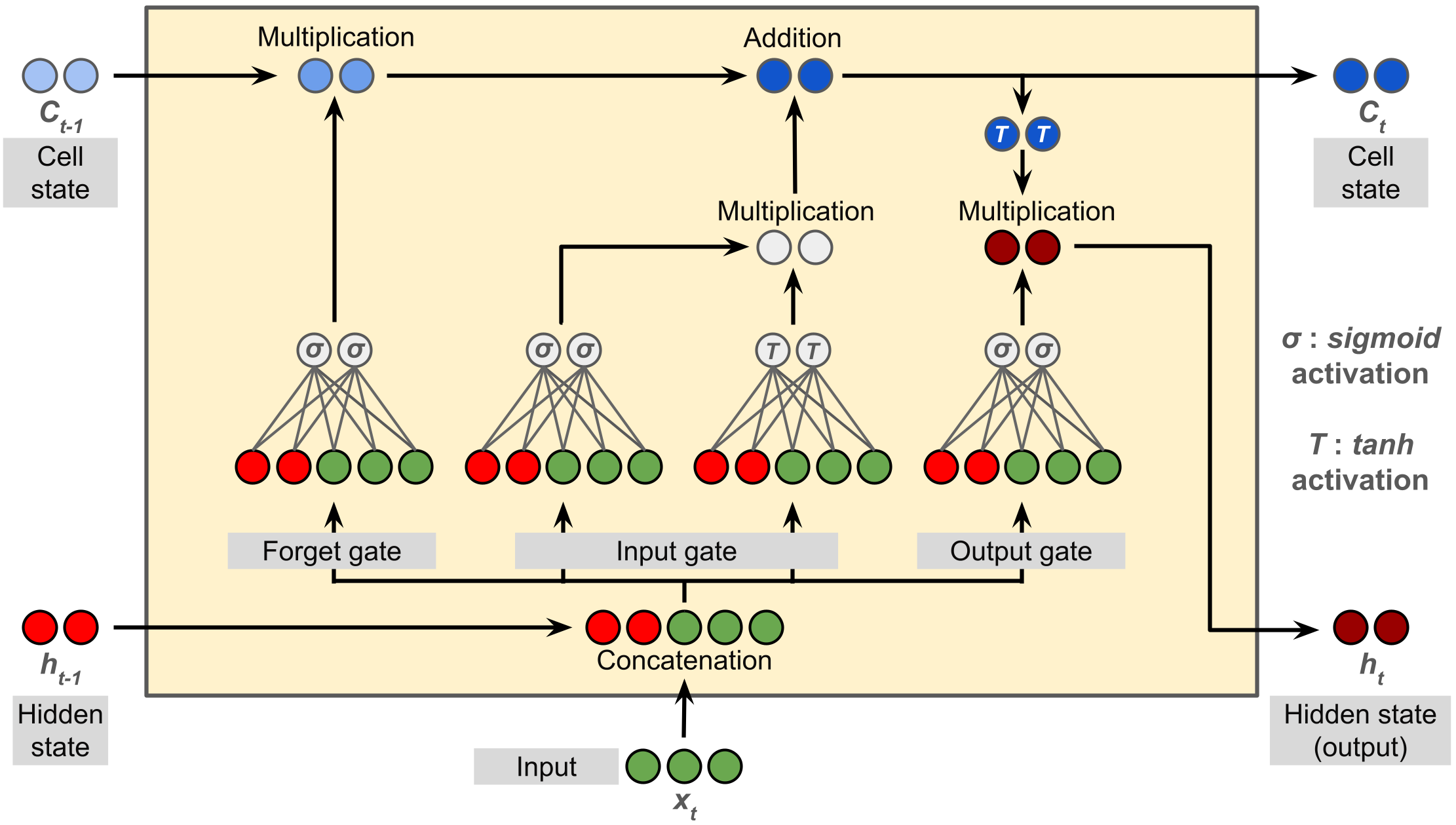}
    \includegraphics[width=0.48\textwidth]{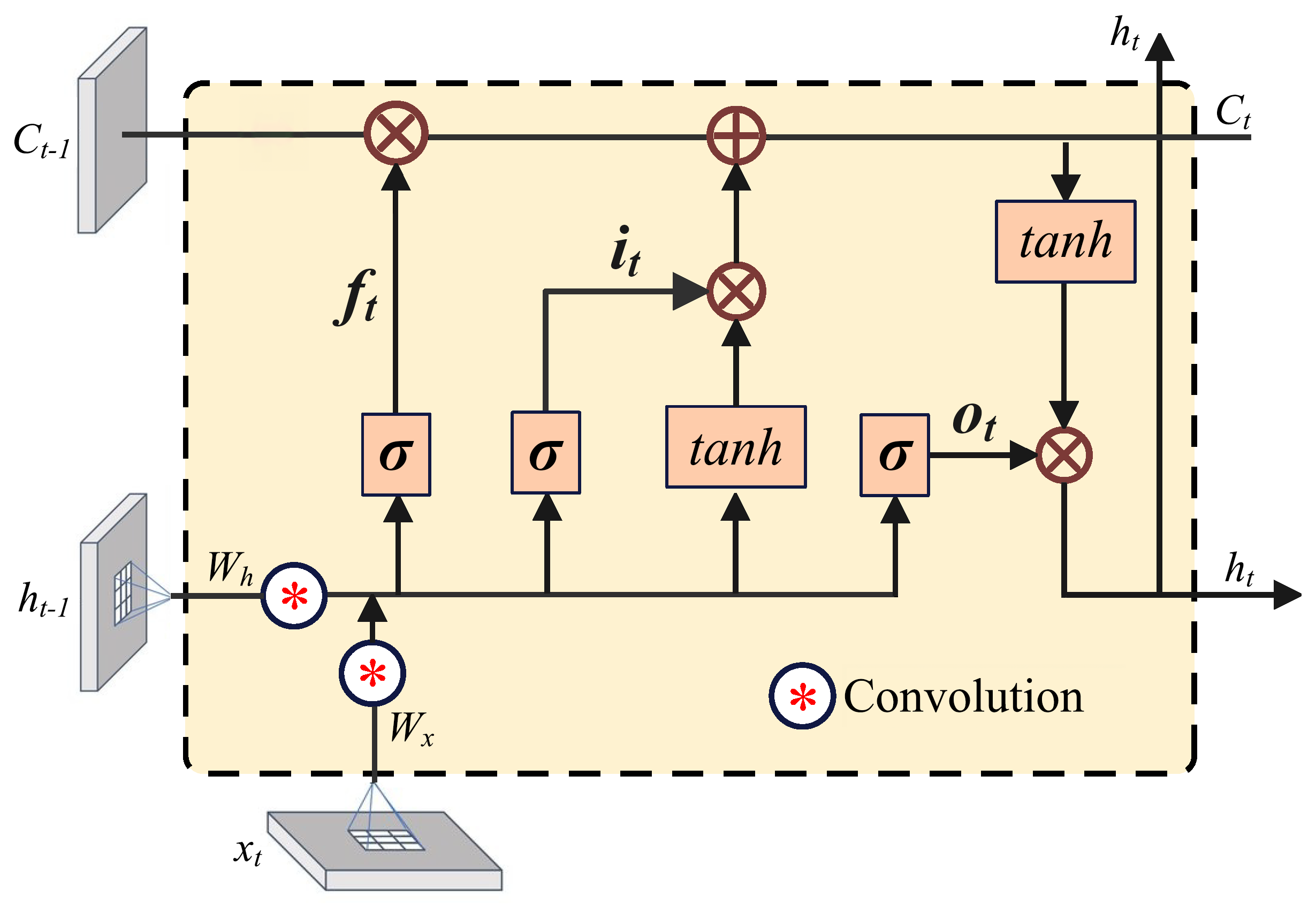}
    \caption{The left panel demonstrates how a 2-unit \texttt{LSTM} layer processes a sequence with 3 features at each time step. For better visual illustration, we represent the sum of two matrix multiplications, in Equations~\eqref{eq:lstm11}–\eqref{eq:lstm14}, as a single matrix multiplication between a concatenated weight matrix of shape $(2,5)$ and a stacked vector of $\mathbf{x_t}$ and $\mathbf{h_{t-1}}$ of shape $(5,1)$, as described in Equation~\eqref{eq:concatenate}. The output $h_t$ has a length of 2. 
    The right panel, adapted from \citet{convlstm2}, illustrates the generalization to \texttt{ConvLSTM2D}, where matrix multiplication (similar to fully-connected operations) is replaced by convolution operations (denoted by *), while handling both the short-term memory and input data, all of which are 2D at each time step. While the gating mechanism remains the same between an \lstm~ and a \convlstm~ cell, the two panels depict them differently to enhance clarity.}
    \label{fig:lstm_convlstm}
\end{figure*}

Although the details of Long Short-Term Memory ({\tt LSTM}) and Convolutional \texttt{LSTM} ({\tt ConvLSTM}) are widely available in the literature, we provide an overview in this section to offer a more comprehensive and complete understanding. We begin with {\tt LSTM}, which was designed to address the limitations of simple {\it Recurrent Neural Networks} (RNNs), specifically the vanishing and/or exploding gradient problem. Particularly, {\tt LSTM} introduces the concept of short and long term memories.

\subsection{An {\tt LSTM} layer}
An {\tt LSTM} layer comprises three gating mechanisms that collectively carry out four core operations at each time step, as outlined below:
\begin{align}
    \mathbf{f_t} =& \sigma\left(\mathbf{W^f_x} \cdot \mathbf{x_t} + \mathbf{W^f_h} \cdot \mathbf{h_{t-1}} + \mathbf{b^f} \right)\;, \label{eq:lstm11}  \\
    \mathbf{i_t} =& \sigma\left(\mathbf{W^i_x} \cdot \mathbf{x_t} + \mathbf{W^i_h} \cdot \mathbf{h_{t-1}} + \mathbf{b^i} \right)\;, \label{eq:lstm12}  \\
    \mathbf{\tilde{C}_t} =& \tanh\left(\mathbf{W^C_x} \cdot \mathbf{x_t} + \mathbf{W^C_h} \cdot \mathbf{h_{t-1}} + \mathbf{b^C} \right) \;,\label{eq:lstm13} \\
    \mathbf{o_t} =& \sigma\left(\mathbf{W^o_x} \cdot \mathbf{x_t} + \mathbf{W^o_h} \cdot \mathbf{h_{t-1}} + \mathbf{b^o} \right) \;. \label{eq:lstm14}
\end{align}
Here, $\sigma$ and $\tanh$ represent the sigmoid and hyperbolic tangent activation functions, respectively. $\mathbf{x_t}$ is the input data at time step $t$, and $\mathbf{h_{t-1}}$ is the hidden state (short-term memory) output from the previous time step. The symbol $\cdot$ denotes the inner product. The superscripts `f', `i', `C' and `o' correspond to the forget gate, input gate, cell state, and output gate, respectively. Meanwhile, $\mathbf{\tilde{C}_t}$ denotes the candidate cell state, which represents the new information proposed to update the cell state. Next, the cell state and hidden state are updated as follows:
\begin{align}
 \mathbf{C_t} =& \mathbf{f_t} \odot \mathbf{C_{t-1}} + \mathbf{i_t} \odot \mathbf{\tilde{C}_t} \label{eq:lstm21} \;,\\
\mathbf{h_t} =&\mathbf{o_t} \odot \tanh\left( \mathbf{C_t} \right) \;,\label{eq:lstm22}
\end{align}
where $\odot$ indicates element-wise (Hadamard) multiplication.
The hidden state $\mathbf{h_t}$  serves as the output of the \texttt{LSTM} layer at the current time step $t$.

Analogous to how a dense layer processes fixed feature sets without modeling internal correlations, an \lstm~layer efficiently models sequences of feature vectors, such as time series of temperature, humidity, and air pressure, where the input at each time step, $\mathbf{x_t}$, is one-dimensional and contains no spatial structure or internal correlations. 

Similar to multiple nodes in a dense layer, an \lstm~ layer can have multiple units, denoted by $N_h$ (which can be invoked by {\tt LSTM($N_h$)} in {\tt TensorFlow}), producing an output $\mathbf{h_t}$ of shape $(N_h, 1)$ assuming the input $\mathbf{x_t}$ is one-dimensional. This is illustrated in the left panel of Figure~\ref{fig:lstm_convlstm}, where, for clarity, we simplify the sum of the two matrix multiplications appearing in Equations \eqref{eq:lstm11}--\eqref{eq:lstm14} as
\begin{equation}\label{eq:concatenate}
\mathbf{W_x} \cdot \mathbf{x_t} + \mathbf{W_h} \cdot \mathbf{h_{t-1}} = \begin{bmatrix} \mathbf{W_x} \mathbf{W_h} \end{bmatrix} \cdot \begin{bmatrix} \mathbf{x_t} \\ \mathbf{h_{t-1}} \end{bmatrix}\;,
\end{equation}
given that the number of rows of $\mathbf{W_x}$ and $\mathbf{W_h}$ are the same ($N_h$) and $\mathbf{x_t}$ and $\mathbf{h_{t-1}}$, being 1D, have matching numbers of columns.

All terms on the left-hand side of equations \eqref{eq:lstm11}-\eqref{eq:lstm22} share this same shape, $(N_h,1)$. Therefore, $N_h$ can be considered the dimensionality of the \lstm~layer. For example, consider a one-dimensional input vector of length $N_x$. The weight matrices for the input, all four $\mathbf{W_x}$, then have the shape $(N_h, N_x)$, while the weight matrices for the hidden state, all four $\mathbf{W_h}$, have the shape $(N_h, N_h)$. Each bias term will have the shape $N_h$. Consequently, the total number of parameters in the \lstm~ layer is given by:

\[
N_{\text{P,\texttt{LSTM}}}=4 \times ((N_h N_x + N_h^2 + N_h)) =4 \times N_h(N_x + N_h + 1).
\]

\subsection{A {\tt ConvLSTM2D} layer}
In \citet{convlstm}, the \lstm~ architecture is extended to capture spatial correlations by replacing inner products ($\cdot$) with convolutional operations ($\ast$). This transformation, which mirrors the shift from dense to convolutional layers, leads to the \convlstm~model. The gating mechanisms remain unchanged, so equations \eqref{eq:lstm11}-\eqref{eq:lstm22} still hold, with the inner product ($\cdot$) replaced by convolution ($\ast$), as shown by the right panel of Figure \ref{fig:lstm_convlstm} which is adapted from \citet{convlstm2}. When we are interested in 2D convolutions, typically needed to capture spatial correlations in 2D data, this is called \convlstmd. As a result, both the memory and output at each time step retain a 2D structure, analogous to that produced by a standard convolutional layer.

Suppose the input sample data have the shape $(N_x,N_y,N_c)$ ($N_c$ = no. of channels in the input data), and the convolution kernel has the spatial dimension $(K_x,K_y)$. With {\tt ConvLSTM2D($N_k$, kernel\_size=$(K_x,K_y)$)} the following properties hold:

\begin{itemize}
    \item Shape of all input kernels ($\mathbf{W_x}$) = $(K_x,K_y,N_C)$.
    \item Shape of all kernels for the hidden state ($\mathbf{W_h}$) = $(K_x,K_y,N_k)$.
    \item Shape of the hidden state ($\mathbf{h}$) = $(N_x, N_y,N_k)$ when using the `same' padding, and $(N_x-K_x+1, N_y-K_y+1,N_k)$ for the `valid' padding.\footnote{Note that for convolution over the hidden state, i.e., for all terms like $\mathbf{W_h} \ast \mathbf{h_{t-1}}$ in the core operations in Equations \ref{eq:lstm11}-\ref{eq:lstm14}, `same' padding is always used to ensure consistent hidden state dimensions across time steps.} This will be the shape of the output (at each time step), the cell state etc, basically all the LHS in the Equations \eqref{eq:lstm11}-\eqref{eq:lstm22} above. 
    \item Considering $4$ core operations in Equations \eqref{eq:lstm11}-\eqref{eq:lstm14}, the total number of parameters in a \convlstmd~ layer is given by:
\begin{align}
    N_{\text{P,\texttt{ConvLSTM2D}}}& =4 \times \left[N_k K_x K_y N_c +N_k K_x K_y N_k +N_k \right] \nonumber \\  &=4 \times N_k\left[K_x K_y \left(N_c +N_k \right) +1 \right]\;.
\end{align}

\end{itemize}

Note that, by construction, when $(K_x, K_y) = (1,1)$, i.e., in the limit where the 2D filter reduces to a scalar, \convlstmd~ simplifies to a standard {\tt LSTM}, where the number of convolutional filters $N_k$ plays the role of the hidden dimension $N_h$. Another interesting limit arises at the beginning of the sequence, where {\tt ConvLSTM} behaves similarly to a standard convolution layer. This occurs when the initial short-term memory (hidden state) is zero-initialized, as is typically the default case, while the cell state is set to a constant-valued (non-zero). Therefore, {\tt ConvLSTM} initially operates like a convolution layer, but as the sequence progresses and the memory mechanism becomes active, it progressively improves its predictions.

\section{Recovery Performance by LSNe Ia Properties}
\label{app:classification_performence_comp}
\begin{figure*}
    \includegraphics[width=\textwidth]{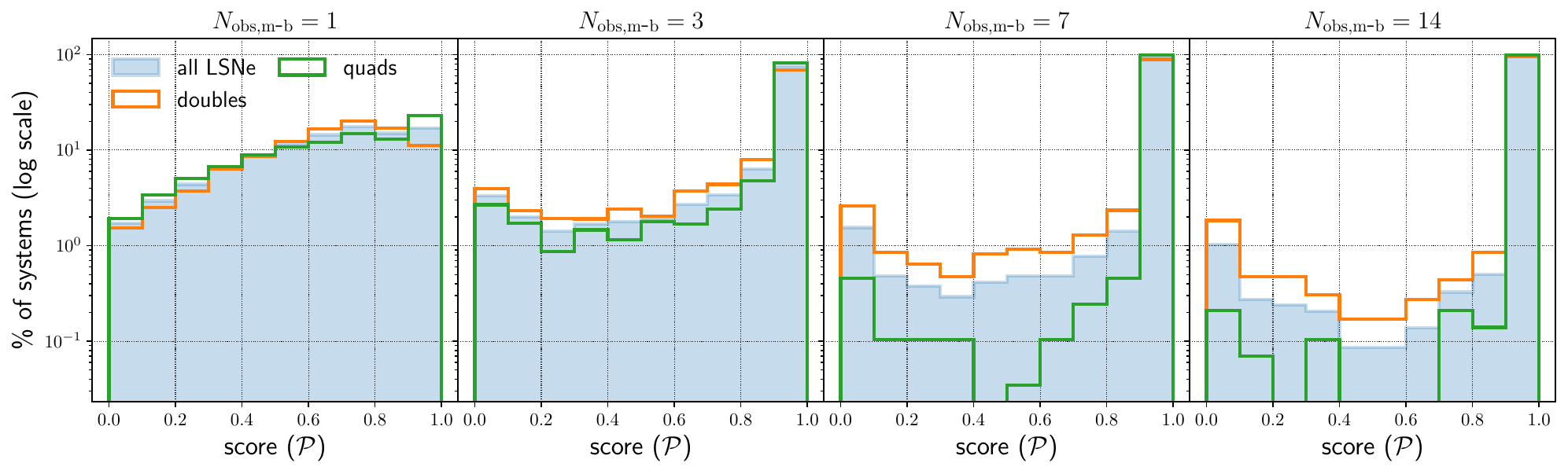}
    \caption{The figure presents the distributions of model-predicted scores ($\mathcal{P}$) for all LSNe Ia, as well as separately for doubles and quads, shown across four panels corresponding to the results after the 1st, 3rd, 7th, and 14th observation epochs. It is evident that quads generally receive higher scores than doubles, indicating that they are, on average, easier to recover.}
    \label{fig:scores_db_qd}
\end{figure*}
\begin{figure*}
    \centering
    \subfigure[]{
        \includegraphics[width=0.485\textwidth]{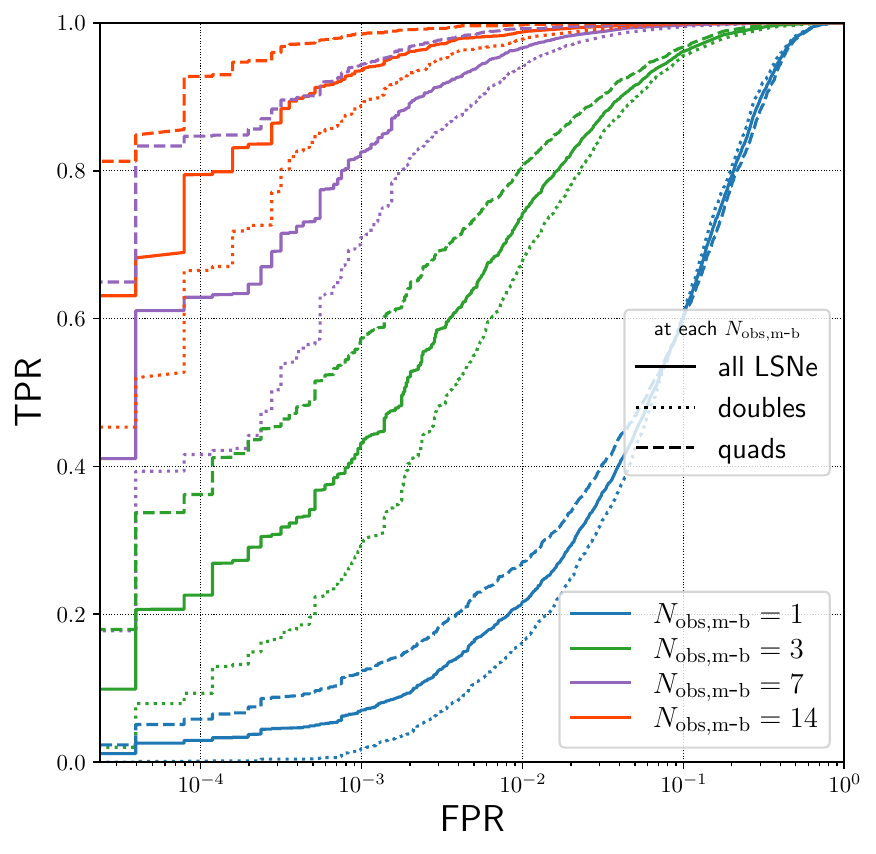}
        \label{fig:roc_dbqd}
    }
    \subfigure[]{
        \includegraphics[width=0.485\textwidth]{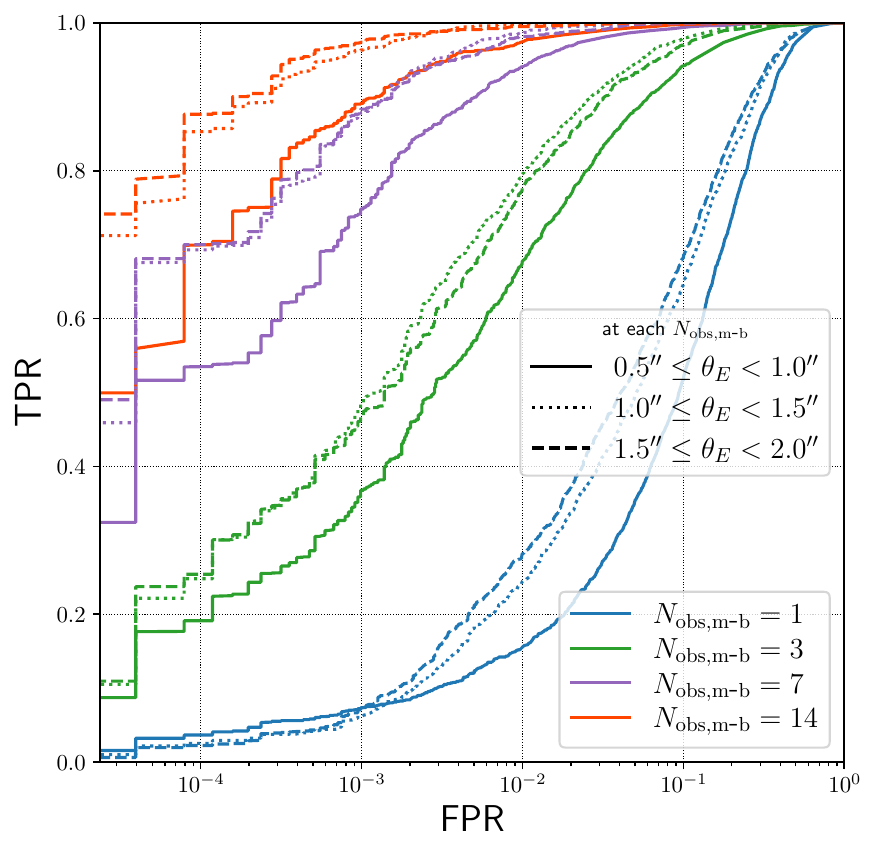}
        \label{fig:roc_re}
    }
    \caption{Comparison of classification performance across different sub-classes of LSNe Ia systems using ROC curves. The left panel compares doubles and quads, while the right panel shows results for systems grouped by angular separation, highlighting the impact of image separation. Results at the 1st, 3rd, 7th, and 14th observation epochs are shown in blue, green, purple, and orange, respectively, in both panels. In the left panel, solid, dotted, and dashed curves of a given color represent all LSNe Ia, doubles, and quads, respectively; in the right panel, the same line styles denote the three $\thE$ bins at the corresponding observation epoch.}
    \label{fig:roc_dbqd_re}
\end{figure*}

\begin{figure*}
    \centering
    \includegraphics[width=\textwidth]{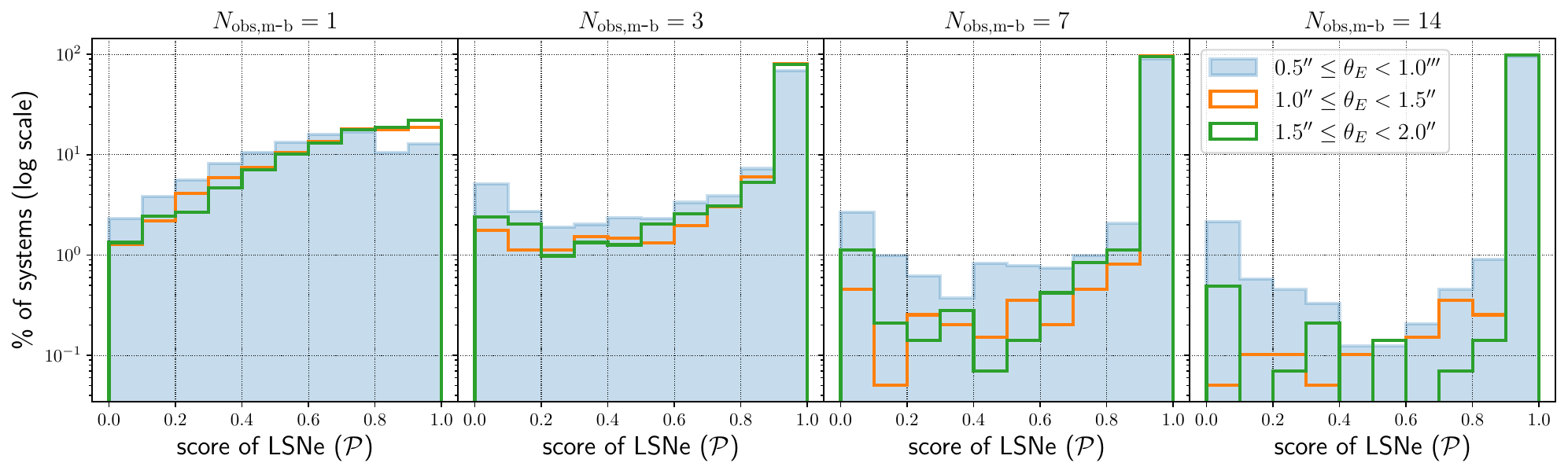}
    \caption{ To investigate the impact of image separation on classification, we divide the LSNe Ia systems into three bins based on Einstein radius: $0.5 \arcsec \leq \thE < 1.0 \arcsec$, $1.0\arcsec \leq \thE < 1.5\arcsec$, and $1.5\arcsec \leq \thE < 2.0\arcsec$. The four panels show the distributions of predicted scores ($\mathcal{P}$) for these three subsets after the 1st, 3rd, 7th, and 14th observation epochs, respectively.}
    \label{fig:scores_re}
\end{figure*}

\subsection{Doubles versus quads}
We first assess the classification performance between doubles and quads among the LSNe Ia samples. Figure~\ref{fig:scores_db_qd} presents the distribution of predicted scores ($\mathcal{P}$) for all LSNe Ia, as well as separately for doubles and quads, in four panels corresponding to results after the 1st, 3rd, 7th, and 14th epochs of observation. The left panel of Figure \ref{fig:roc_dbqd_re} displays the corresponding ROC curves -- different colors represent the four observation epochs, while solid, dotted, and dashed curves (within each color) denote all mock LSNe Ia, doubles, and quads, respectively. 

It is clearly evident from Figure~\ref{fig:scores_db_qd} that quads generally receive higher scores than doubles. This is expected, as quads with four images exhibit more profound spatial features compared to doubles with only two. Additionally, quads typically have shorter time delays between images, causing multiple images to appear closer together in time. Together, these factors result in stronger spatial and temporal correlations in quads, making them easier for the model to classify correctly. The relatively easier recoverability of quads compared to doubles is also evident in the ROC curves shown in the left panel of Figure \ref{fig:roc_dbqd_re}.

\subsection{Separation of multiple SN images}
In Figure~\ref{fig:scores_re}, we examine how the classification accuracy varies with the image separation of LSNe Ia. To do this, we divide the LSNe Ia samples into three bins based on their Einstein radius: $0.5 \arcsec \leq \thE < 1.0 \arcsec $, $1.0\arcsec  \leq \thE < 1.5 \arcsec$, and $1.5\arcsec  \leq \thE < 2.0 \arcsec$. The four panels of Figure~\ref{fig:scores_re} show the distributions of predicted scores ($\mathcal{P}$) for these three subsets after the 1st, 3rd, 7th, and 14th observation epochs, respectively. The right panel of Figure \ref{fig:roc_dbqd_re} presents the corresponding ROC curves -- colors indicate the observation epoch, while solid, dotted, and dashed lines represent the three $\thE$ bins at each $\Ne$.

We find that systems with lower separations ($0.5 \arcsec  \leq \thE < 1.0 \arcsec $) generally receive lower scores compared to systems with larger separations, and are therefore relatively harder to recover as evident from the ROC curves in the right panel of Figure \ref{fig:roc_dbqd_re}. This is also expected, as smaller separations increase the chance of multiple images blending together due to seeing/PSF effects, or merging with the light from the lens galaxy, which makes them harder to detect and classify confidently.

Interestingly, there is no significant difference in score distributions between the intermediate and large separation bins ($1.0 \arcsec \leq \thE < 1.5\arcsec$ and $1.5 \arcsec \leq \thE < 2.0 \arcsec$). A plausible explanation is that, although higher separation generally makes the system easier to resolve, it can also come with challenges: such systems often correspond to higher source redshifts, leading to fainter images, and very large separations can sometimes place images far enough from the lens to be affected by nearby bright objects in the field. These competing effects may balance out, resulting in comparable classification performance for the two higher-separation bins as evident in Figure \ref{fig:scores_re} and the right panel of Figure \ref{fig:roc_dbqd_re}.

\subsection{Time-delay distributions of high-score mock lenses}
\begin{figure*}
    \centering
    \includegraphics[width=0.485\textwidth]{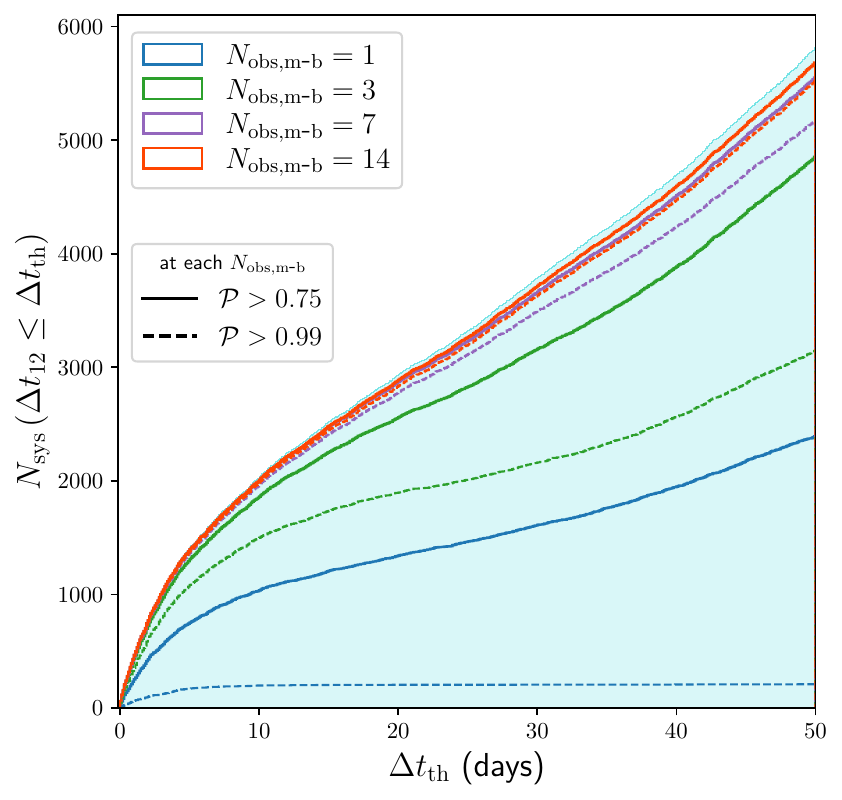}
    \includegraphics[width=0.485\textwidth]{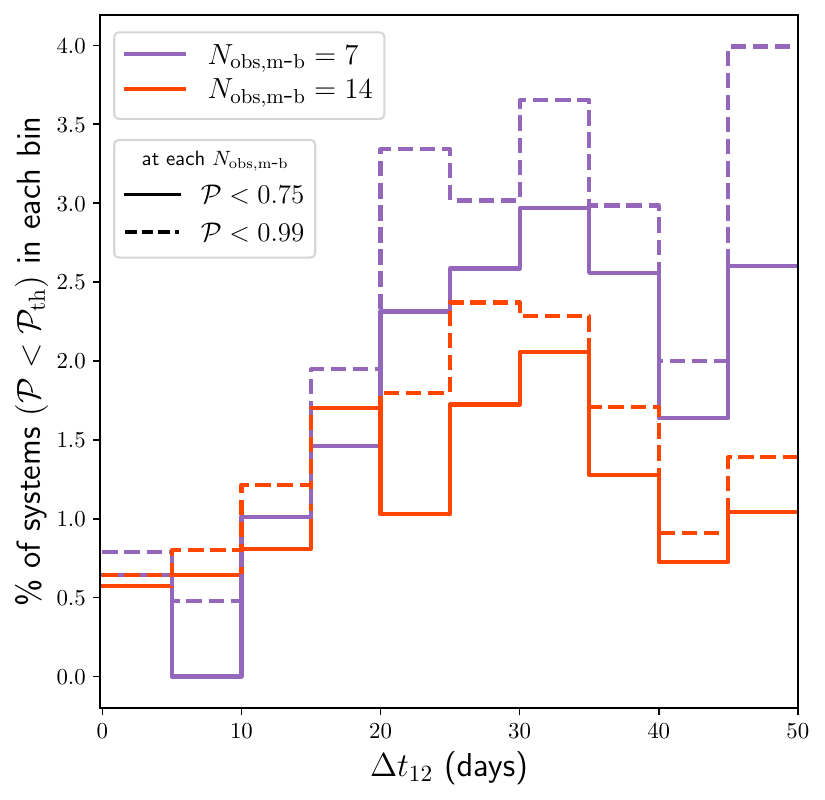}
    \caption{Distribution of the time delay between the first and second arriving images ($\Delta t_{12}$) for mock lensed systems. The left panel shows the cumulative distribution, i.e., the number of systems with $\Delta t_{12}$ smaller than a threshold $\Delta t_{\rm th}$, plotted along the x-axis. The filled cyan histogram represents the full test sample of 5,811 systems. Blue, green, purple, and red curves correspond to the distributions after the 1st, 3rd, 7th, and 14th observations, respectively. For these curves, solid and dashed line-styles indicate the subsets with scores $\mathcal{P}>0.75$ and $\mathcal{P}>0.99$. The right panel presents the $\Delta t_{12}$ distributions for lensed systems with scores below those thresholds, i.e., with $\mathcal{P}<0.75$ and $\mathcal{P}<0.99$, using solid and dashed curves, respectively. The y-axis gives the fraction of such systems in each $\Delta t_{12}$ bin as a percentage. Purple and red curves again correspond to results after the 7th and 14th observations.}
    \label{fig:TD_dist}
\end{figure*}

The score threshold ($\mathcal{P}{\rm th}$) corresponding to a given FPR varies with the observation epoch ($\Ne$) as the scores of individual samples evolve. In this initial work, we do not fix a detection threshold, since doing so would require careful calibration of the false alarm rate, which depends on the true proportions of positive and negative samples. Our training set will also be refined in future work. At this stage, our goal is to demonstrate the method and explore how model architecture and data preprocessing affect the performance, rather than to benchmark it under fully realistic conditions. Nevertheless, we examine the time-delay distribution of mock lensed systems receiving high scores in this section, focusing on subsets defined by thresholds $\mathcal{P}{\rm th}=0.75$ and $0.99$.

The left panel of Figure \ref{fig:TD_dist} shows the cumulative distribution of the time delay between first and second arriving images, $\Delta t_{12}$. The filled cyan distribution represents the full test sample of $5811$ mock lensed systems. The blue, green, purple, and red curves correspond to the distributions after the 1st, 3rd, 7th, and 14th observations, respectively. For these colored curves, the solid and dashed line-styles indicate systems with $\mathcal{P}>0.75$ and $\mathcal{P}>0.99$, highlighting subsets of high-score mock lensed systems.

As expected, systems with smaller $\Delta t_{12}$ tend to receive high scores earlier, that is, after fewer observations (i.e. at lower $\Ne$). The arrival of the second image generally strengthens the overall signal, both spatially and temporally, while also making these systems more distinct from normal SNe Ia in LRGs. In addition, the gap between the dashed and solid curves narrows as more epochs are observed, indicating that the model becomes increasingly confident with additional data, consistent with Figure \ref{fig:scores}.

On the other hand, the right panel of Figure \ref{fig:TD_dist} shows the distribution of $\Delta t_{12}$ for systems with scores below these thresholds, i.e., $\mathcal{P}<0.75$ and $\mathcal{P}<0.99$, using solid and dashed line styles. Purple and red curves correspond to results after the 7th and 14th observations, highlighting the distributions at later stages of the time series. The distributions are normalized, with the y-axis showing the fraction (in percentage) of systems in each $\Delta t_{12}$ bin receiving scores below the thresholds. As expected, systems with larger $\Delta t_{12}$ initially tend to receive lower scores until the second image is observed, as indicated by their higher representation in the larger $\Delta t_{12}$ bins.

By construction of our time series, the second image always appears by the final observation ($\Ne=14$), see Appendix~\ref{app:sim_pipe}. As a result, the higher $\Delta t_{12}$ bins no longer show an excess at $\Ne=14$. In fact, systems with larger $\Delta t_{12}$ often have wider angular separations and more distinct lensing features, making them easier to recognize once the second image is observed.

A potential complication arises for high–$\Delta t_{12}$ systems when the second image, being widely separated from the first, is missed during its bright phase due to sparse time sampling, as in our HSC-based simulations. In such cases, while many systems still achieve high scores, some may be misidentified. This issue is expected to be far less significant for LSST, which will provide a much denser cadence than HSC.

\end{document}